\documentclass[sigconf]{acmart}

\AtBeginDocument{%
  \providecommand\BibTeX{{%
    \normalfont B\kern-0.5em{\scshape i\kern-0.25em b}\kern-0.8em\TeX}}}

\usepackage{subfigure}
\usepackage{graphicx}
\usepackage{multirow}
\usepackage{algorithm}
\usepackage{algorithmicx}
\usepackage{booktabs}
\usepackage{amsmath}

\usepackage{algpseudocode}
\usepackage{url} 

\begin{document}

\title{CHIEF: Clustering with Higher-order Motifs in Big Networks}


\author{Feng Xia}
\affiliation{%
\department{School of Engineering, IT and Physical Sciences}
\institution{Federation University Australia}
\city{Ballarat}
\postcode{VIC 3353}
\country{Australia}}
\email{f.xia@ieee.org}

\author{Shuo Yu}
\affiliation{%
\department{School of Software}
\institution{Dalian University of Technology}
\city{Dalian}
\postcode{116620}
\country{China}}
\email{y\_shuo@outlook.com}

\author{Chengfei Liu}
\affiliation{%
\department{Department of Computer Science and Software Engineering}
\institution{Swinburne University of Technology}
\city{Melbourne}
\postcode{VIC 3122}
\country{Australia}}
\email{cliu@swin.edu.au}

\author{Jianxin Li}
\affiliation{%
\department{School of IT}
\institution{Deakin University}
\city{Melbourne}
\postcode{VIC 3125}
\country{Australia}}
\email{jianxin.li@deakin.edu.au}

\author{Ivan Lee}
\affiliation{%
\department{STEM}
\institution{University of South Australia}
\city{Adelaide}
\postcode{SA 5001}
\country{Australia}}
\email{ivan.lee@unisa.edu.au}

\renewcommand{\shortauthors}{Xia et al.}

\begin{abstract}
 Clustering a group of vertices in networks facilitates applications across different domains, such as social computing and Internet of Things. However, challenges arises for clustering networks with increased scale. This paper proposes a solution which consists of two motif clustering techniques: standard acceleration CHIEF-ST and approximate acceleration CHIEF-AP. Both algorithms first find the maximal $k$-edge-connected subgraphs within the target networks to lower the network scale, then employ higher-order motifs in clustering. In the first procedure, we propose to lower the network scale by optimizing the network structure with maximal $k$-edge-connected subgraphs. For CHIEF-ST, we illustrate that all target motifs will be kept after this procedure when the minimum node degree of the target motif is equal or greater than $k$. For CHIEF-AP, we prove that the eigenvalues of the adjacency matrix and the Laplacian matrix are relatively stable after this step. That is, CHIEF-ST has no influence on motif clustering, whereas CHIEF-AP introduces limited yet acceptable impact. In the second procedure, we employ higher-order motifs, i.e., heterogeneous four-node motifs clustering in higher-order dense networks. The contributions of CHIEF are two-fold: (1) improved efficiency of motif clustering for big networks; (2) verification of higher-order motif significance. The proposed solutions are found to outperform baseline approaches according to experiments on real and synthetic networks, which demonstrates CHIEF's strength in large network analysis. Meanwhile, higher-order motifs are proved to perform better than traditional triangle motifs in clustering.
\end{abstract}





\maketitle

\section{Introduction}
\label{sec:introduction}
Networks are everywhere. As an effective way to represent complex connections among entities, networks are widely used in various domains such as social computing, Internet of Things, transportation, and bioinformatics. In particular, networks have become an indispensible approach to the composition, management, and delivery of smart services in social computing applications~\cite{camacho2020four,ranjan2020tsc}. With the rapid development of modern information technology like Internet of Things, recent years have been witnessing the emergence of large scale networks accompanied with big data~\cite{bedru2020csr,yu2019access}. Such progress increases the scale and complexity of the underlying big networks, subsequently raises unprecedented challenges in data analytics, including clustering, anomaly detection, association-rule mining, and prediction.

In social computing applications, for instance, nodes and relations are organized under certain rules to form clusters. Clustering algorithms have been regarded as powerful network mining tools, and various large-scale computing methods have been proposed for analyzing big network data~\cite{li2017motif}. Some prior studies investigate network connectivity within the cohesive graph substructures to present solutions for clustering. Leskovec \emph{et al.}~\cite{leskovec2007graph} and Akrida \emph{et al.}~\cite{akridaMSZ20} showed that network density generally increases superlinearly over time. Although real social networks have increase dense grow denser over time, they are still relative sparse from the aspect of data analytics. Benson \emph{et al.}~\cite{benson2016higher} proposed an effective approach to identify clusters of network motifs in networks, which offers a new insight into the organization of social networks.

However, the scales of social networks are in rapid expansion, which makes the computational analysis intractable~\cite{yin2017local}. Motifs are needed to be firstly detected, thus consuming much computational time when implementing motif clustering methods. Therefore, a major challenge of network research lies on the reduction of network complexities while improving the efficiency of motif clustering. Consequently, network topological structures appear to be particularly important in the context of big networks. To address this problem, it is essential to understand the fundamental structures of social networks, especially when big data have already enlarged the scale~\cite{mora2019team,kotsios2019analysis}. Therefore, we aim to develop a dimensionality reduction algorithm with strong applicability for motif clustering in this work. Meanwhile, theoretical proof as well as experimental verification are both required to expand application scenarios of the proposed method.

In this paper, we propose a clustering algorithm CHIEF (Clustering with HIgher-ordEr motiFs) to address the above mentioned challenge in big (social) networks. CHIEF mainly contains two procedures: (1) finding maximal $k$-edge-connected subgraphs (abbreviated as $k$-connected subgraphs for the rest of this paper) in the target network; (2) employing higher-order motifs for clustering. Specifically, CHIEF contains two approaches: CHIEF-ST and CHIEF-AP. CHIEF-ST is the standard acceleration, which improves the algorithmic efficiency in large-scale networks without compromising the motif clustering results. CHIEF-AP is an approximate acceleration approach, which retains the eigenvalues of the adjacency matrix and the Laplace matrix meanwhile at the highest degree. CHIEF is novel and innovative in the following aspects.

\begin{itemize}
  \item \textbf{High algorithmic efficiency:} CHIEF employs maximal $k$-connected subgraphs finding approach, which yields higher efficiency than the baseline approaches. We conduct experiments on 18 top conferences and journals networks and 5 different synthetic social networks. Experimental results show that CHIEF is the most efficient algorithm as compared with the baseline methods.
  \item \textbf{Optional acceleration approaches:} CHIEF includes two acceleration approaches: CHIEF-ST and CHIEF-AP. CHIEF-ST provides an accuracy acceleration whereas CHIEF-AP provides an approximate optimal approach. Theoretically, both approaches can accelerate motif clustering, especially for scenarios of extremely large networks.
  \item \textbf{Higher-order motifs:} CHIEF employs heterogeneous higher-order motifs. Compared with triangle motifs, CHIEF achieves better clustering results using heterogenous four node motifs.
\end{itemize}

We run CHIEF-AP with different $k$ on two large-scale academic networks as well as on five synthetic networks. The two large-scale academic networks cover two different disciplines, including the computer science data from Microsoft Academic Graph (MAG) and the entire data from American Physical Society (APS). To employ larger size motifs in clustering, we use heterogeneous four-node motifs to explore higher order organizations. To examine the effectiveness, we calculate the CII (Collaboration Intensity Index)~\cite{yu2017team} of vertices within each identified cluster. We find that four-node motifs cluster performs better than heterogeneous triangle motifs cluster under certain circumstances. This illustrates that higher-order motifs may be more applicable than triangle motifs due to the degree of network density. We generate five synthetic networks in different scales, and compare the proposed solution against two baseline algorithms, the MovCut algorithm~\cite{jeub2014think} and the motif clustering algorithm~\cite{benson2016higher}.

The rest of the paper is organized as follows. Related works are introduced in the next section. Section III gives the preliminaries, including definitions of network motifs and $k$-connected subgraphs, and the formulation of the problem. Section IV illustrates the design of CHIEF. Section V presents the basic theories of standard acceleration and approximate acceleration. Section VI discusses experimental details, including data pre-processing, experiment design, baseline methods, and evaluation indices. Section VII analyzes the experimental results. Finally, Section VIII concludes this paper.

\section{Related Work}
Applied clustering algorithms for analyzing and categorizing larger number of applications have been under on-going development~\cite{Feng2014Exploiting,sahoo2019classification}. Previous studies developed a categorizing framework for clustering algorithms from a theoretical perspective. When selecting clustering algorithms, three criteria should be considered, including data size, data dimension, and presence of outliers. Variety is the ability of a clustering algorithm to handle different types of data, including numerical, categorical, and hierarchical data~\cite{song2019exploring}. Velocity refers to the processing speed of a clustering algorithm. Algorithmic complexity and execution time can be used in guiding the selection of a suitable clustering algorithm~\cite{wu2018energy}. Generally, clustering algorithms can be categorized into five types, including partitioning-based, hierarchical-based, density-based, grid-based, model-based, and motif-aware clustering algorithms~\cite{fahad2014survey}.

Partitioning-based clustering algorithms divide network data objects into many partitions. Each partition represents a different cluster, wherein clusters should meet two criteria: (1) each cluster should at least contain one object; and (2) each object must belong to exactly one group without any overlap~\cite{nie2019k}. Hierarchical-based clustering algorithms organize data in a hierarchical manner, which depends on the medium of proximity~\cite{alonso2020hierarchical}. In such algorithms, individual data is presented by leaf nodes. While propagating through the hierarchy, the initial cluster will be divided into several clusters gradually. Density-based clustering algorithms separate the object data set according to their regions of density, connectivity, and boundary. A cluster is usually defined as a connected dense subgraph or component, which will grow in any direction that the density leads to~\cite{lu2017effective}. Grid-based clustering algorithms divide the data objects into the form of grids. The accumulated grid-data makes grid-based clustering techniques independent of the number of data objects that employ a uniform grid to collect regional statistical data. Then cluster process will perform on the grid instead of the whole data set directly~\cite{hireche2020grid}. Model-based clustering algorithms assume that the data is generated by a mixture of underlying probability distributions. Model-based methods optimize the fitting functions, which represent the relationships between the given data set and some predefined mathematical models~\cite{liang2017remold}.

Differ from algorithms mentioned above, motif-aware clustering algorithms rely on network motifs, which have been recognized as the basic units of networks in network topological structure~\cite{ xia2019accesssurvey,fu2020local,yu2020csrsurvey,piao2021predicting}. Network motifs refer to those induced subgraphs with a significant high frequency corresponding to small patterns that appear in networks. In many kinds of networks such as academic networks and social networks, network motifs have been used to describe sorts of entities in various networks. The wide use of network motifs makes the methods of identifying and analyzing motif clusters especially important~\cite{lai2019distributed}. As one of the most fundamental units in networks, motifs can be used to uncover the basic building blocks of most networks. Motifs have been applied in solving many different kinds of problems in the application scenarios of social computing~\cite{zhang2020crowdsourcing,li2018novel}. Among these applications, motif-aware clustering is the most typical one.

It is proved that motif-based clustering algorithms always contain a procedure of motif detection, which leads to high computational complexity of such kind of algorithms~\cite{li2020community}. Before Benson \textit{et al.}~\cite{benson2016higher} proposed a fast motif-based clustering algorithm, the efficiency of motif-aware clustering algorithms has not been improved substantially. Afterwards, many studies work on this kind of clustering algorithms, e.g.,~\cite{wenhong2019local} and~\cite{pei2019edmot}. The main challenge of motif-aware clustering algorithms is how to enlarge the basic clustering unit from nodes to motifs in networks and reduce the computational complexity at the same time.

Current motif-aware clustering algorithms are generally based on spectral clustering, which aims at minimizing the motif conductance when clustering motifs. Yin \textit{et al.}~\cite{yin2017local} proposed a graph diffusion method called approximate personalized PageRank (APPR) that ``spreads" mass from seed set with minimal motif conductance to identify the cluster, and it naturally handles directed graphs clustering. Tsourakakis \textit{et al.}~\cite{tsourakakis2017scalable} focused on triangle motifs within graphs. The notion of triangle motif conductance is presented to appropriate random walk on the graph and to generalize graph expansion based on triangles motif. Zhou \textit{et al.}~\cite{zhou2017local} first computed the distribution of high-order random walk and clustered motifs, which starts with a seed vertex and iteratively explores its neighborhood until a small motif conductance is found. Huang \textit{et al.}~\cite{huang2018harmonic} designed a novel higher-order structure termed harmonic motif to integrate higher-order structural information, which is taken as the auxiliary information for discovering the multi-layer community.

Network connectivity occupies a decisive position as well as graph connectivity. However, most existing defined structures focus on node degrees instead of network connectivity within the cohesive graph substructures. To address the computational complexity issue in motif-aware clustering methods, we propose CHIEF in this work to first partition the network with $k$-connected graphs and then conduct motif clustering.

\section{Preliminaries and Problem Formulation}

This section introduces network motifs and $k$-edge-connected subgraphs in networks, and gives the formal definition of the research problem.

\subsection{Network Motifs}
\label{sec:network motifs}
Network motifs appear repeatedly among various networks, which reflect underlying functional properties within networks~\cite{yu2019rum}. Each of these motifs are defined by a particular pattern of interactions between graph nodes. Therefore, motifs have a notable significance in uncovering structural design principles of complex networks. Generally, motifs are organized in different sizes. Herein, we introduce two kinds of motifs in undirected networks.

\begin{itemize}
  \item A three-node motif, i.e., triangle motif, is a connected graph, which is comprised of three nodes and at least two edges that connect nodes.
  \item A four-node motif, e.g., bi-fan motif, is a connected graph, which is comprised of four nodes and at least three edges.
\end{itemize}

\begin{figure}[htbp]
  \centering
  \includegraphics[width=0.35\textwidth]{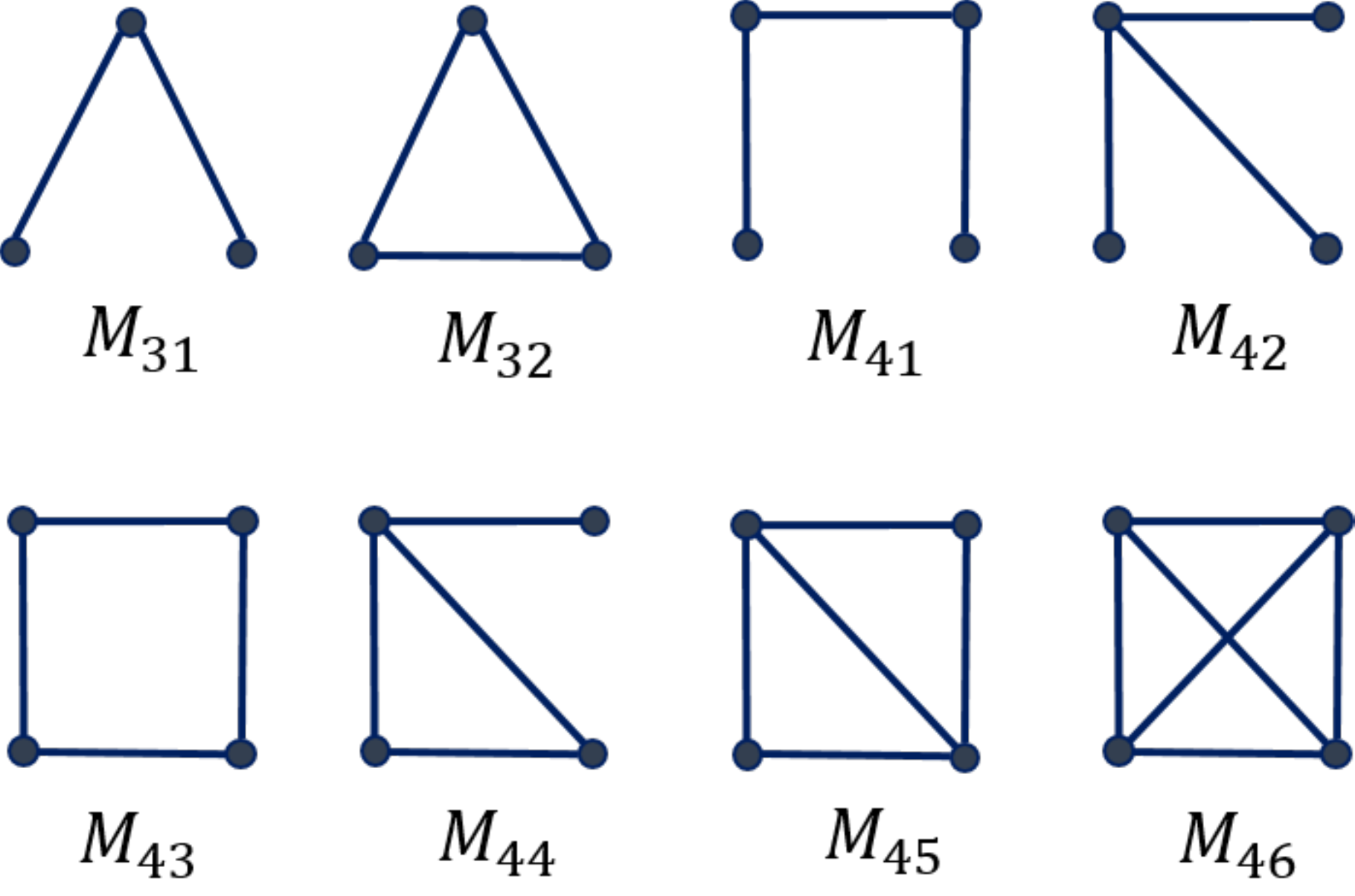}
  \caption{Heterogeneous structures of triangle motifs and four-node motifs in undirected graphs.}
  \label{fig:motif}
\end{figure}

Higher-order motifs are also called graphlets, subgraphs, or components, etc~\cite{piao2021predicting}. Some researchers distinguish these concepts depending on their scale. In this paper, we use motif to avoid ambiguity. The structure of the motif generally becomes more complicated when the number of vertices increases. In CHIEF, both triangle and four-node motifs are used to identify clusters in different networks. Since there are several structural forms of motifs, we label each kind of heterogeneous motif in Figure.~\ref{fig:motif}. Since $M_{31}$ and $M_{41}$ are homogeneous with paths in undirected graphs, and these paths are generally of little significance in social networks. Therefore, these two kinds of motifs are removed in our experiments.

\subsection{$k$-connected Subgraphs}
Connectivity is a fundamental subject in graph theory. Graphs are used to represent relationships between entities in real world. Effective $k$-connected subgraph finding algorithms can be used to identify closely related individuals. However, as a fundamental subject in graph theory, checking graph connectivity is more time consuming than checking node degrees~\cite{sood2019towards,wang2020g}. Thus, in this paper, we use an efficient approach to discover all maximal $k$-connected subgraphs. Here, we model network data as a simple and undirected graph $G=(V,E)$, where $V$ is a set of nodes and $E$ is a set of edges.

\begin{itemize}
  \item A graph $G$ is $k$-connected graph refers to the connected graph that cannot be disconnected by removing less than $k$ edges~\cite{holberg1992decomposition}.
  \item A graph $G$ is maximal $k$-connected refers that the $k$-connected subgraphs are not contained in other $k$-connected subgraphs~\cite{zhou2012finding}.
\end{itemize}

\begin{figure}[htbp]
  \centering
  \includegraphics[width=0.5\textwidth]{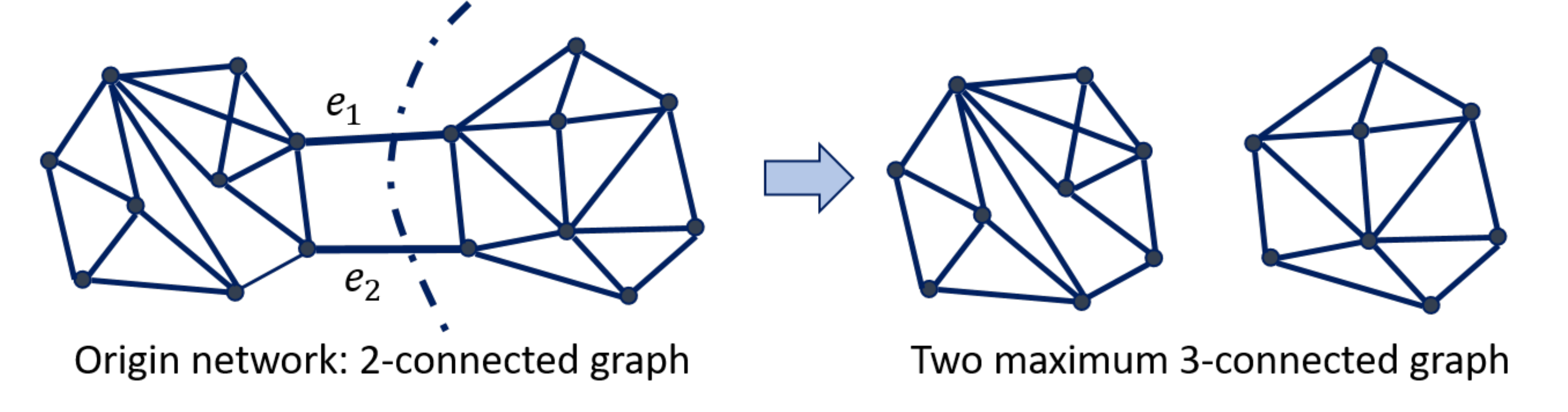}
  \caption{An example of $2$-connected subgraph. If we remove two edges $e_1$, $e_2$ of a 2-connected graph, then we get two maximal 3-connected subgraphs of original graph.}
  \label{fig:k}
\end{figure}

Figure.~\ref{fig:k} shows a 2-connected graph and its maximal 3-connected subgraphs, respectively. Under many circumstances, it is a better approach to use $k$-connected subgraphs to model vertex clusters. Several methods are proposed to speed up $k$-connected finding, edge reduction, and cut pruning. In this paper, we will use edge reduction methods to improve algorithm execution efficiency.

\subsection{Problem Formulation}
\textbf{Consider} a connected, undirected, large-scale network $G=(V,E)$, wherein $V$ is a set of vertices, $E$ is a set of edges, $n$ and $m$ are the number of vertices and edges in $G$, respectively. To a given network motif $M$, recognize a cluster of vertices, $S$, with the following two constraint conditions. (1) Nodes in $S$ should participate in many motifs with the same structure to $M$. (2) The set $S$ should avoid cutting $M$. This occurs when only a subset of the vertices from a motif are in the set $S$. \textbf{Find} a cluster that minimizes the ratio $\phi_M(S)$, wherein $\phi_M(S)$ is defined as shown in Equation~\eqref{eq:conductance}.

\begin{equation}
\label{eq:conductance}
\phi_M(S)={\rm cut}_M(S,\overline{S})/{\rm min}[{\rm Vol}_M(S),{\rm Vol}_M(\overline{S})],
\end{equation}
wherein, $\overline{S}$ refers to the complement of $S$, cut$_M(S,\overline{S})$ represents the number of motif $M$ with at least one node in $S$ and one in $\overline{S}$, and Vol$_M(S)$ is the number of nodes in $M$ that reside in $S$. $\phi_M(S)$ refers to the motif conductance of $S$ with respect to $M$.

\begin{figure}
  \centering
  \includegraphics[width=0.4\textwidth]{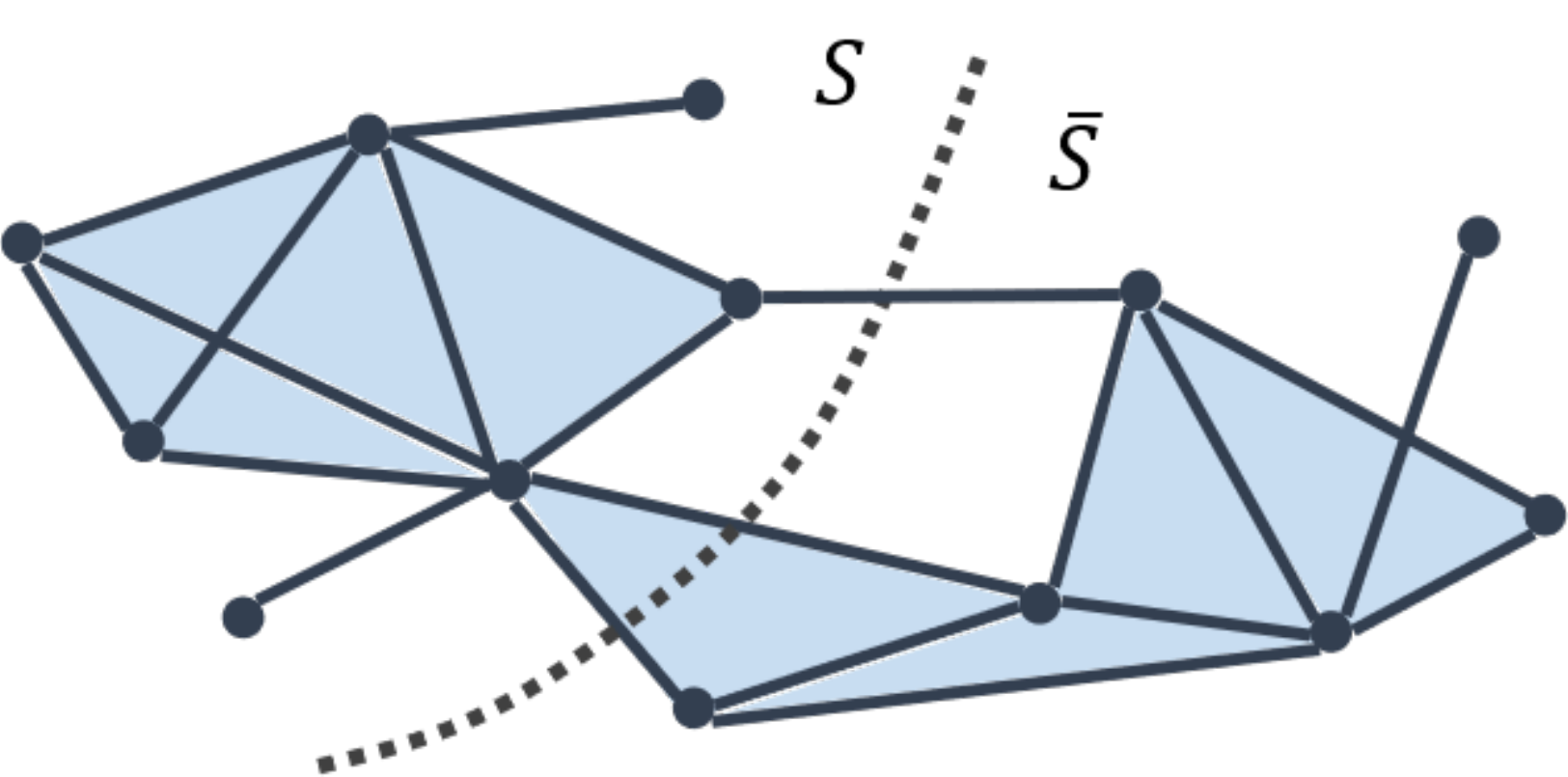}\\
  \caption{An example of motif conductance calculation.}\label{fig:exp-mc}
\end{figure}

Motif conductance is proposed to guide motif clustering. When clustering motifs, we aim at minimizing motif conductance. A lower motif conductance means that less motifs are cut, thus leading to less loss in motif clustering. We give a simple example to illustrate the calculation of motif conductance. If we choose the triangle motif as $M$, then there exist totally 9 instances of $M$. in Figure~\ref{fig:exp-mc}. Only one motif instance is cut in this example. Therefore, motif conductance is $1/9$.

Motif conductance minimization is an NP-hard problem~\cite{benson2016higher}. Therefore, the main difficulty that needs to be addressed in this problem is how to reduce computing complexity. To solve this problem, we first find maximal $k$-connected subgraphs in the network to reduce computing complexity from the perspective of network connectivity. Then we cluster higher-order motifs in the network.

\section{The Design of CHIEF}

This section introduces the CHIEF framework. The first part introduces the main idea of finding the maximal $k$-connected subgraphs. The second part introduces the procedure of higher-order motif clustering. The last part introduces the integral process of CHIEF.

\subsection{Finding $k$-connected Subgraphs}

Graph connectivity is usually used to measure the reliability of the network when the network topology structure is represented by graph. Graph connectivity is significant in many problems. In this problem, graph connectivity can be used to decide which edges or vertices should be removed. This process aims at disconnecting an originally connected graph with less loss. By convention, if a vertex is removed, then all edges joining it will be removed as well. However, the converse may not establish. Here, we introduce graph connectivity and vertex connectivity as follows.

To a given graph $G$, if removing a set of edges $E_0(G)$ can make $G$ unconnected, then $E_0(G)$ is a disconnecting set of edges of $G$. The smallest disconnecting set is called an edge cut set. Similarly, if removing a set of vertices $V_0(G)$ can make $G$ unconnected, then $V_0(G)$ is disconnecting set of vertices of $G$. The smallest disconnecting set is called a vertex cut set. The edge cut set or vertex cut set which contains the least number of edges or vertices is called \textbf{minimum-cut set}.

\begin{algorithm}[bp]
  \caption{SW minimum-cut algorithm}
  \begin{flushleft}
    \textbf{Input:}~a graph $G=(V,E)$;\\
    \textbf{Output:}~the minimum-cut edge set $E_{cut}$;\\
  \end{flushleft}
  \begin{algorithmic}[1]
    \State initialize $E_{cut}=E$;
    \State initialize $A$=$\{$a randomly chosen vertex $V\}$;
    \While{$A \neq V$}\\
    \State add the most tightly connected vertex of $V$ into $A$;
    \EndWhile
    \State merging the last two vertices added into $A$;
    \State $E'_{cut}=$edges which connect the last added vertex to other vertices;
    \While{$|V|>1$}
    \If{$|E_{cut}|>|E'_{cut}|$}
    \State $E_{cut}=E'_{cut}$;
    \EndIf
    \EndWhile\\
    \Return~$E_{cut}$
  \end{algorithmic}
  \label{alg:SW}
\end{algorithm}

\subsubsection{Minimum-cut Based Approach}

In this paper, we use a reasonably efficient and low-complexity Stoer-Wagner (SW) algorithm, which is introduced briefly in this section. The SW algorithm was proposed by Mechthild Stoer and Frank Wagner~\cite{stoer1997jacm}. In Algorithm~\ref{alg:SW}, it first selects a seed vertex, and then repeatedly takes out other vertices from $V$ to merge with the seed vertex. In each loop, the vertex which owns the highest connectivity with the seed set is selected and the selected vertex should be removed from $V$. In Step 7, $E'_{cut}$ contains the edges which connect the last added vertex to other vertices.

The SW algorithm can be used to accelerate maximal $k$-connected subgraph finding. This is because that this algorithm solves the minimum-cut problem by using $|V|-1$ minimum $s-t$ cut computations. A $s-t$ cut is defined as a partition of $V$, which separates vertices $s$ and $t$ into two different components. $|E_{cut}|$ is calculated as cut-off threshold in maximal $k$-connected subgraph finding, which will be explain in Section~\ref{sec:maximal}.

\subsubsection{Finding Maximal $k$-connected Subgraphs}
\label{sec:maximal}
To reduce the computational complexity, we employ the arithmetic idea from the $k$-connected subgraph finding algorithm, which is first proposed in~\cite{zhou2012finding}. Though the algorithm framework proposed in~\cite{benson2016higher} has been proved to identify higher-order structure in high efficiency, the computational complexity still can be reduced by partitioning large-scale networks. For a $k$-connected subgraph $G$, we can introduce some precomputed maximal $k'$-connected subgraphs as bases to explore $k$-connected subgraphs.

\begin{algorithm}[hbp]
  \caption{Finding maximal $k$-connected subgraphs}
  \begin{flushleft}
    \textbf{Input:} a graph $G=(V,E)$, $k$;\\
    \textbf{Output:} a set of maximal $k$-connected subgraphs $R$;
  \end{flushleft}
  \begin{algorithmic}[1]
    \State initialize $R_0=\{G\}$;
    \For {each subgraph $G_a(V_a,E_a)$($|V_a|\neq1$) in $R_0$}
    \State find SW minimum-cut of $G_a$ with cut set $E_{cut}$;
    \If {$|E_{cut}|<k$}
    \State cut $G_a$ into $G_{a_1}$, $G_{a_2}$;
    \State $R_0 = R_0 \cup \{G_{a_1},G_{a_2}\}-\{G_a\}$;
    \Else
    \State $R=R\cup\{G_a\}$;
    \EndIf
    \EndFor\\
    \Return $R$
  \end{algorithmic}
  \label{alg:k}
\end{algorithm}

If a maximal $k'$-connected subgraph $G'$ has $k' \geq k$, then it can be referred that $G'$ is also $k$-connected. However, $G'$ may not be maximal at $k$. If all maximal $k'$-connected subgraphs that $k'>k$, then these $k'$-connected subgraphs can be regarded as a whole subgraph according to Lemma 1. Compared with the original graph, the size of the resulting graph is then significantly reduced.

If a maximal $k'$-connected subgraph $G'$ has $k'<k$, then $G'$ may contain induced $k$-connected subgraphs. Therefore, we can find all maximal $k$-connected subgraphs from $G'$ directly. If all maximal $k'$-connected subgraphs if obtained (when $k'<k$), we can start searching from these $k'$-connected subgraphs without resorting to the original graph due to the fact that a $k$-connected subgraph is also $k'$-connected.

In this work, we first use the SW algorithm to calculate the values of $k$ for each subgraph. Then we use connectivity threshold $k$ to find a set of maximal $k$-connected subgraphs. The logical structure of the algorithm is shown in Algorithm~\ref{alg:k}. There are two facts worth mentioning, which are illustrated as follows.
\begin{itemize}
  \item When implementing global minimum cut algorithm, the lowest value among the $|V|-1$ $s-t$ cuts. That is, if any $|E_{cut}|$ among all these $|V|-1$ cuts having $|E_{cut}|<k$, we can stop finding other $s-t$ cuts on $G_a$ and then separate $G_a$ by $E_{cut}$. This mechanism provides an early-stop property for our method. Besides, the SW algorithm has a much lower computational complexity as compared with those flow-based algorithms. That is the reason why the SW algorithm is chosen.
  \item The size of resulting graph is significantly reduced when a maximal $k'$-connected subgraph $G'$ has $k'\geq{k}$. Under some circumstances, some neighbor vertices of $G_a$ may be directly removed by our method. Here, a neighbor vertex refers to a vertex not in $G_a$ but is incident on an edge which has the other end in $G_a$. To avoid loss in maximal $k$-connected subgraph finding, the expanding process can be introduced in this procedure. The expanding process will first expand $G_a$, and then contract it.
  The main idea of subgraph expanding is to let $G_a$ absorb neighbor vertices while keeping $G_a$ connected, end the absorbing process when $G_a$ is not growing fast. However, the expanding process will take much more time than directly adding $G_a$ to $R$ in the step 8 of Algorithm~\ref{alg:k}.
\end{itemize}
\begin{algorithm}[tp]
  \caption{Motif Clustering}
  \begin{flushleft}
    \textbf{Input:}a set of maximal $k$-connected subgraphs $R$, motif $M_{ij}$;\\
    \textbf{Output:}a set of subgraphs $S$;\\
  \end{flushleft}
  \begin{algorithmic}[1]
    \State initialize $S=\emptyset$;
    \For {$k'$-connected subgraph $S_{k'}$ in $R$}
    \State calculate adjacent matrix $A_{M_{ij}}$ and diagonal matrix $D_{M_{ij}}$;
    \State get the re-weighted graph $S_{k'}$;
    \State calculate Laplacian matrix $L_M$;
    \State $\sigma_i$=the index of $i_{\rm {th}}$ smallest value of $D^{-\frac{1}{2}}_Mz$;
    \State calculate the conductance of each $\sigma_i$;
    \State cut $S_{k'} = S_1 \cup \overline{S_1}$ by min$\phi_M(S_{k'})$;
    \State add arg min$\{|S|,|\overline{S}|\}$ to $R_0$;
    \EndFor\\
    \Return $S$
  \end{algorithmic}
  \label{alg:motif}
\end{algorithm}

\subsection{Clustering with Higher-order Motifs}
Benson \emph{et al.}~\cite{benson2016higher} use higher-order network structures to identify clusters in big networks. To a given network $G$ with $n$ vertices, and certain motif type of interest $M$, construct the motif adjacency matrix $W_M$, wherein, $(W_M)_{ij}$ refers to the co-occurrence counts of nodes $i$ and $j$ in the motif $M$. Here, the algorithm firstly calculate the spectral ordering $\sigma$ of the vertices of the nodes from normalized motif Laplacian matrix constructed via $W_M$, followed by finding the subset of $\sigma$ with smallest motif conductance, as illustrated in the following two steps:

\subsubsection{Spectral Ordering $\sigma$}
To a given graph or network $G$ having $n$ vertices, the Laplacian matrix $L=[L_{ij}]$, also called admittance matrix or Kirchhoff matrix, is defined as follows.

\begin{equation}
\label{eq:l}
L_{ij} =
\begin{cases}
k_i   &\rm if \ i=j    \\
-1    &\rm if \  i \neq j,\  v_i \  {\rm adjacent} \  v_j,  \\
0     &\rm otherwise
\end{cases}
\end{equation}

wherein, $k_i$ is the vertex degree of $v_i$, $i,j=1,...,n$.

To a certain motif $M$, we use $L_M$ to represent the normalized Laplacian matrix in this work. The normalized Laplacian matrix is calculated according to the adjacency matrix $W_M$ and the diagonal matrix $D_M$. $L_M$ is defined as follows.

\begin{equation}
\label{eq:nl}
L_M=I-D^{-\frac{1}{2}}_MW_MD^{-\frac{1}{2}}_M,
\end{equation}

It is worth mentioning that $L_M$ is always symmetrical and semipositive definite. Besides, the following theorem establishes.\\
\textbf{Lemma 1:} The Laplacian matrix of a given graph $G$ is denoted as $L$. $p$ is the number of disjoint subgraphs in the graph (or components in the network). Then $L$ has real eigenvalues $\lambda_1\leq\lambda_2\leq...\leq\lambda_p$. Therefore, if the graph (or network) is connected, which also means that $p=1$, then $0=\lambda_1<\lambda_2\leq...\leq\lambda_n$. Herein, $\lambda_2$ is called the spectral gap of the corresponding Laplacian matrix and the ordering of $\{\lambda_i\}$ is called spectral ordering~\cite{brouwer2011spectra}.

\subsubsection{Motif Conductance}
Motif conductance is used as the threshold value in this step. To minimize $\phi_M(S)$ and identify motif clusters, we use an optimization method that can find near-optimal clusters. The clustering methodology is extended based on the eigenvalues and eigenvectors of matrices, to account for higher-order structures. Specifically, a cluster $S$ is identified when $\phi_M(S)\leq2\sqrt{\phi*_M}$. Wherein, $\phi*_M(S)=\min_{T\subset{V}}(T)$ is the most proper motif conductance over all sets $T$. The logical structure of the algorithm in this procedure is shown in Algorithm~\ref{alg:motif}.

\subsection{Integral Process of CHIEF}
We give the integral process of CHIEF in Algorithm~\ref{alg:combined}. In CHIEF, we first find all of the maximal $k$-connected subgraphs to partition networks, which assists speeding up higher-order motifs based clustering algorithm in Algorithm~\ref{alg:motif}. Then we implement the clustering approach.

\begin{algorithm}[bp]
  \caption{CHIEF}
  \begin{flushleft}
    \textbf{Input:} a graph $G=(V,E)$, motif $M_{ij}$, $k$;\\
    \textbf{Output:} a set of subgraphs $S$;\\
  \end{flushleft}
  \begin{algorithmic}[1]
    \State initialize $R_0$=$\{$precomputed $k'$-connected subgraphs;$\}$
    \State \textbf{remove} infrequent edges and nodes;
    \If {$deg(v)<k$}
    \State \textbf{remove} $v$;
    \EndIf
    \For {$k'$-connected subgraph $S_{k'}$ in $R_0$}
    \If  {$|V_{S_{k'}}|<k$}
    \State \textbf{remove} $S_{k'}$;
    \EndIf
    \If {$k'\geq k$}
    \State add $S_{k'}$ to $R$;
    \Else
    \State $E_{cut}$=MinimumCut($S_{k'}$);
    \State \textbf{remove} $E_{cut}$ from $S_{k'}$;
    \EndIf
    \EndFor
    \State $S$=MotifCluster($R$,$M_{ij}$);\\
    \Return $S$
  \end{algorithmic}
  \label{alg:combined}
\end{algorithm}

Algorithm~\ref{alg:combined} shows the whole process of CHIEF. Firstly, we initialize $R_0$ as the pre-computed $k'-$connected subgraphs. To reduce computing complexity and improve algorithm efficiency, we remove infrequent edges and nodes by degree in Step 2 to Step 5. For the $k'-$connected subgraphs in $R_0$, if $|V_{S_{k'}}|$, i.e., the vertices number of $S_{k'}$ is less than $k$, then remove $S_{k'}$. If not, add $S_{k'}$ to $R$ when $k'{\geq}k$. When $k'{\leq}k$, implement MinimunCut function on $S_{k'}$ and remove $E_{cut}$ from $S_{k'}$. Finally, implement MotifCluster function with parameters $R$ and $M_{ij}$. Let $S$ equal MotifCluster($R,M_{ij}$) and then return $S$.

\subsection{Input Analysis of CHIEF}
The main process of CHIEF have been introduced. Actually, the main procedures of CHIEF-ST and CHIEF-AP are the same, but they end up with different clustering results when input parameters are different. Based on the discussion in Section~\ref{sec:thechief}, we analyze the accuracy of CHIEF-ST and CHIEF-AP below.
\subsubsection{CHIEF-ST}
According to Theorem 1 in Section~\ref{sec:thechief}, when the network motif is $k$-connected, the first procedure will not affect the second procedure at all. Therefore, when input target motif and $k$ satisfy that the target motif is $k$-connected, CHIEF is an accuracy acceleration algorithm, i.e., CHIEF-ST.
\subsubsection{CHIEF-AP}
When the input target motif is not $k$-connected, the proposed algorithm, i.e., CHIEF-AP, achieves approximate optimal clustering results. Theorem 3 gives the boundaries of difference between minimal eigenvalues of $\mathcal{L}_k$ and that of $\mathcal{L}_G$, which indicates that the impact of first procedure on clustering result is acceptable.

\section{Theoretical Proof  for CHIEF}
\label{sec:thechief}
In this section, we discuss some theoretical proofs and analyze the accuracy of CHIEF. We will examine CHIEF-ST first and then CHIEF-AP.

The following lemma guarantees the consistence of $k$-connectivity in the contracted graph and the original graph, which has been proved in~\cite{zhou2012finding}.\\
\\
\textbf{Lemma 2:} Given a graph $G=(V,E)$, let $G_{k}=(V_{k},E_{k})$ be a $k$-connected subgraph of $G$, let $G'=(V', E')$ be the graph produced from $G$ by contracting $G_{k}$ into a vertex $v_{new}$. For any vertex, define $image(v) \in V'$ as:
\begin{enumerate}
  \item $image(v)=v_{new}$, if $v \in V_{k}$;
  \item $image(v)=v$, if $v \in V \backslash V_{k}$,
\end{enumerate}
then for any vertices $v_1, v_2 \in V$, $v_1,v_2$ are $k$-connected in $G$, if and only if either $image(v_1)=image(v_2)=v_{new}$ or $image(v_1)$ and $image(v_2)$ are $k$-connected in $G'$.

\subsection{CHIEF-ST}
Herein, we give theoretical proofs and analysis for CHIEF. We prove that if the network motif is $k$-connected, then the first procedure will not affect the second procedure.\\
\\
\textbf{Theorem 1:} For a target network $G_{tar}=(V_{tar}, E_{tar})$, $S_{k}=\{K_{1},\dots,K_{m}\}$ is a set containing all maximal $k$-connected subgraphs of $G_{tar}$. Suppose all target motifs are in the set $M_{tar}=\{M_1,\dots,M_n\}$, wherein $M_i=(V_{M_i},E_{M_i})$. Then $M_{tar}\subseteq{S_{k}}$ when ${M_i}$ is $k$-connected, $i=1,\dots,n$.\\
\\
\textbf{Proof:} 
If the cut set $E_{cut}\subseteq{E(M_i)}$ after $k$-connected subgraph finding procedure, then this must be because that $M_i$ is $k$-connected. Wherein, $i=1,\dots,n$. As we know that $degree(E_{cut})\leq{k}$, therefore the $k$-connected subgraph finding procedure cannot cut off any motif in $E(M_i)$. From this, we can infer that $G_{tar}$ cannot be cut off as well. After finding SW minimum-cut, it can be divided into following two cases.
\begin{enumerate}
  \item There exists $degree(E_{cut})=k_i<k-1$ and $E_{cut}\cap{E(M_i)}=\Phi, i=1,\dots,n$. Then the rest $k-1-k_i$ edges of $E_{cut}$ are from $E(M_i)$. These $k-1-k_i$ edges cannot be cut because that $M_i$ is $k$-connected. The minimal node degree of $M_i$ is no less than $k$, i.e., $degree(V_{M_i})\geq{k}$.
  \item When $E_{cut}\cap{E(M_i)}\neq\Phi, i=1,\dots,n$, the edges belong to $E_{cut}\cap{E(M_i)}$ should be deleted. Otherwise, this is not the minimum cut.
\end{enumerate}
To sum up, all $M_i$ will not be cut. This fact indicates that the first procedure of CHIEF cannot affect the second one at all. 
q.e.d.

Take $M_{46}$ in Figure~\ref{fig:motif} as an example, all of $M_{46}$ in the network will be maintained after finding maximal 3-connected subgraphs in the network. However, if we aim at finding maximal 4-connected subgraphs or 4 higher connected subgraphs, some of motifs isomorphic to $M_{46}$ may be broken.

From the perspective of probability, the nodes with lower degree are generally not contained in many motifs. That is, the node with low node degree is more difficult to form motif than the node with high node degree. Therefore, not many broken motifs exist when finding the maximal $k$-connected subgraphs.

The aforementioned analysis illustrates that when we find maximal $k$-connected subgraphs before motif clustering, we can choose $k$ according to the minimal node degree of target motif in order to accelerate motif clustering accurately.

\subsection{CHIEF-AP}

In order to accelerate the motif clustering algorithm in large-scale network, we employ the maximal $k$-connected subgraphs finding algorithm. Herein, we prove that the finding procedure is an approximate optimal approach.

Herein, we introduce a lemma below. This lemma illustrates an important conclusion between two Hermitian matrices~\cite{meyer2000matrix}.\\
\\
\textbf{Lemma 3:} Suppose $A,B\in{C^{n\times{n}}}$ are two Hermitian matrices and their eigenvalues are ranked in ascending order as shown below.
\begin{equation}
\centering
\begin{matrix}
\lambda_1(A)  &\leq&{\lambda_2(A)}  &\leq&\dots&\leq&{\lambda_n(A)},\\
\lambda_1(B)  &\leq&{\lambda_2(B)}  &\leq&\dots&\leq&{\lambda_n(B)},\\
\lambda_1(A+B)&\leq&{\lambda_2(A+B)}&\leq&\dots&\leq&{\lambda_n(A+B)}.
\end{matrix}
\end{equation}
then there exists
\begin{equation}
\left\{\begin{matrix}
\lambda_i(A)+\lambda_1(B)\\
\lambda_{i-1}(A)+\lambda_2(B)\\
\dots\\
\lambda_1(A)+\lambda_i(B)
\end{matrix}\right.\leq\lambda_i(A+B)\leq\left\{\begin{matrix}
\lambda_i(A)+\lambda_n(B)\\
\lambda_{i+1}(A)+\lambda_{n-1}(B)\\
\dots\\
\lambda_n(A)+\lambda_i(B)
\end{matrix}\right.
\end{equation}\\
\\
\textbf{Theorem 2:} To the given network $G=(V,E)$, let $A_G$ be the adjacency matrix of $G$. After finding maximal $k$-connected subgraphs in $G$, $G_k$ is the network that contains all maximal $k$-connected subgraphs and $G_{\widetilde{k}}$ is the rest part of $G$. Let the adjacency matrix of $G_k$ be $A_k$. Then the eigenvalues of above mentioned networks satisfy $|\lambda_{\min}(A_k)-\lambda_{\min}(A_G)|\leq{\delta}$, wherein, $\lambda_{\min}>0$, $\delta=\sqrt{\lambda_{\max}(A^T_{\widetilde{k}}A_{\widetilde{k}})}$~\cite{zhou2012finding}.\\
\\
\textbf{Proof:} It is evident that $A_k,A_{\widetilde{k}},A_G\in{C^{n\times{n}}}$ are Hermitian matrices, wherein the adjacency matrix of $G_{\widetilde{k}}$ be $A_{\widetilde{k}}(i,j)$. Therefore, the original adjacency matrix $A_G$, the  $A_G=A_k+A_{\widetilde{k}}$. According to Lemma 3, if the eigenvalues are ranked in the ascending order as shown below,
\begin{equation}
\centering
\begin{matrix}
\lambda_1(A_k)\leq{\lambda_2(A_k)}\leq\dots\leq{\lambda_n(A_k)},\\
\lambda_1(A_{\widetilde{k}})\leq{\lambda_2(A_{\widetilde{k}})}\leq\dots\leq{\lambda_n(A_{\widetilde{k}})},\\
\lambda_1(A_G)\leq{\lambda_2(A_G)}\leq\dots\leq{\lambda_n(A_G)}.
\end{matrix}
\end{equation}
then there exists
\begin{equation}
\left\{\begin{matrix}
\lambda_i(A_k)+\lambda_1(A_{\widetilde{k}})\\
\lambda_{i-1}(A_k)+\lambda_2(A_{\widetilde{k}})\\
\dots\\
\lambda_1(A_k)+\lambda_i(A_{\widetilde{k}})
\end{matrix}\right.\leq\lambda_i(A_G)\leq\left\{\begin{matrix}
\lambda_i(A_k)+\lambda_n(A_{\widetilde{k}})\\
\lambda_{i+1}(A_k)+\lambda_{n-1}(A_{\widetilde{k}})\\
\dots\\
\lambda_n(A_k)+\lambda_i(A_{\widetilde{k}})
\end{matrix}\right.
\end{equation}
It can be inferred that
\begin{equation}
\lambda_{\max}(A_{\widetilde{k}})\geq{\lambda_{\min}(A_G)-\lambda_{\min}(A_k)}\geq\lambda_{\min}(A_{\widetilde{k}})
\end{equation}
wherein, $i=1,2,\dots,n$.

Therefore, $|\lambda_{min}(A_G)-\lambda_{min}(A_k)|<\sqrt{\lambda_{max}(A^T_{\widetilde{k}}A_{\widetilde{k}})}$.

The theorem below discusses the perturbation of Laplacian matrix.

\textbf{Theorem 3:} To the given network $G=(V,E)$, let $\mathcal{L}_G$ be the Laplacian matrix of $G$. After finding maximal $k$-connected subgraphs in $G$, $G_k$ is the network that contains all maximal $k$-connected subgraphs and $G_{\widetilde{k}}$ is the rest part of $G$. Let the Laplacian matrix of $G_k$ be $\mathcal{L}_k$ and the adjacency matrix of $G_k$ be $A_k$. The diagonal matrix of $G$ is denoted as $D_G$. Then the following results are obtained.
\begin{equation}
\label{eq:geq}
\begin{split}
\lambda_{\min}(\mathcal{L}_k)&-\lambda_{\min}(\mathcal{L}_G)\geq  \\
&\lambda_{\max}(D^{-1}_G{A_G})-\lambda_{\max}(D^{-1}_k{A_G}) +\lambda_{\max}(D^{-1}_k{A_{\widetilde{k}}})
\end{split}
\end{equation}
\begin{equation}
\label{eq:leq}
\lambda_{\min}(\mathcal{L}_k)-\lambda_{\min}(\mathcal{L}_G)\leq\frac{\lambda_{\max}(A_G)}{\min(D_{G_{ii}})}-\lambda_{\max}(D^{-1}_kA_k)
\end{equation}
wherein, $D_G$ and $D_k$ refers to the diagonal matrix of $A_G$ and $A_k$, respectively. $D_{G_{ii}}$ refers to the diagonal elements of $G$'s diagonal matrix, which is defined as $D_{G_{ii}}=\sum^n_{j=1}{A(i,j)}$. $\delta$ is the spectral radius of $A_{\widetilde{k}}$.\\
\\
\textbf{Proof:} First we give the proof of Equation~\eqref{eq:geq}. According to the definition of Laplacian matrix, we can get that $\mathcal{L}_G=I-D^{-\frac{1}{2}}_G{A_G}D^{-\frac{1}{2}}_G$. Therefore, the eigenvalues satisfy $\lambda(\mathcal{L}_G)=1-\lambda(D^{-\frac{1}{2}}_G{A_G}D^{-\frac{1}{2}}_G)$. This indicates that when $\lambda(D^{-\frac{1}{2}}_G{A_G}D^{-\frac{1}{2}}_G)$ is at the maximum value, $\lambda(L_G)$ is at the minimum value. We denote the minimum value of $\lambda(L_G)$ as $\lambda_{\min}(L_G)$, which is
\begin{equation}
\label{eq:lg}
\lambda_{\min}(\mathcal{L}_G)=1-\lambda_{\max}(D^{-\frac{1}{2}}_G{A_G}D^{-\frac{1}{2}}_G)
\end{equation}
Since we have illustrated that $A_G=A_k+A_{\widetilde{k}}$. Therefore, the Laplacian matrix of $G_k$ satisfies Equation~\eqref{eq:lk}.
\begin{equation}
\label{eq:lk}
\begin{split}
\mathcal{L}_k = I-D^{-\frac{1}{2}}_k{A_k}D^{-\frac{1}{2}}_k =I-D^{-\frac{1}{2}}_k(A_G-A_{\widetilde{k}})D^{-\frac{1}{2}}_k
=I-F
\end{split}
\end{equation}
wherein, $F=D^{-\frac{1}{2}}_k{A_G}D^{-\frac{1}{2}}_k-D^{-\frac{1}{2}}_k{A_{\widetilde{k}}}D^{-\frac{1}{2}}_k$.
Therefore, there exists $\lambda(\mathcal{L}_k)=1-\lambda(F)$. According to the aforementioned discussion, Equation~\eqref{eq:1-f} establishes.
\begin{equation}
\label{eq:1-f}
\lambda_{\min}(\mathcal{L}_k)=1-\lambda_{max}(F)
\end{equation}
As it is known that the norm of difference between two matrices is less or equal to the absolute value of the difference between the norms of the two matrices. Then Equation~\eqref{eq:f} establishes.
\begin{equation}
\label{eq:f}
\lambda_{\max}(F)\leq|\lambda_{\max}(D^{-\frac{1}{2}}_k{A_G}D^{-\frac{1}{2}}_k)-\lambda_{\max}{D^{-\frac{1}{2}}_k{A_{\widetilde{k}}}D^{-\frac{1}{2}}_k}|
\end{equation}
According to Equations~\eqref{eq:1-f} and \eqref{eq:f}, we can achieve

\begin{equation}
\label{eq:minlk}
\begin{split}
\lambda_{\min}(\mathcal{L}_k)&=1-\lambda_{\max}(F)\\
&\geq1-|\lambda_{\max}(D^{-\frac{1}{2}}_k{A_G}D^{-\frac{1}{2}}_k)-\lambda_{\max}{D^{-\frac{1}{2}}_k{A_{\widetilde{k}}}D^{-\frac{1}{2}}_k}|\\
&=1-\lambda_{\max}(D^{-\frac{1}{2}}_k{A_G}D^{-\frac{1}{2}}_k)+\lambda_{\max}{D^{-\frac{1}{2}}_k{A_{\widetilde{k}}}D^{-\frac{1}{2}}_k}
\end{split}
\end{equation}

If we multiply both sides of this equation with -1 of Equation~\eqref{eq:lk} and then we add it with Equation~\eqref{eq:minlk}, we can achieve
\begin{equation}
\begin{split}
\lambda_{\min}&(\mathcal{L}_k)-\lambda_{\min}(\mathcal{L}_G)\\
&\geq\lambda_{\max}(D^{-\frac{1}{2}}_G{A_G}D^{-\frac{1}{2}}_G)-\lambda_{\max}(D^{-\frac{1}{2}}_k{A_G}D^{-\frac{1}{2}}_k)\\
&+\lambda_{\max}(D^{-\frac{1}{2}}_k{A_{\widetilde{k}}}D^{-\frac{1}{2}}_k)\\
&=\lambda_{\max}(D^{-1}_G{A_G})-\lambda_{\max}(D^{-1}_k{A_G})+\lambda_{\max}(D^{-1}_k{A_{\widetilde{k}}})
\end{split}
\end{equation}

Then we give the proof of Equation~\eqref{eq:leq}. According to Equations~\eqref{eq:lg} and \eqref{eq:1-f}, it can be inferred that
\begin{equation}
\begin{split}
\lambda_{\min}&(\mathcal{L}_k)-\lambda_{\min}(\mathcal{L}_G)\\
&=1-\lambda_{\max}(F)-(1-\lambda_{\max}(D^{-1}_G)A_G)\\
&=\lambda_{\max}(D^{-1}_G)A_G)-\lambda_{\max}(D^{-1}_kA_k)\\
&\leq{\lambda_{\max}(D^{-1}_G)\cdot\lambda_{\max}(A_G)}-\lambda_{\max}(D^{-1}_kA_k)\\
&=\frac{\lambda_{\max}(A_G)}{\min(D_{G_{ii}})}-\lambda_{\max}(D^{-1}_kA_k)
\end{split}
\end{equation}
q.e.d.

\section{Experiments}
Since theoretical proof shows that CHIEF-ST offers accurate acceleration, we only examine the effectiveness and efficiency of CHIEF-AP in this section. We evaluate CHIEF-AP on both real-world networks and synthetic networks. Since synthetic networks require no pre-processing, the pre-processing procedure and the experiment design only apply to real-world academic networks. Then we introduce baseline methods and evaluation indices.

\begin{figure*}[htbp]
  \centering
  \includegraphics[width=1.0\textwidth]{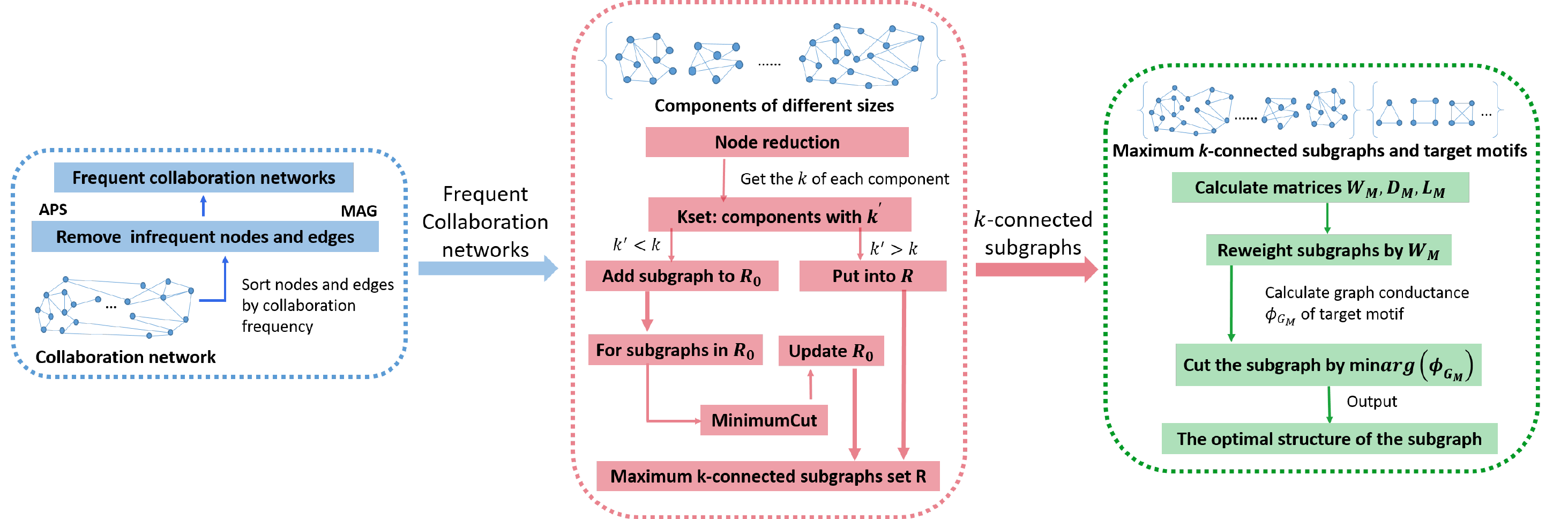}
  \caption{The framework of CHIEF-AP in co-author networks.}
  \label{fig:framework}
\end{figure*}

\subsection{Data Preprocessing}
We use both the MAG and the APS datasets for experiment. Specifically, we retrieve publication data in computer science discipline from MAG and those in physics discipline from the APS dataset. In this work, we examine papers published in the year period of 2009 to 2013. All experiments run on a server with 2.6 GHz Intel Xeon E5-2620 v4 processor and 128GB of main memory. The pre-processing of the datasets includes the following two steps.

\begin{enumerate}
  \item \textbf{Name Disambiguation:} Data pre-processing is required since there are (almost) no name disambiguation mechanisms in these two data sets. We distinguish names mainly in two procedures. First, we distinguish names by their co-authors and then distinguish authors who share no common co-authors by institutions. This method can distinguish most names except scholars who share one name and work in the same institution.
  \item \textbf{Removing Inactive Scholars:} In academic networks, the purpose of motif clustering is to find scholar clusters with high collaborative relationships~\cite{xia2014MVCWalker}. Thus, we remove relationships with lower collaboration frequencies. By statistics, more than 98\% of collaborative relationships are one-time collaboration in MAG. And more than 78\% of collaborative relationships are one-time or two-time collaboration in APS. We remove scholars with limited collaboration history with less than two times.
\end{enumerate}

After the above mentioned data pre-processing, we obtain 159,725 authors as nodes and 35,216,158 collaborative relationships in APS so as to build a collaboration network of the physic field. In MAG, we obtain 194,909 authors and 23,776,184 collaborative relationships to construct a collaboration network of the computer science field. All of source codes can be found in github\footnote{\url{https://github.com/yushuowiki/CHIEF}}.

\subsection{Experiment Design}
We implement CHIEF-AP on the two co-author networks from MAG and APS dataset in the following steps, respectively. Figure~\ref{fig:framework} shows the detailed flow of our experiments.\\
\\
\textbf{Step 1: Network Construction}\\
Construct a collaboration network $G=(V,E)$, whose vertices set $V$ represent scholars and edges $E$ represent collaboration relationships between two certain scholars. Each edge is weighted by the collaboration frequency between the two vertices who collaborates with each other. Since most of scholars have limited collaborators, the adjacency matrix of the collaboration network is a large-scale sparse matrix. Thus, we filter out edges with lower weights and reconstruct a new collaboration network $G_1=(V_1,E_1)$, wherein $V_1$ is the new vertices set and $E_1$ is the new edge sets. The edges are weighted by CII. To reduce the computing complexity and improve the algorithmic efficiency, we take network connectivity into consideration and partition $G_1$ into $k$-connected subgraphs in the next step.\\
\\
\textbf{Step 2: $k$-connected Subgraphs Finding}\\
In this step, we partition the collaboration network into a set of maximal $k$-connected subgraphs from the viewpoint of network connectivity. Therefore, we take no account of weights and set weights of all edges to 1. Before $k$-connected subgraph partition, we calculate the values of $k$ for each component in $G_1$. Next, we use different values of $k'$ to find maximal $k'$-connected subgraphs in each component. For any $k$-connected component, if $k'<k$, add this component directly into the result set $R$. If $k'>k$, find $k'$-connected subgraphs by edge reduction within this component, and the results are added into the result set $R$.\\
\\
\textbf{Step 3: Higher-order Motifs Clustering}\\
In this step, we cluster triangle motifs and heterogeneous four-node motifs. For a given motif $M$, this step aims to recognize a set of nodes $S$, that minimizes motif conductance $\phi_M(S)$. We employ triangle motifs (i.e., $M_{32}$) and heterogeneous four-node motifs (i.e., $M_{42}, M_{43}, M_{44}, M_{45}, M_{46}$) to construct Laplacian matrices for each subgraphs. Then, we calculate the conductance of each subgraphs in $R$ according to the eigenvalues of each Laplacian matrices, and then we cut the motifs with the smallest motif conductance.

\subsection{Baseline Methods}
For comparison, we also employ two baseline methods on our data in two procedures, respectively. We first use MovCut algorithm in the procedure of choosing the most proper $k$, which is mentioned in~\cite{jeub2014think}. MovCut is introduced to verify whether $k$-connected subgraph finding process can achieve the best modularity, so that we can ensure the most proper $k$ has been chosen. We then use higher-order clustering method without $k$-connected subgraph finding procedure as the first contrast algorithm in clustering process~\cite{benson2016higher}. This is used to verify whether the $k$-connected subgraph finding process can accelerate motif clustering.

Here we give a brief introduction to MovCut. To a given graph $G=(V,E)$, a vertex $i$, and a positive integer $k$, MovCut aims at finding a set of nodes $S \subseteq V$ that achieves the minimum conductance among all sets of nodes that contain $i$ and $|S| \leq k$. The solution $\vec{x}*$ of the optimization problem is of the form in Equation~\ref{eq:solution}.

\begin{equation}
\label{eq:solution}
\vec{x}*=a(L_G-\gamma{D_G})^+D_{G^{\vec{S}}},
\end{equation}
wherein, $\gamma{\in}(-\infty,\lambda_2(G))$, $\gamma=\frac{\alpha-1}{\alpha}$, and $a\in{[0,\infty]}$ is a normalization constant.

In \cite{jeub2014think}, $k$ is chosen as 50 in the experiments of collaboration network. Besides, $\alpha$ is set to be 20 values of equal space intervals in between $[0.7, (1-\lambda_2)^{-1}-10^{-10}]$, where $(1-\lambda_2)^{-1}$ is the theoretical maximum for $\alpha$.

In our experiments, we set $k=50$ as well in order to achieve similar performance. We set $\alpha$ to 20 values of equal space intervals in between $[0.7,\frac{1}{1-e}]$, where $e$ is the eigenvector of $L_G$.

\subsection{Evaluation Indices}
We introduce two indices, network modularity and collaboration intensity index (CII), to evaluate the collaboration intensity within co-author networks. We also select two indices, cluster compactness (CCP) and cluster separation (CSP), from the perspective of graph structure to verify experimental results of the proposed algorithm. In this section, the formal definitions of above indices are given.

\subsubsection{Network Modularity}
Since we use academic social networks for our experiments, we introduce network modularity as one of evaluation metrics. Network modularity is first proposed to quantify performance of community detection algorithm. A higher network modularity value refers that the community is of higher cohesion~\cite{xia2015tc}. In this paper, we calculate network modularity as Equation~\eqref{eq:modularity} according to Newman's definition in ~\cite{newman2006modularity}.

\begin{equation}
\label{eq:modularity}
Q=\dfrac{1}{4m}\sum\limits_{ij}{(A_{ij}-\dfrac{k_ik_j}{2m})}(s_is_j+1)
\end{equation}

Herein, $Q$ is the value of network modularity of a certain subgraph; $m$ is the number of edges within the subgraph; $i,j$ are any two nodes within or subgraph. Let $s_i=1$ if node $i$ belongs to subgraph 1, otherwise $s_i=-1$.

\subsubsection{Collaboration Intensity Index}
We introduce the CII index~\cite{yu2017team} to evaluate the cluster results of co-author network. CII is calculated based on both the collaboration frequency and the number of papers two scholars published. CII is proposed to evaluate the collaboration intensity between two scholars, which is calculated according to Equation~\eqref{eq:cii}.

\begin{equation}\label{eq:cii}
CII=\frac{{\Delta}_{t_2-t_1}{k_{ij}^2}}{{\Delta}_{t_2-t_1}{k_{i}}{\Delta}_{t_2-t_1}{k_{j}}}
\end{equation}

In Equation~\eqref{eq:cii}, ${\Delta}_{t_2-t_1}{k_{i}}$ refers to the number of papers published between year $t_1$ and $t_2$ by scholar $i$, and ${\Delta}_{t_2-t_1}{k_{ij}^2}$ refers to the number of papers that scholar $i$ and $j$ co-authored between year $t_1$ and $t_2$.

CII is a relative evaluation index that can be used to represent collaboration relationships between scholars. In contrast to CF (Collaboration Frequency), CII reflects how close two scholars collaborate.

\begin{figure}[htbp]
  \centering
  \includegraphics[width=0.4\textwidth]{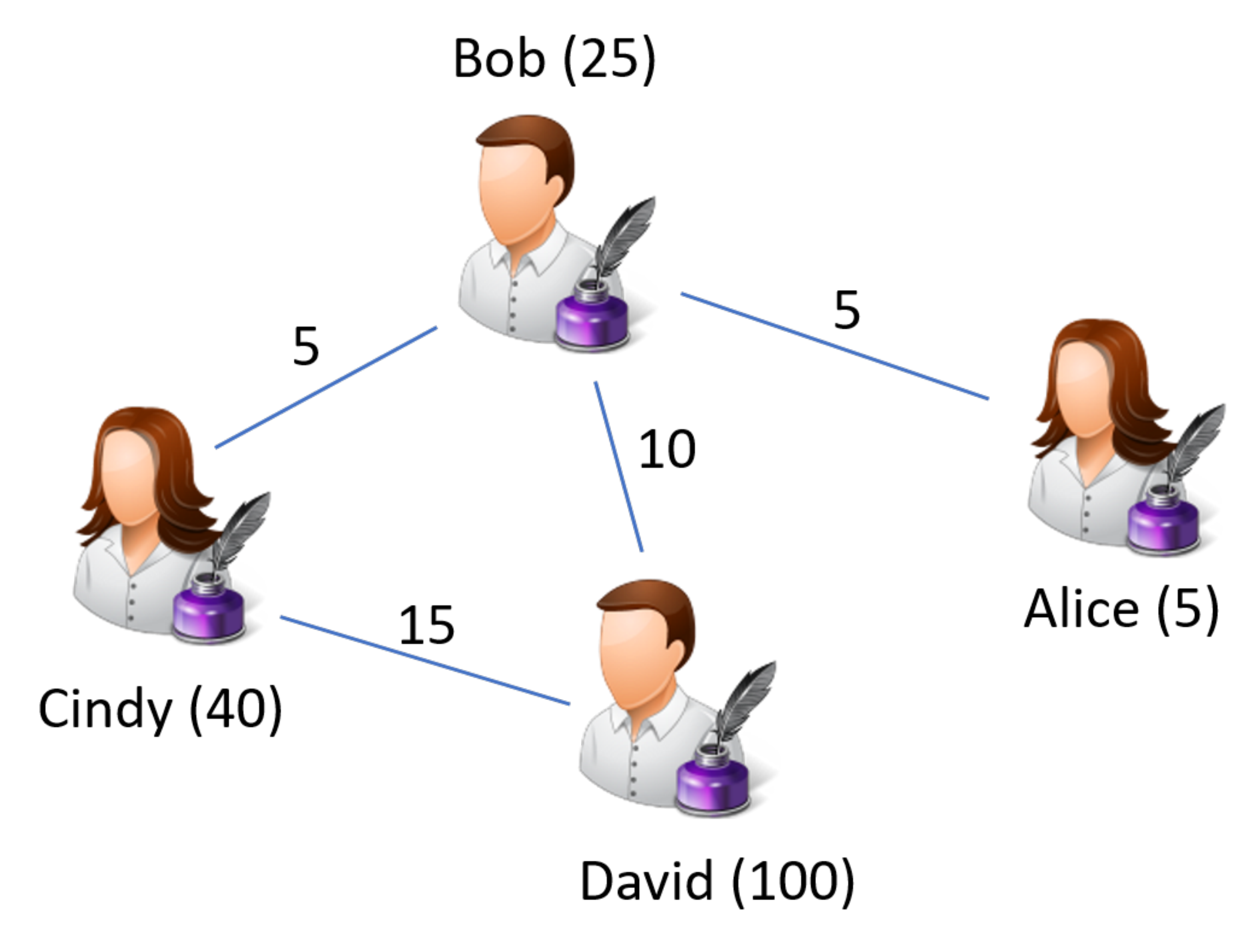}\\
  \caption{An example of co-author network. Each edge weight represents the number of collaborations between two scholars.}\label{fig:cii}
\end{figure}

As an example, Figure~\ref{fig:cii} shows the co-author relationships of four scholars, and each edge weight represents the number of co-authored papers between the co-author pair. The number beside scholar names represent the total number of papers published by the scholar. It can be seen that Alice and Bob have collaborated 5 times, which equals that of Bob and Cindy. However, Bob is the only scholar collaborated with Alice, while Alice is not the only one for Bob. CII can reflect such kind of distinctions. The CII between Bob and Cindy, i.e., CII$_{BC}$, is calculated by $\frac{5^2}{25\times40}$, which equals 0.025. But CII$_{BA}$, i.e., the CII between Bob and Alice, is calculated by $\frac{5^2}{25\times5}$, which equals 0.2. CII$_{BA}$ is much higher than CII$_{BC}$. This fact represents that the collaboration intensity between Alice and Bob is much stronger.

\subsubsection{Cluster Compactness}
Cluster compactness (CCP)~\cite{huang2013extensions}is proposed to evaluate whether the distances within cluster are compact or not. CCP is the average value of the distances between each vertex and the center vertex in one certain cluster. A lower value of CCP illustrates that the cluster is tighter. Meanwhile, a cluster with lower CCP is also generally believed to be a better cluster. For vertex $i$ in the cluster, the compactness of $i$ and cluster center, i.e., $\overline{CCP_{i}}$, is defined in Equation~\eqref{eq:cpi}.

\begin{equation}
\label{eq:cpi}
\overline{CCP_{i}}=\frac{1}{|\Omega_{i}|}\sum_{x_{i}\in{\Omega_{i}}}\|x_i-\omega_i\|
\end{equation}

In Equation~\eqref{eq:cpi}, $\Omega_{i}$ is the vertex set of the cluster and $\omega_i$ is the cluster center vertex. In this work, the vertex center is chosen as the vertex with maximum degree in the cluster. When there exist more than one vertex owning maximum degree, we randomly choose one of them. Besides, the distance is defined as the shortest path in the cluster. Suppose there are $n$ clusters totally, then the formal definition of $\overline{CCP}$ is shown in Equation~\eqref{eq:cp}.

\begin{equation}
\label{eq:cp}
\overline{CCP}=\frac{1}{n}\sum_{i=1}^{n}\overline{CCP}_i
\end{equation}

CCP reflects the distance between each vertex and cluster center vertex. However, CCP can only reflect the compact degree within the cluster. Therefore, we introduce another index to evaluate the cluster algorithm, which is called cluster separation.

\subsubsection{Cluster Separation}
Cluster separation (CSP)~\cite{praveen2019inter} is proposed to evaluate the distance between clusters. CSP is the average distance between center vertices of clusters. A higher value of CSP refers to a farther distance between two clusters, which indicates that the cluster results is better. The formal definition of CSP is given in Equation~\eqref{eq:sp}.

\begin{equation}
\label{eq:sp}
\overline{CSP}=\frac{2}{k^2-k}\sum_{i=1}^{k}\sum_{j=i+1}^{k}\parallel{\omega_i-\omega_j}\parallel_2
\end{equation}

In Equation~\eqref{eq:sp}, $k$ is the number of clusters, and $\omega_i, \omega_j$ represent the center vertex of cluster $i$ and $j$, respectively. CSP can reflect the distances between clusters at a macro level. However, CSP ignores the closeness within the clusters. Therefore, CSP and CCP are always used together to evaluate cluster results.

\section{Results and Discussion}
In this section, we discuss about the experimental results from mainly two perspectives. First, we discuss and evaluate the cluster results according to experiments on collaboration networks. Then, we analyze the cluster effectiveness according to experiments on synthesized networks.

\subsection{Implementing CHIEF on Collaboration Networks}
We respectively analyze two procedures, i.e., (1) selecting the most proper value of $k$ in finding the maximal $k$-connected subgraphs, and (2) higher order motif clustering. Selecting the most proper $k$ for networks is a vital process in the whole algorithm. Therefore, we analyze this process by itself in particular.

\begin{figure}[!tp]
  \centering
  \includegraphics[width=0.5\textwidth]{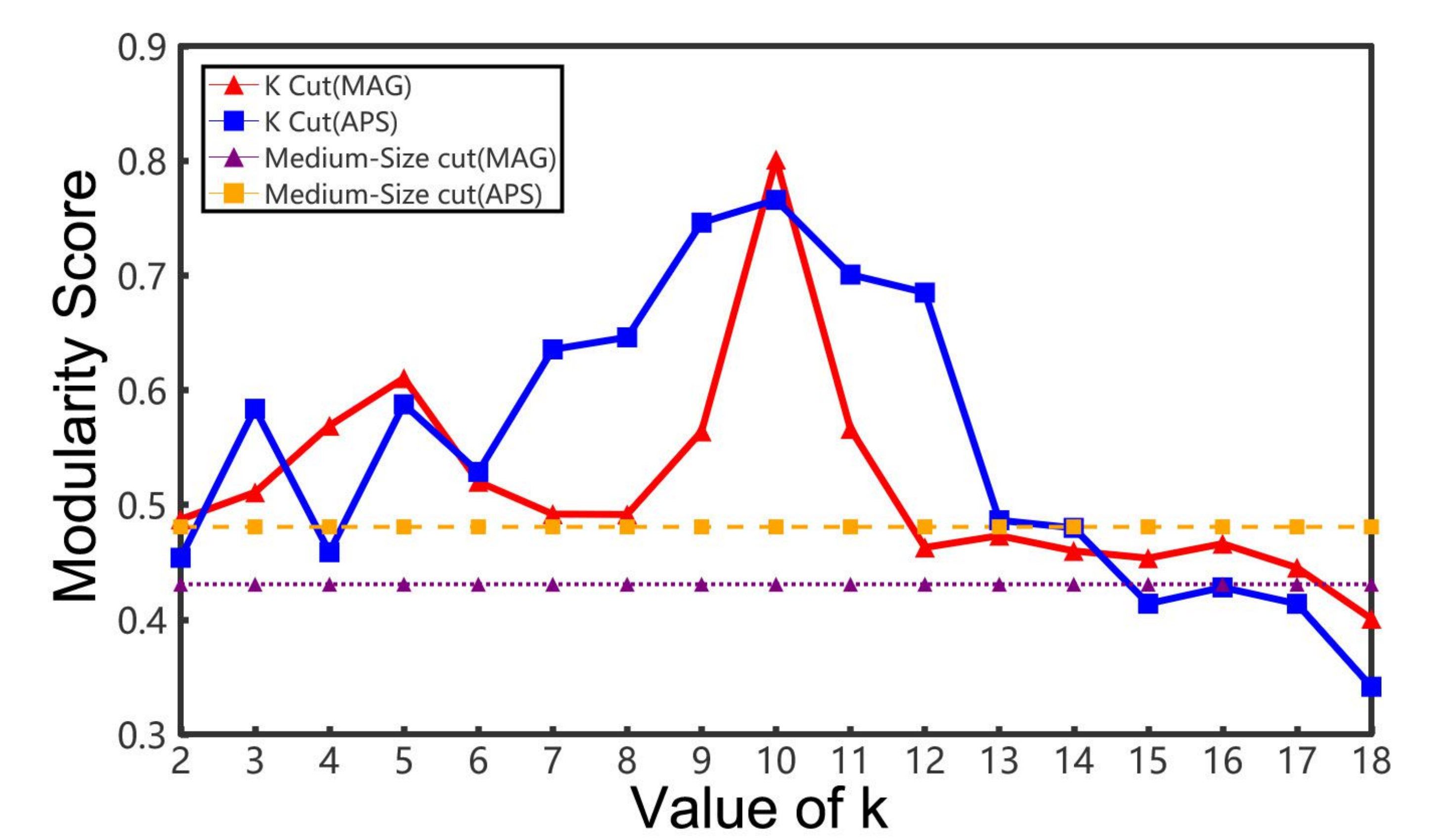}\\
  \caption{Experimental results of $k$-connected subgraph based cut and MovCut. Network modularity of each network partitioned by different $k$ are shown.}\label{fig:modu}
\end{figure}

\subsubsection{Selection of the Most Proper $k$}
In order to reduce computational complexity, we first find the maximal $k$-connected subgraphs. By taking this step, the computational complexity of network can be significantly reduced. At the same time, it can keep the network connectivity up to the hilt, which ensures that this step will maintain the original network structure as much as possible. However, there still exist some differences when using different values of $k$ in finding the maximal $k$-connected subgraphs. Different values of $k$ will affect the network connectivity. Therefore, to achieve the best network partition, we need to find the most suitable value of $k$.

\begin{figure*}[!htp]
  \centering
  \subfigure[$k=2,3,4,5$]{
    \includegraphics[width=0.22\textwidth]{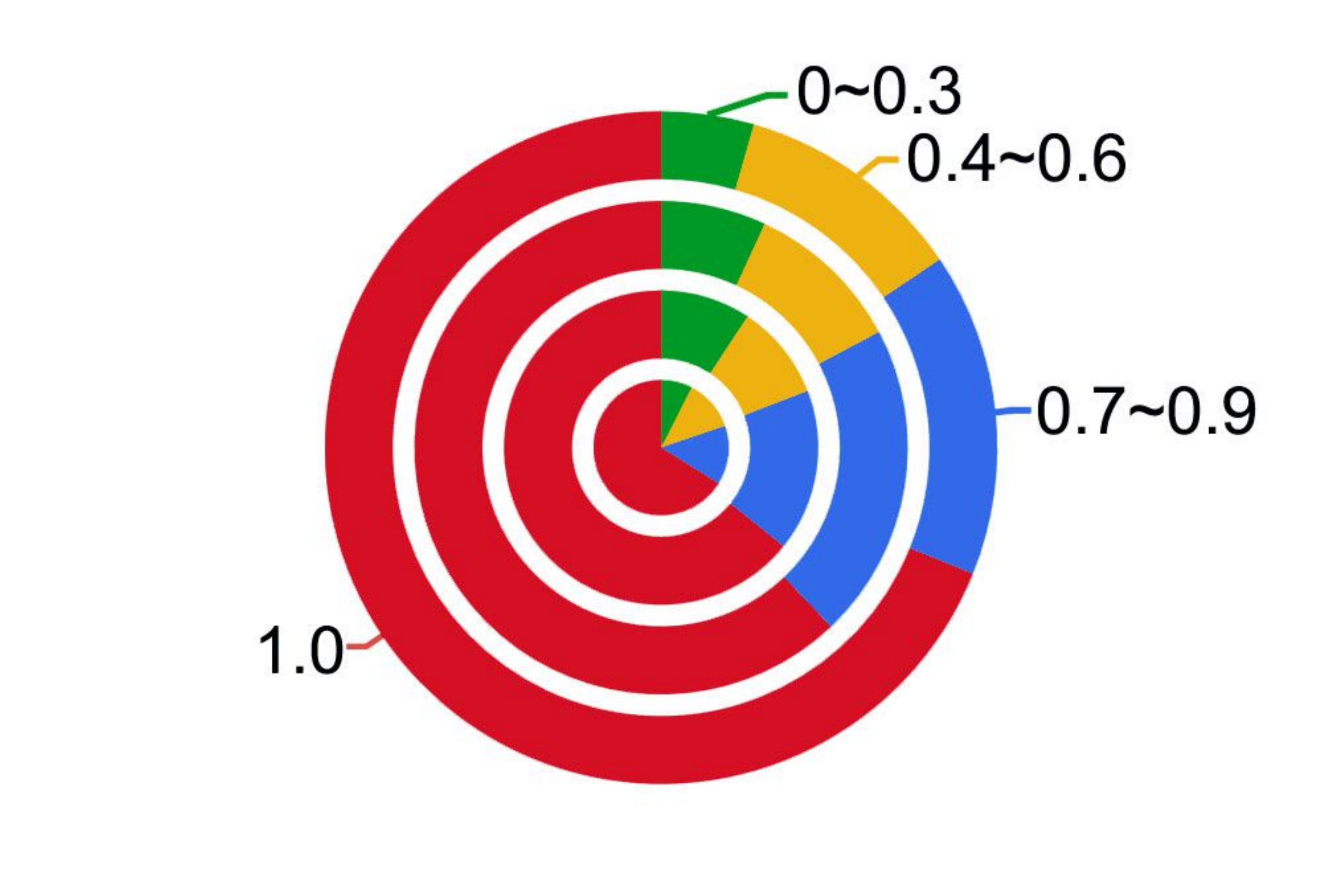}
  }
  \subfigure[$k=6,7,8,9$]{
    \includegraphics[width=0.22\textwidth]{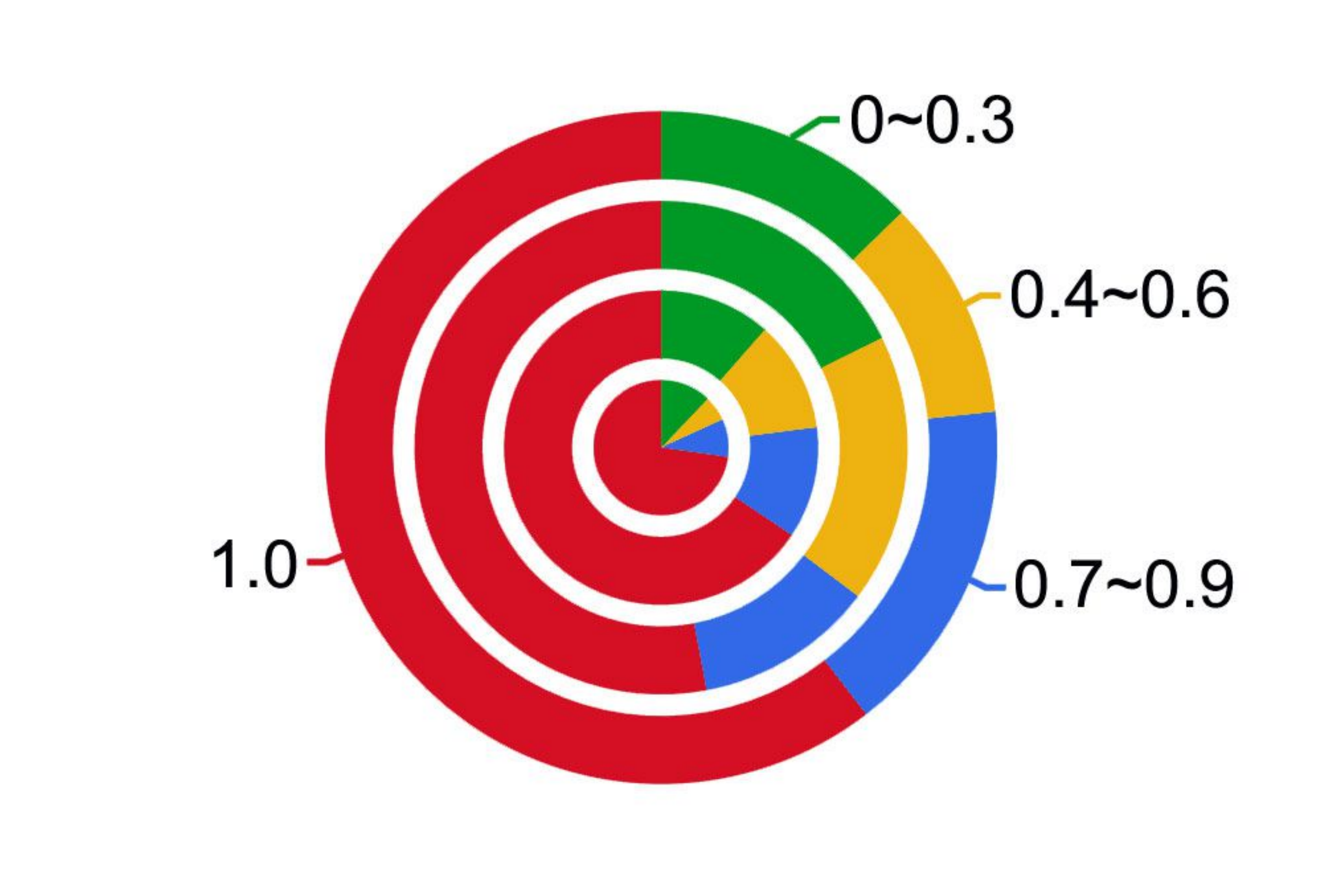}
  }
  \subfigure[$k=10,11,12,13$]{
    \includegraphics[width=0.22\textwidth]{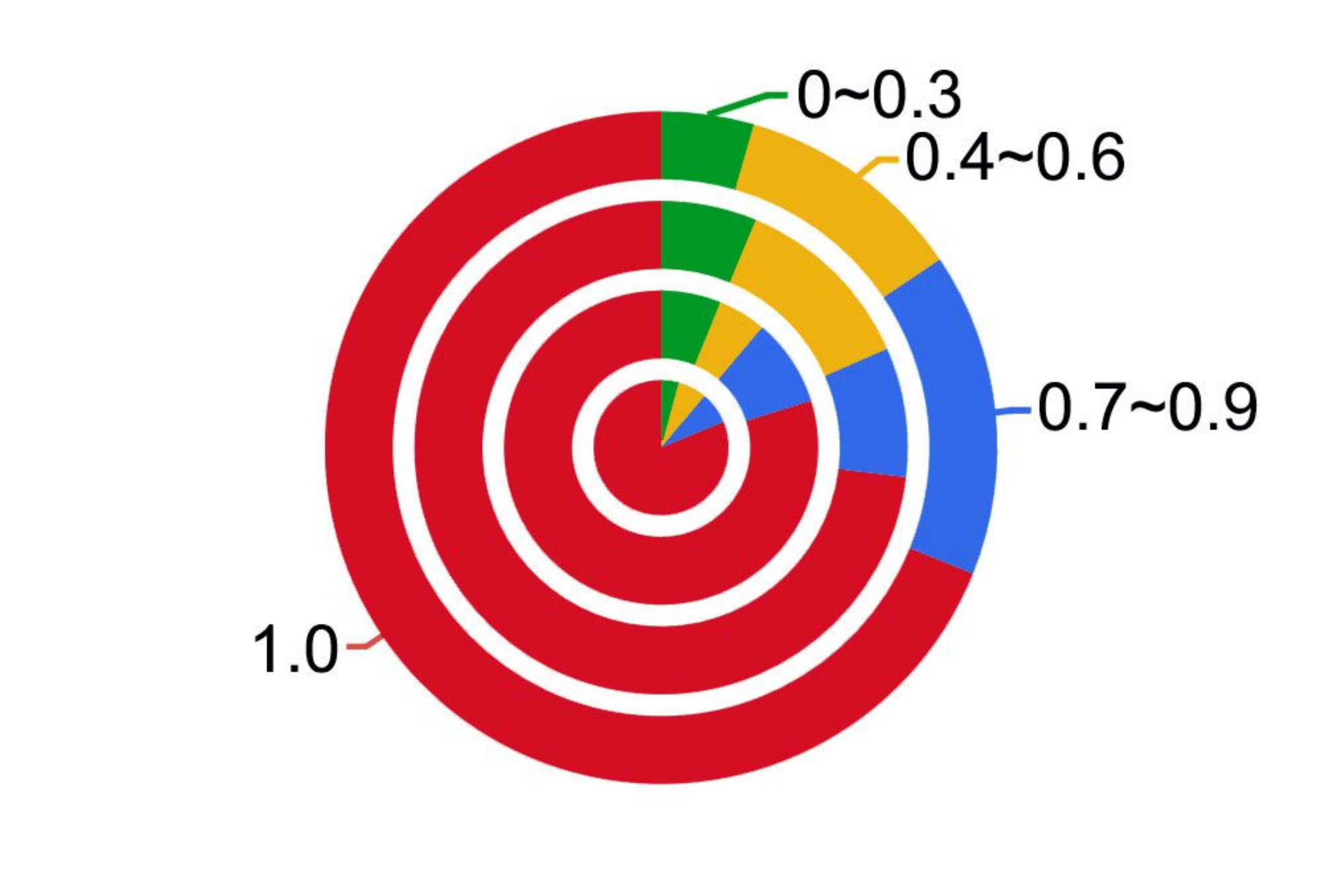}
  }
  \subfigure[$k=14,15,16,17$]{
    \includegraphics[width=0.22\textwidth]{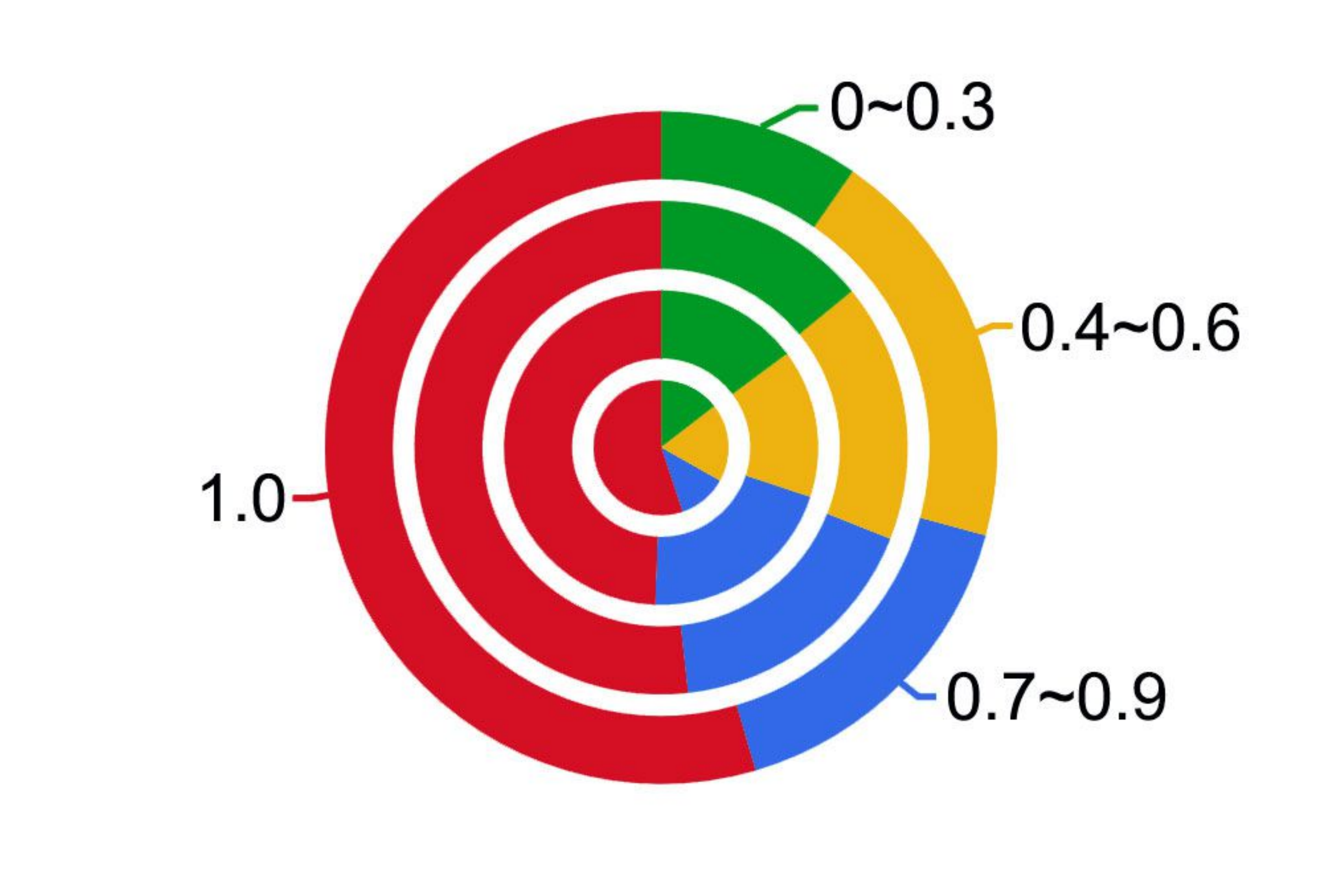}
  }
  \caption{CCP of the network partitioned by different maximal $k$-connected subgraphs in MAG.}
  \label{fig:id-MAG}
\end{figure*}

\begin{figure*}[!htp]
  \centering
  \subfigure[$k=2,3,4,5$]{
    \includegraphics[width=0.22\textwidth]{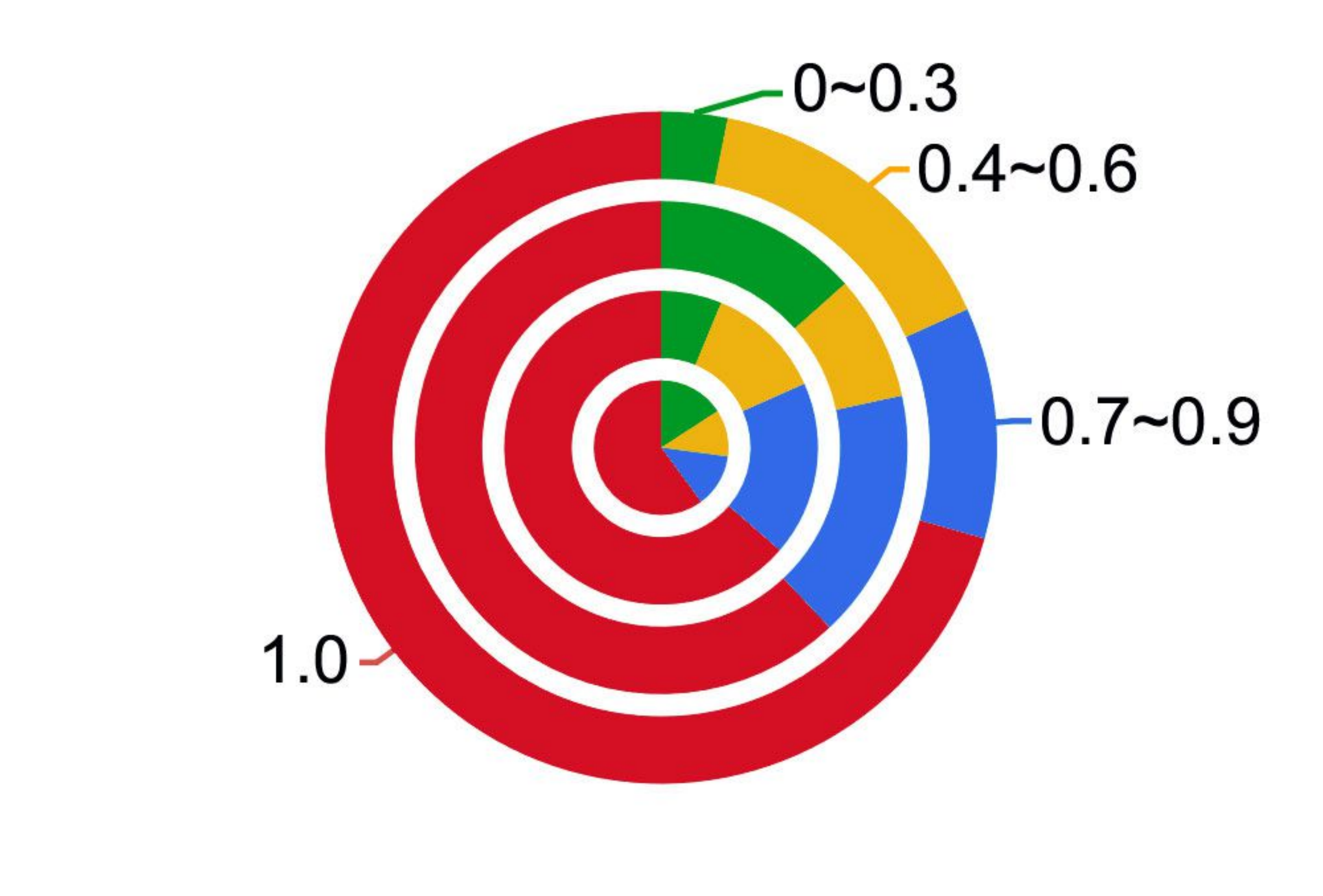}
  }
  \subfigure[$k=6,7,8,9$]{
    \includegraphics[width=0.22\textwidth]{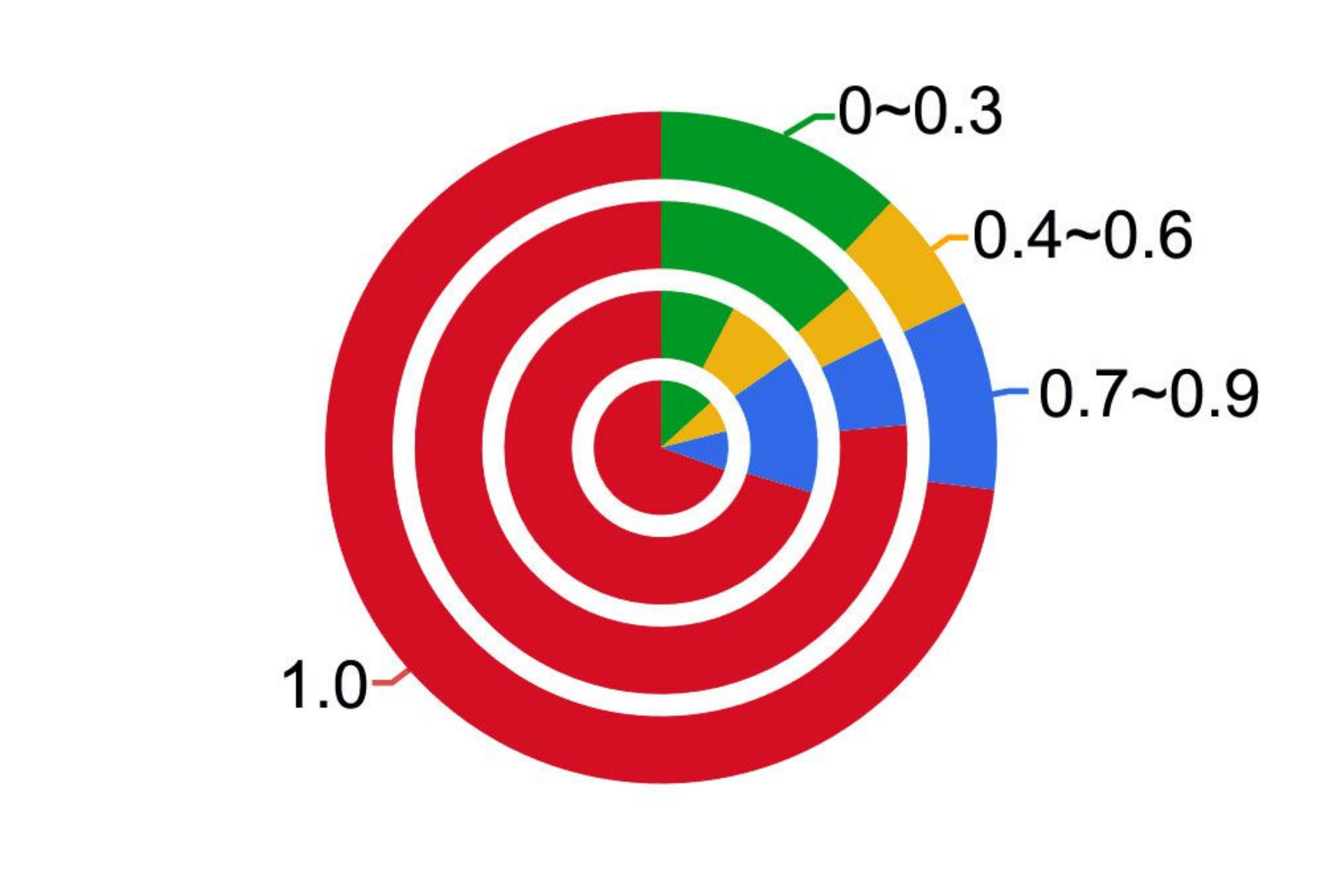}
  }
  \subfigure[$k=10,11,12,13$]{
    \includegraphics[width=0.22\textwidth]{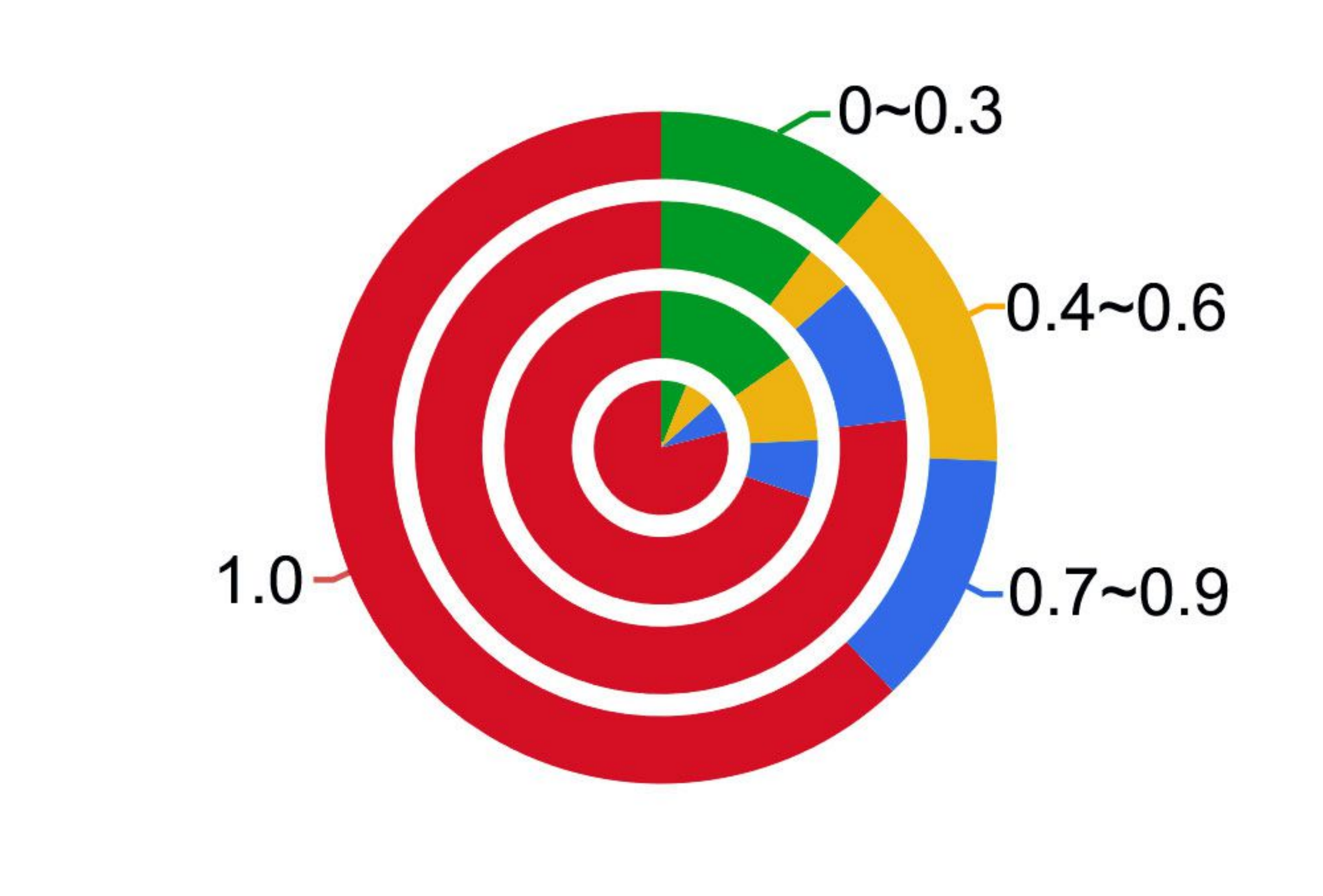}
  }
  \subfigure[$k=14,15,16,17$]{
    \includegraphics[width=0.22\textwidth]{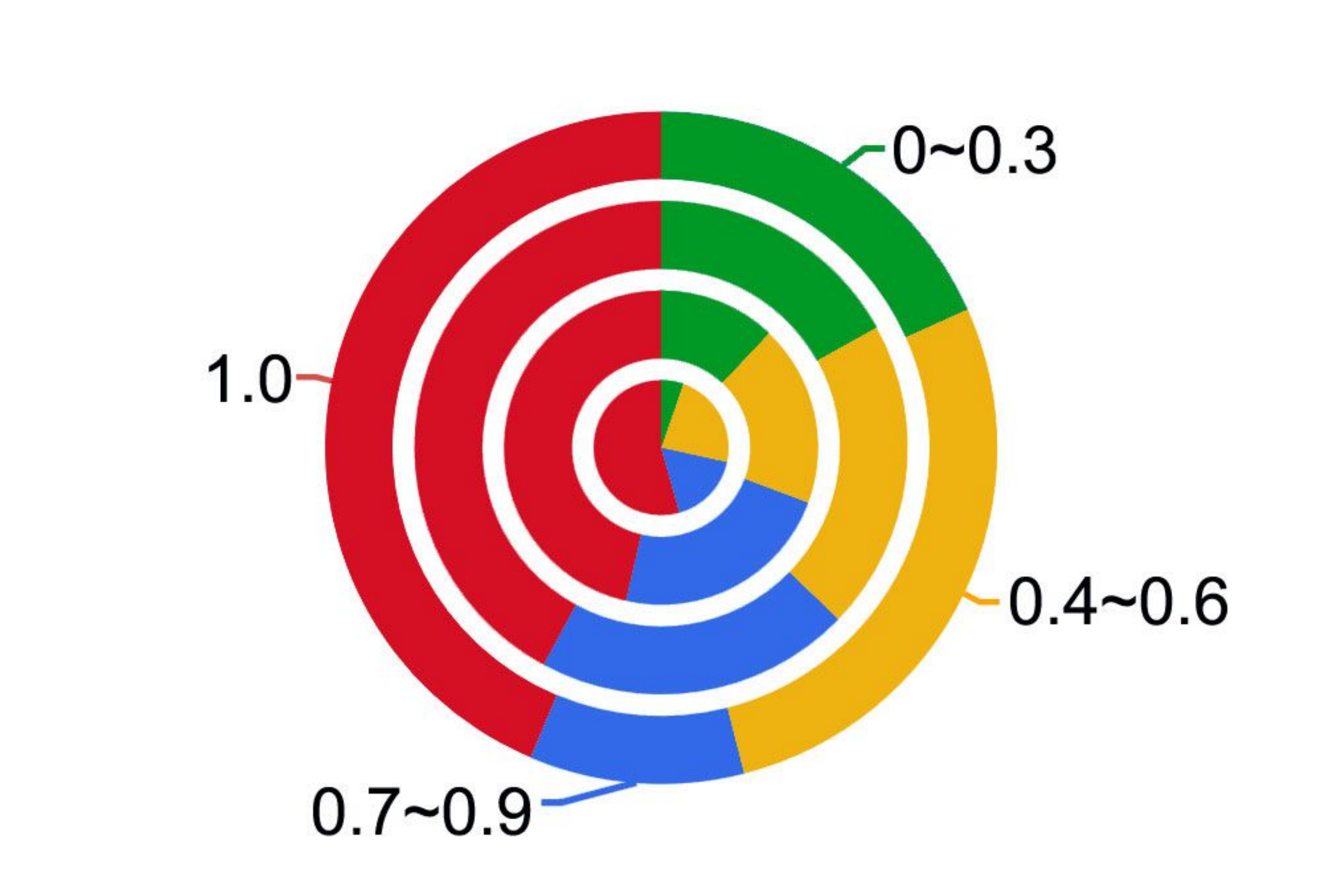}
  }
  \caption{CCP of the network partitioned by different maximal $k$-connected subgraphs in APS.}
  \label{fig:id-APS}
\end{figure*}

We implement experiments to select the most proper $k$ in MAG and APS, respectively. Figure~\ref{fig:id-MAG} and Figure~\ref{fig:id-APS} show the values of CCP in MAG and APS, respectively. Both in Figure~\ref{fig:id-MAG} and Figure~\ref{fig:id-APS}, most values of CCP equal 1. However, differences can be obviously seen from these figures. In all these figures, each circle represents the values of a set of $k$, in which the inner circle corresponds to a lower value of $k$. That is, take Figure~\ref{fig:id-MAG} (a) as an example, $k=2,3,4,5$ correspond to the circles from inside to outside, respectively. Different color represents different proportions for each CCP value.

\begin{table*}[htbp]
  \centering
  \caption{Basic Information of Academic Conference Networks}
  \resizebox{\textwidth}{!}{
    \begin{tabular}{ccccccccccc}
      \toprule
      Academic Conference Name & JCDL  & CSCW  & SIGIR & KDD   & CIKM  & ICDE  & INFOCOM & WWW   & IJCAI & AAAI \\
      \midrule
      Number of Scholars & 3182  & 4491  & 5531  & 6443  & 8086  & 8333  & 11533 & 10472 & 11322 & 16286 \\
      Number of Papers & 1713  & 2322  & 4101  & 3236  & 4356  & 4357  & 8252  & 5567  & 7982  & 11030 \\
      Number of Vertices & 7763  & 11975 & 12906 & 18750 & 18222 & 21583 & 28802 & 24508 & 21052 & 36480 \\
      Optimal Value of $k$ & 3     & 5     & 4     & 6     & 6     & 4     & 5     & 8     & 6     & 7 \\
      Modularity Score & 0.574 & 0.479 & 0.543 & 0.57  & 0.486 & 0.583 & 0.631 & 0.533 & 0.615 & 0.468 \\
      \bottomrule
  \end{tabular}}
  \label{tab:m}
\end{table*}

\begin{table*}[htbp]
  \centering
  \caption{Basic Information of Academic Journal Networks}
  \resizebox{\textwidth}{!}{
    \begin{tabular}{ccccccccc}
      \toprule
      Academic Journal Name & Nature & STM   & Trans & Pro   & AI    & SIAM  & IEEE  & ACM \\
      \midrule
      Number of Scholars  & 2519 & 6146 & 8090 & 6934 & 10299& 25742& 278813& 31325 \\
      Number of Papers    & 655  & 5125 & 3326 & 2323 & 6187 & 25368& 270673& 21712 \\
      Number of Vertices  & 60921& 10150& 14910& 20892& 19566& 41858& 902769& 68758 \\
      Optimal Value of $k$& 3    & 6    & 5    & 8    & 8    & 6    & 7     & 5 \\
      Modularity Score    & 0.578& 0.585& 0.605& 0.471& 0.607& 0.559& 0.476 & 0.613 \\
      \bottomrule
  \end{tabular}}
  \label{tab:j}
\end{table*}
Patterns can be obviously found in both Figure~\ref{fig:id-MAG} and Figure~\ref{fig:id-APS}. One of the most obvious patterns is that different values of $k$ do have an influence on the network partition results. Higher values of $k$ may over partition the network, which leads to the situations in Figure~\ref{fig:id-MAG}(d) and in Figure~\ref{fig:id-APS}(d). In these two figures, it can be seen that there exist considerable large proportion of $[0.7, 0.9]$ when $k=$13,14,15,16,17. Reason behind this phenomenon is because of an inappropriate value of $k$. Higher values of $k$ will make $k$-connected subgraphs extremely compact, leading to the fact that some edges with high connectivity are cut off. This is why the CCPs in $[0.7,0.9]$ occupy a large proportion in Figure~\ref{fig:id-MAG}(c) and Figure~\ref{fig:id-MAG}(d). Choosing higher values of $k$ will result in over partition, while choosing lower values of $k$ will give rise to a negative network partition as well. From Figure~\ref{fig:id-MAG} and Figure~\ref{fig:id-APS}, we can see that when $k$ is between 2 and 9, similar patterns can be also found in Figure~\ref{fig:id-MAG}(a), (b) and Figure~\ref{fig:id-APS}(a), (b). The proportions of $[0.7, 0.9]$ are large when $k=$2,3,4,5,6,7,8,9. Apparently, smaller values of $k$ may lead to looser structure of $k$-connected subgraphs. This fact makes smaller values of $k$ perform worse when partitioned network since edges with lower connectivity are not cut off. There is still a slight difference between MAG and APS. When $k>14$, the proportion of CCP=1 in APS is much less than that in MAG. In other 3 subgraphs, this difference does not obviously exist. This phenomenon may be caused by disciplinary differences. In computer science area, scholars may collaborate within smaller groups or teams. Moreover, scholars from computer science area generally collaborate with higher CII than scholars from physics. Results about CII in Figure~\ref{fig:ciiresult} prove this as well.

\begin{figure*}[!htp]
  \centering
  \subfigure[$k=2,3,4,5$]{
    \includegraphics[width=0.22\textwidth]{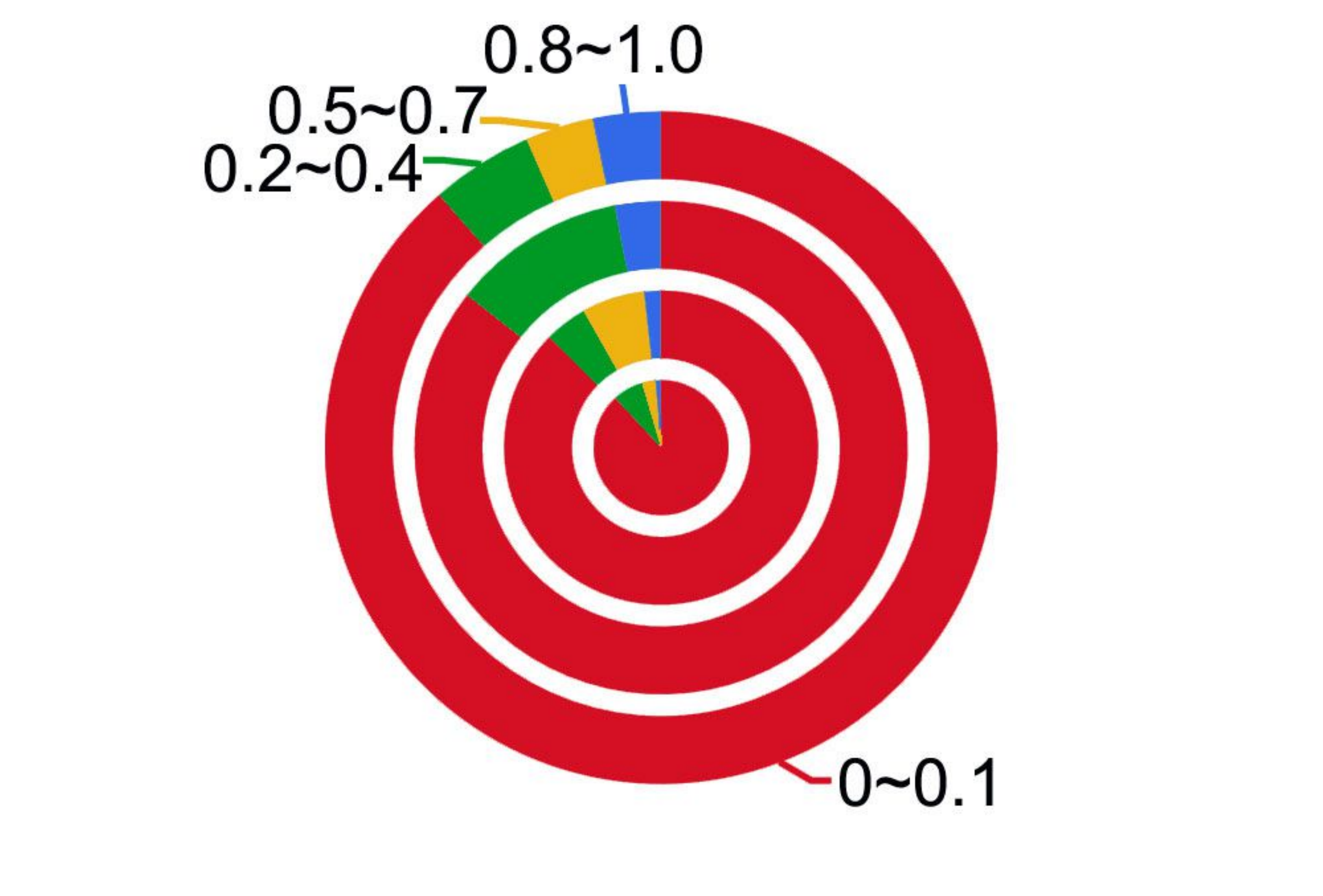}
  }
  \subfigure[$k=6,7,8,9$]{
    \includegraphics[width=0.22\textwidth]{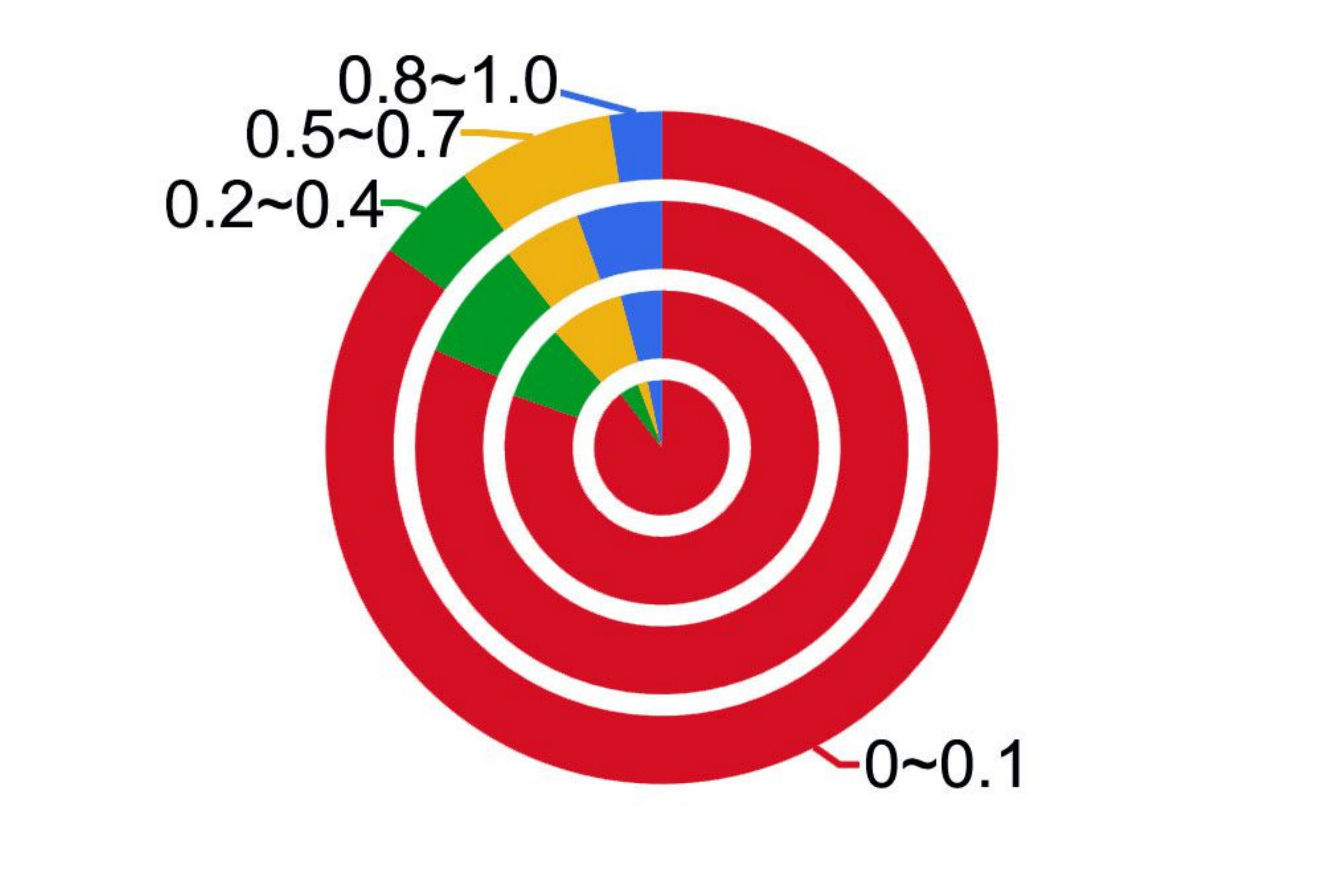}
  }
  \subfigure[$k=10,11,12,13$]{
    \includegraphics[width=0.22\textwidth]{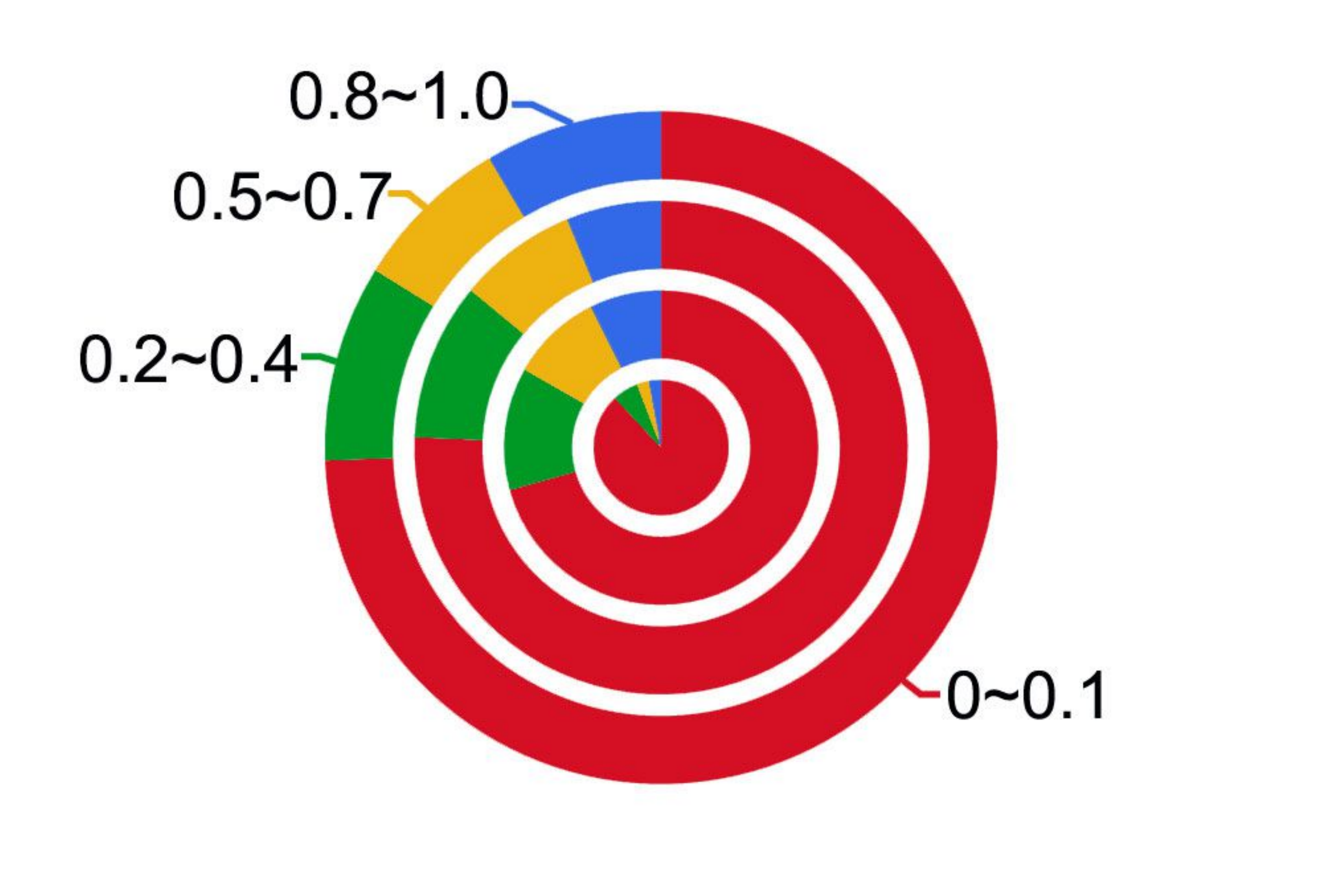}
  }
  \subfigure[$k=14,15,16,17$]{
    \includegraphics[width=0.22\textwidth]{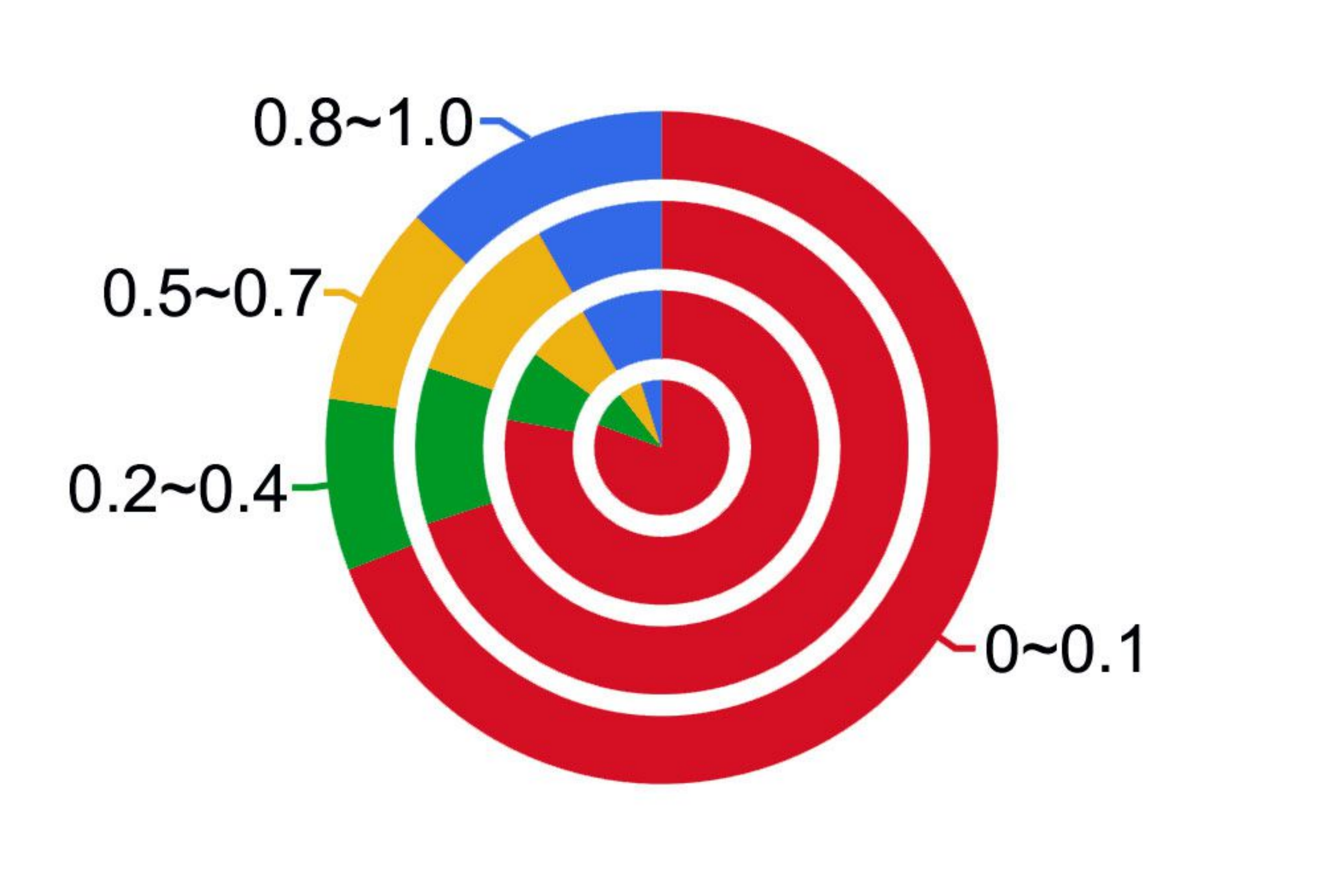}
  }
  \caption{CSP of the network partitioned by different maximal $k$-connected subgraphs in MAG.}
  \label{fig:id-MAG-E}
\end{figure*}
\begin{figure*}[!htp]
  \centering
  \subfigure[$k=2,3,4,5$]{
    \includegraphics[width=0.22\textwidth]{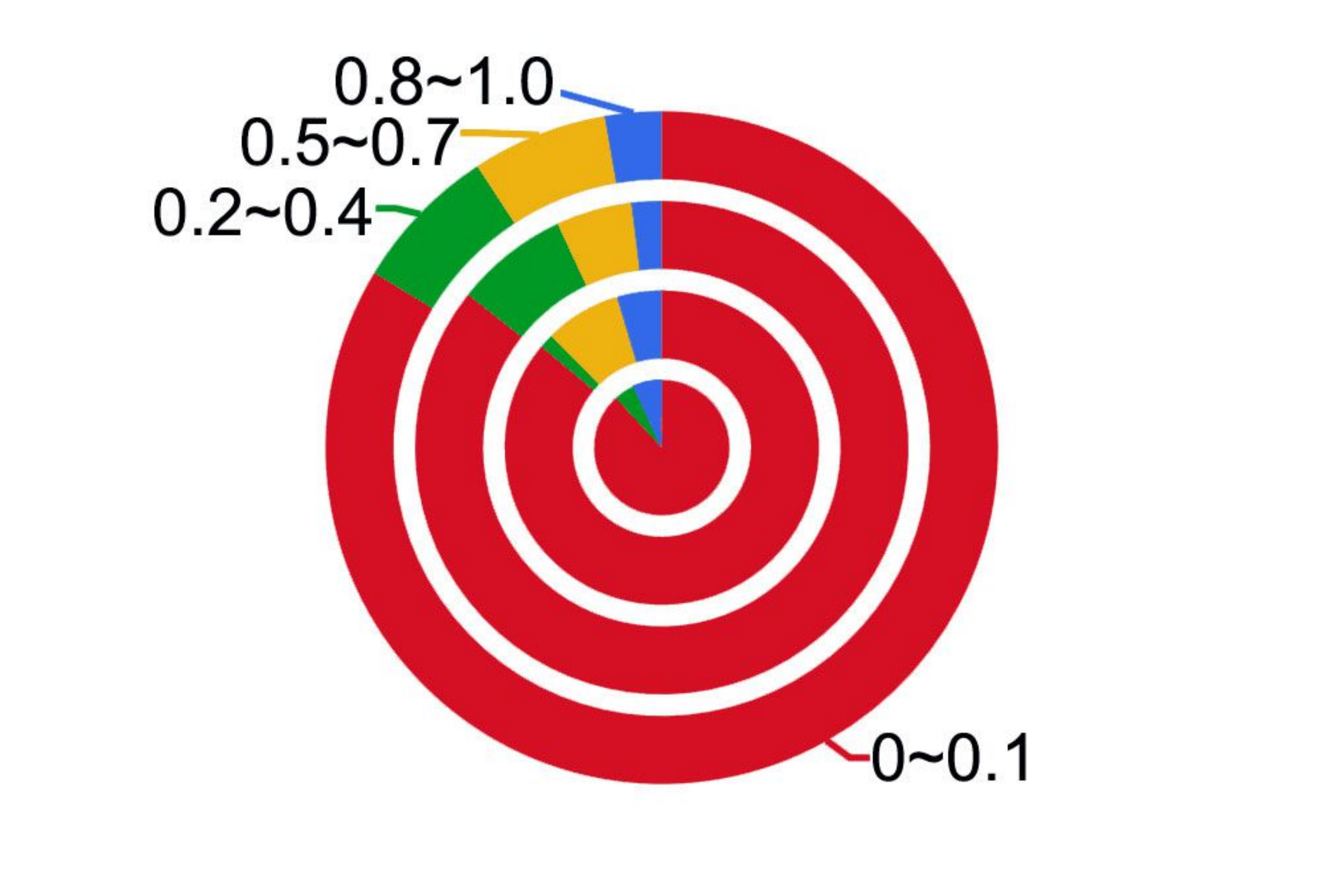}
  }
  \subfigure[$k=6,7,8,9$]{
    \includegraphics[width=0.22\textwidth]{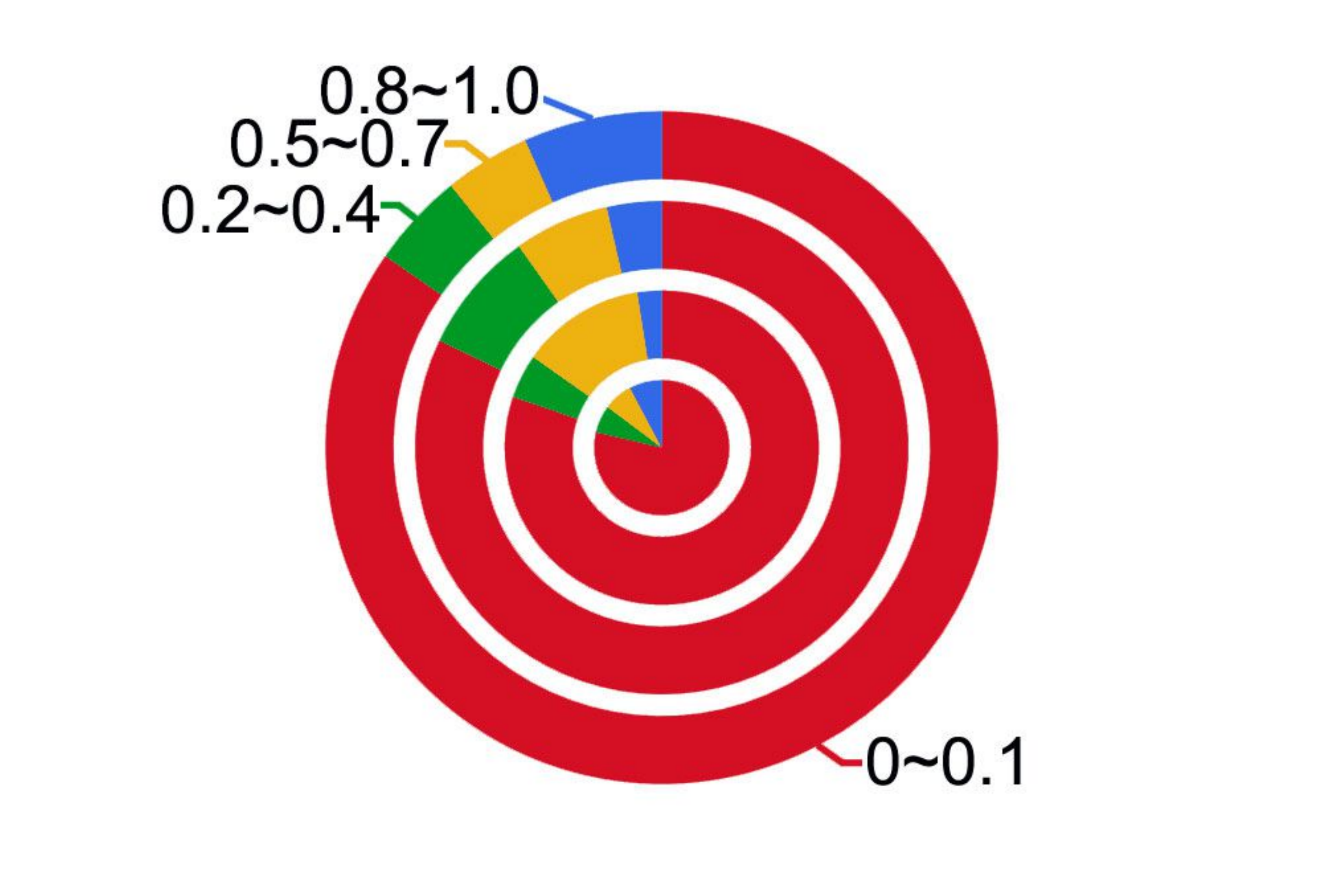}
  }
  \subfigure[$k=10,11,12,13$]{
    \includegraphics[width=0.22\textwidth]{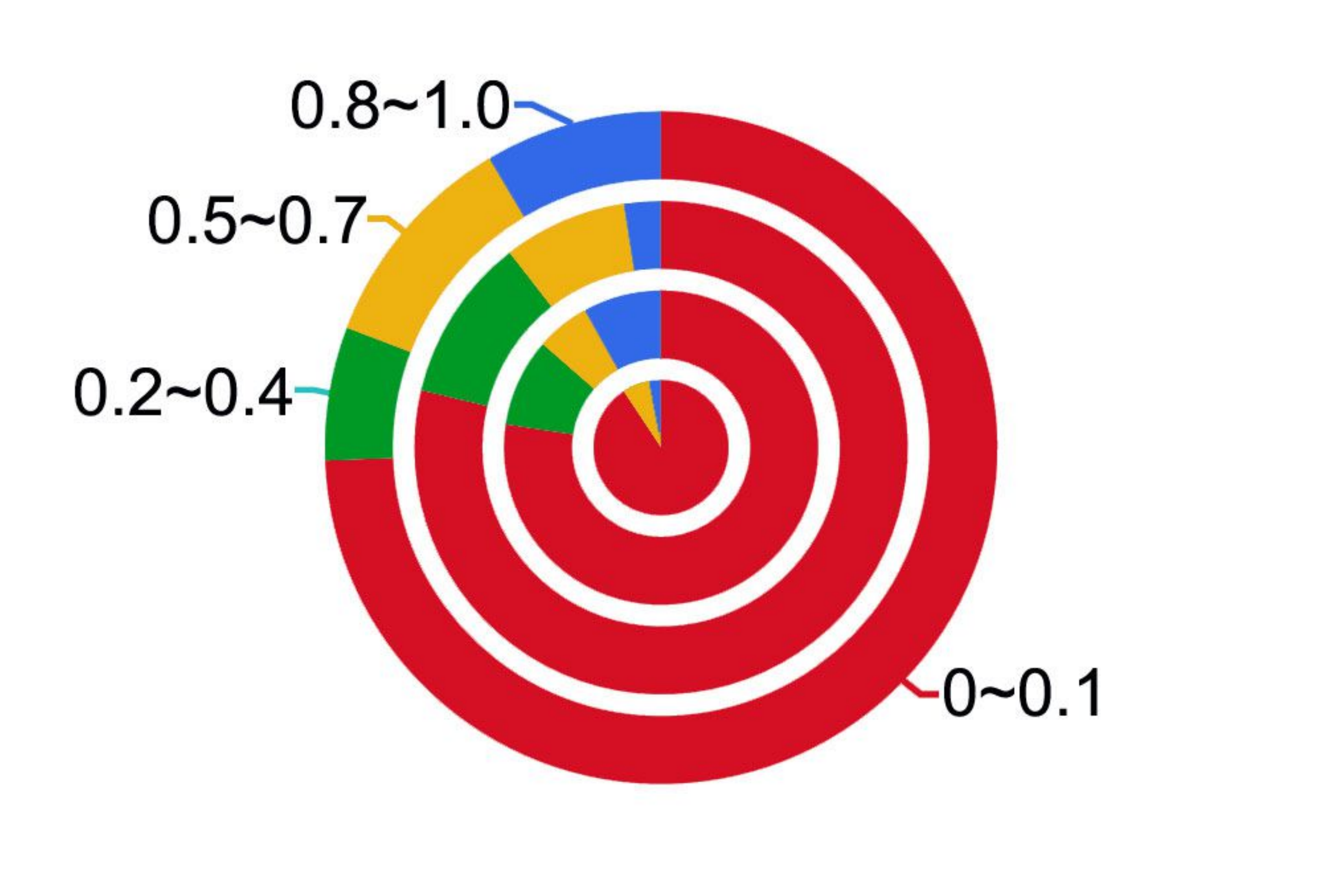}
  }
  \subfigure[$k=14,15,16,17$]{
    \includegraphics[width=0.22\textwidth]{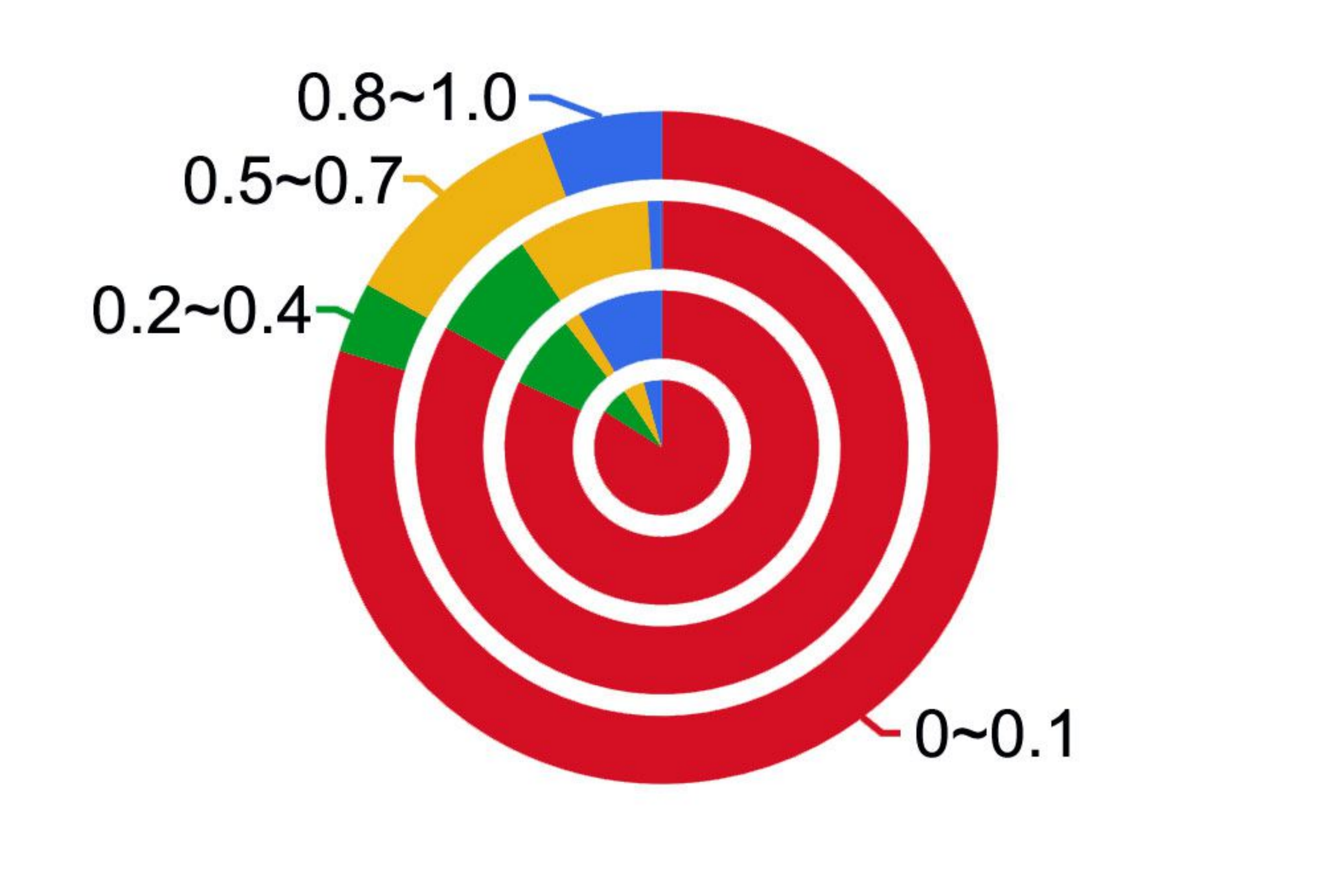}
  }
  \caption{CSP of the network partitioned by different maximal $k$-connected subgraphs in APS.}
  \label{fig:id-APS-E}
\end{figure*}

Figure~\ref{fig:id-MAG-E} and Figure~\ref{fig:id-APS-E} show the CSP of MAG and APS respectively. As mentioned above, the more CSP approaches 0, the better $k$ is. Overall, CSP mostly approaches 0 no matter in MAG or APS. This may be caused by the differences of the network properties. While co-author network is considered a kind of social network, and it exhibits similar attributes of being sparse. Thus, CSP values are generally equal or approach 0. This is the reason that those subgraphs in Figure~\ref{fig:id-MAG-E} and Figure~\ref{fig:id-APS-E} share similar CSP distributions.

Considering several values of $k$, it would be better choosing a $k$ value from ${8, 9, \dots, 14}$. Thus we calculate network modularity for each network partitioned with different $k$ values. Network modularity is shown in Figure~\ref{fig:modu}. The value of network modularity mainly depends on network vertices distribution. It is widely used in quantifying network community classification. When network modularity approaches 1, it is proved that the network community structure intensity is strong. Therefore, the optimal network partition can be obtained by maximizing network modularity. It can be seen from Figure~\ref{fig:modu}, both APS and MAG obtain the highest network modularity when $k=10$.

In general, choosing an optimal value $k$ is vital in CHIEF. We use CCP and CSP as indicators in order to choose a more proper value of $k$. To specifically make a better decision on choosing $k$, we then calculate network modularity to ensure that network modularity gets as close to 1 as possible. This procedure is quite important since different $k$ effect differently on partitioning networks. Cutting off edges with lower connectivity is quite meaningful, especially in clustering methods. With this procedure, network scale can be sharply reduced without reducing connectivity of the network. At the same time, the accuracy of clustering results would not be reduced. Based on our experimental results, the value of $k$ in finding maximal $k$-connected subgraphs is set to 11 for both MAG and APS (datasets).

\subsubsection{Motif Clustering}
\begin{figure*}[tbp]
  \centering
  \subfigure[$M_{42}$]{
    \includegraphics[width=0.195\textwidth]{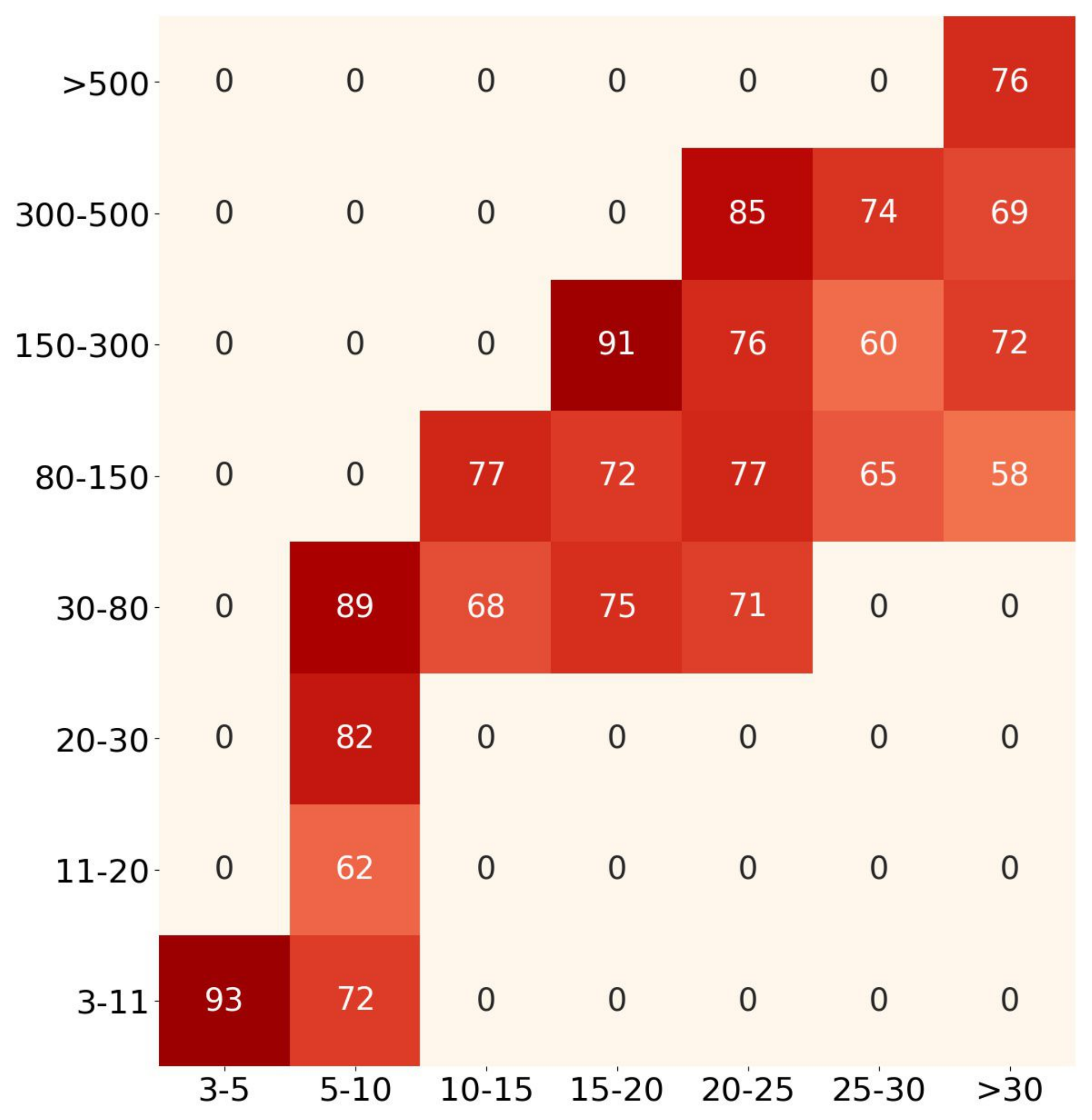}
  }
  \subfigure[$M_{43}$]{
    \includegraphics[width=0.17\textwidth]{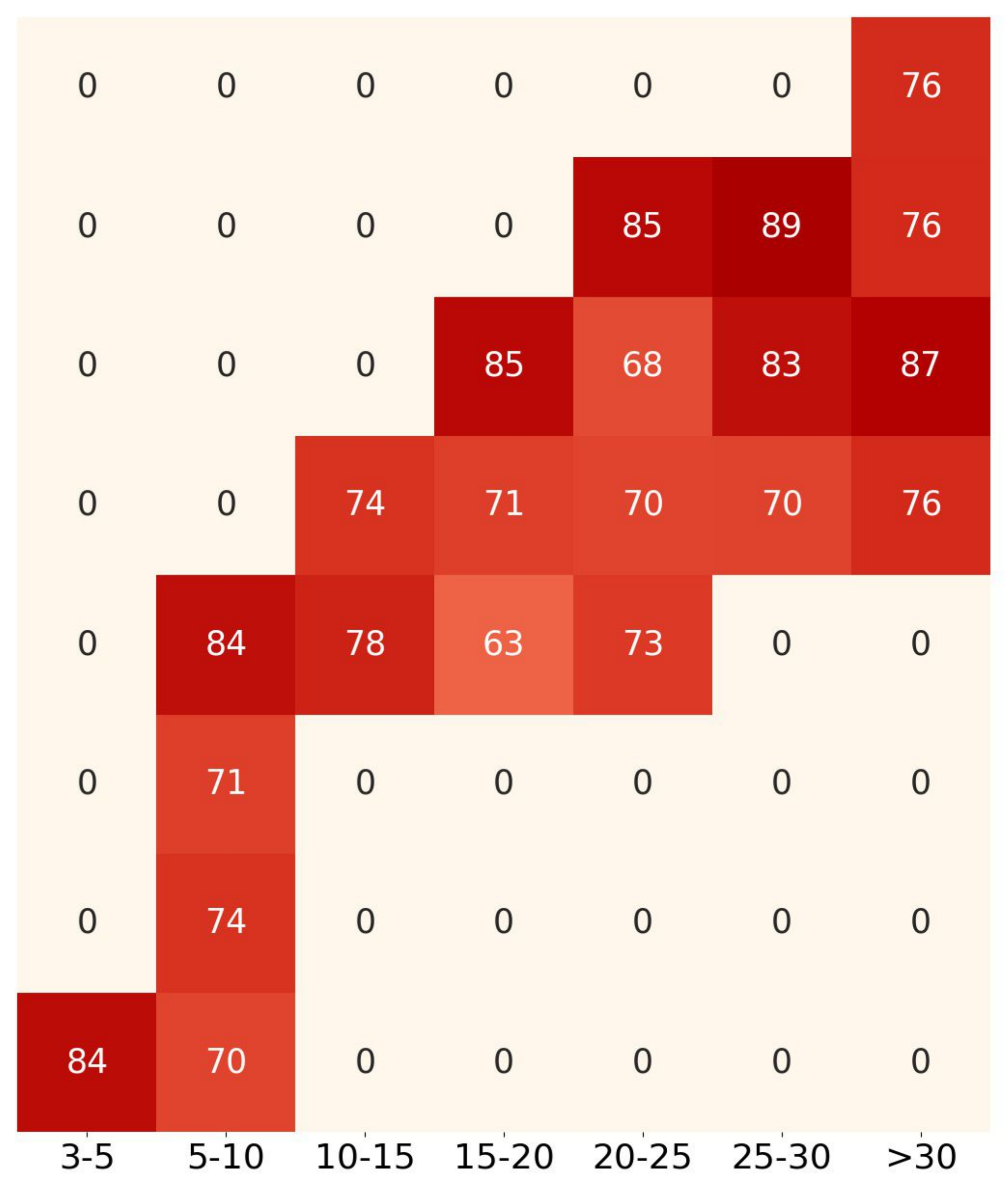}
  }
  \subfigure[$M_{44}$]{
    \includegraphics[width=0.17\textwidth]{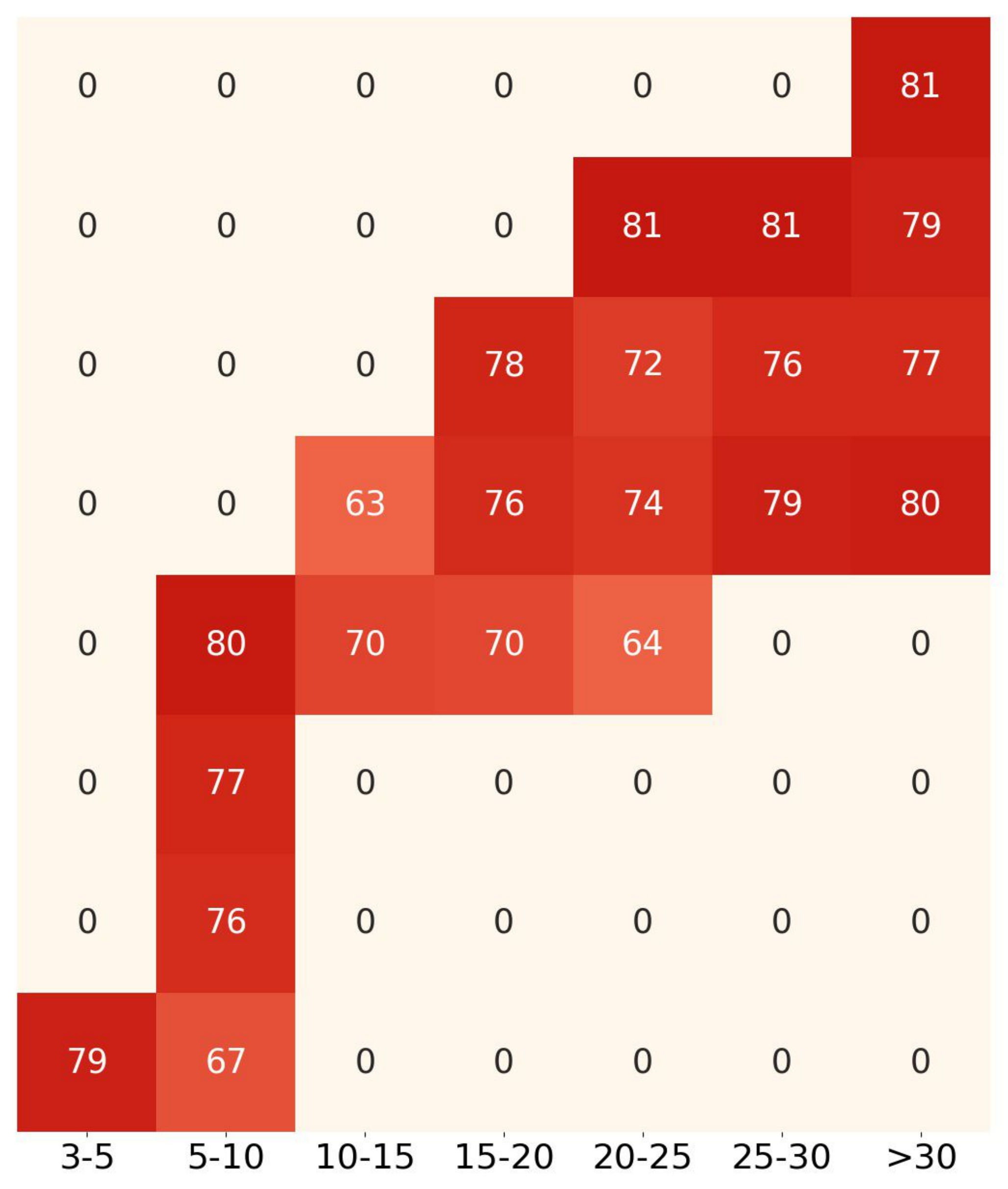}
  }
  \subfigure[$M_{45}$]{
    \includegraphics[width=0.17\textwidth]{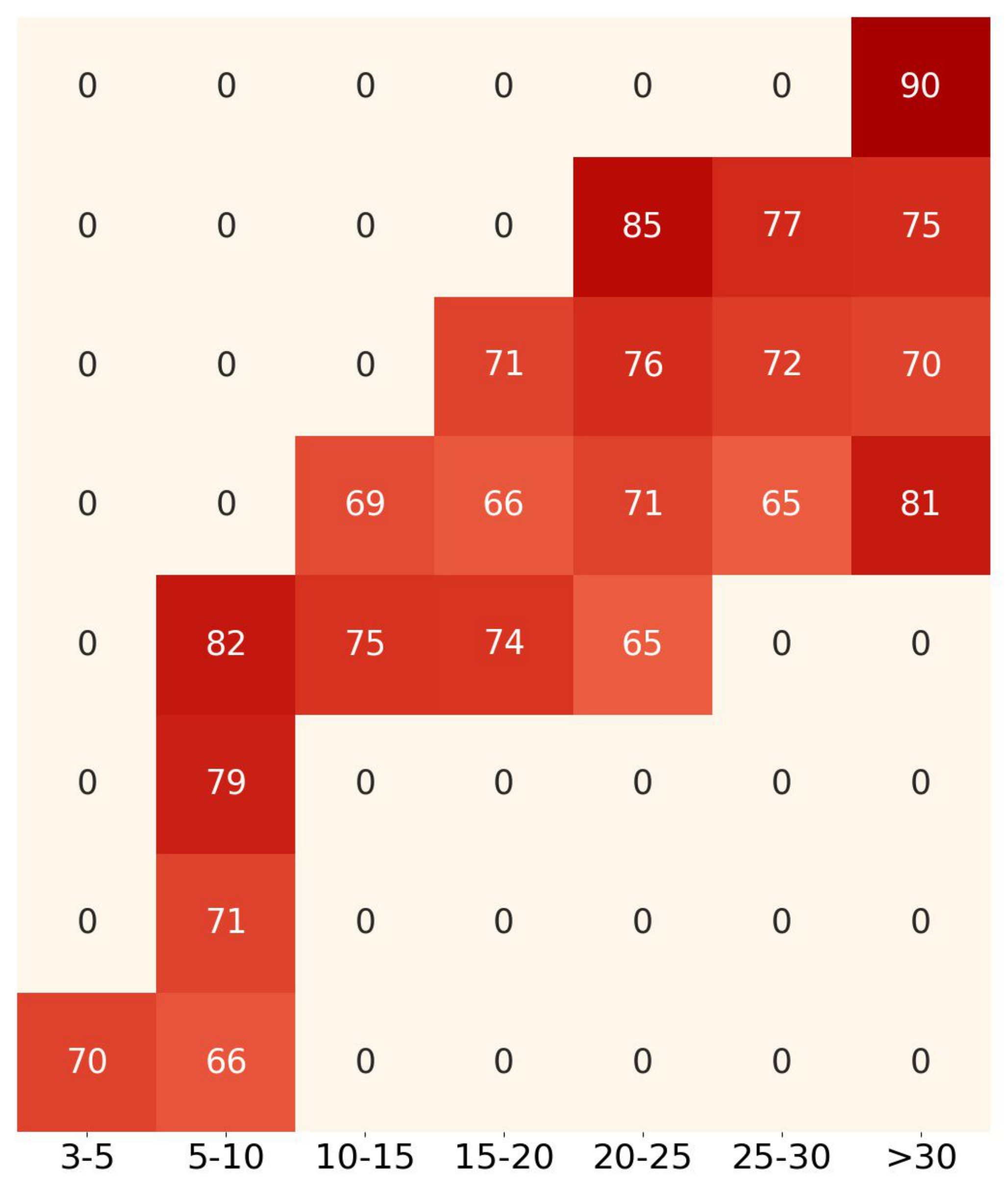}
  }
  \subfigure[$M_{46}$]{
    \includegraphics[width=0.17\textwidth]{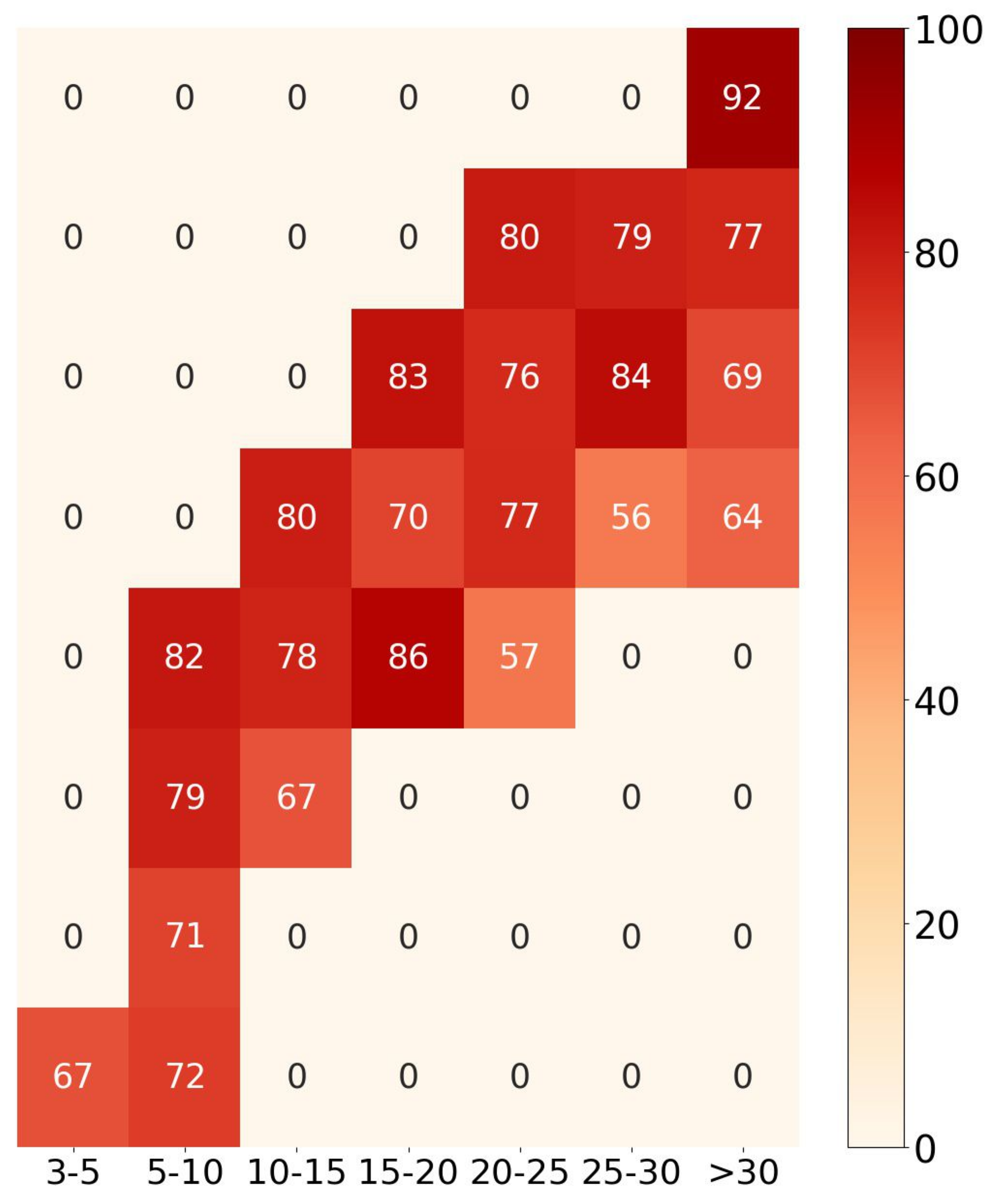}
  }\\
  \subfigure[$M_{42}$]{
    \includegraphics[width=0.195\textwidth]{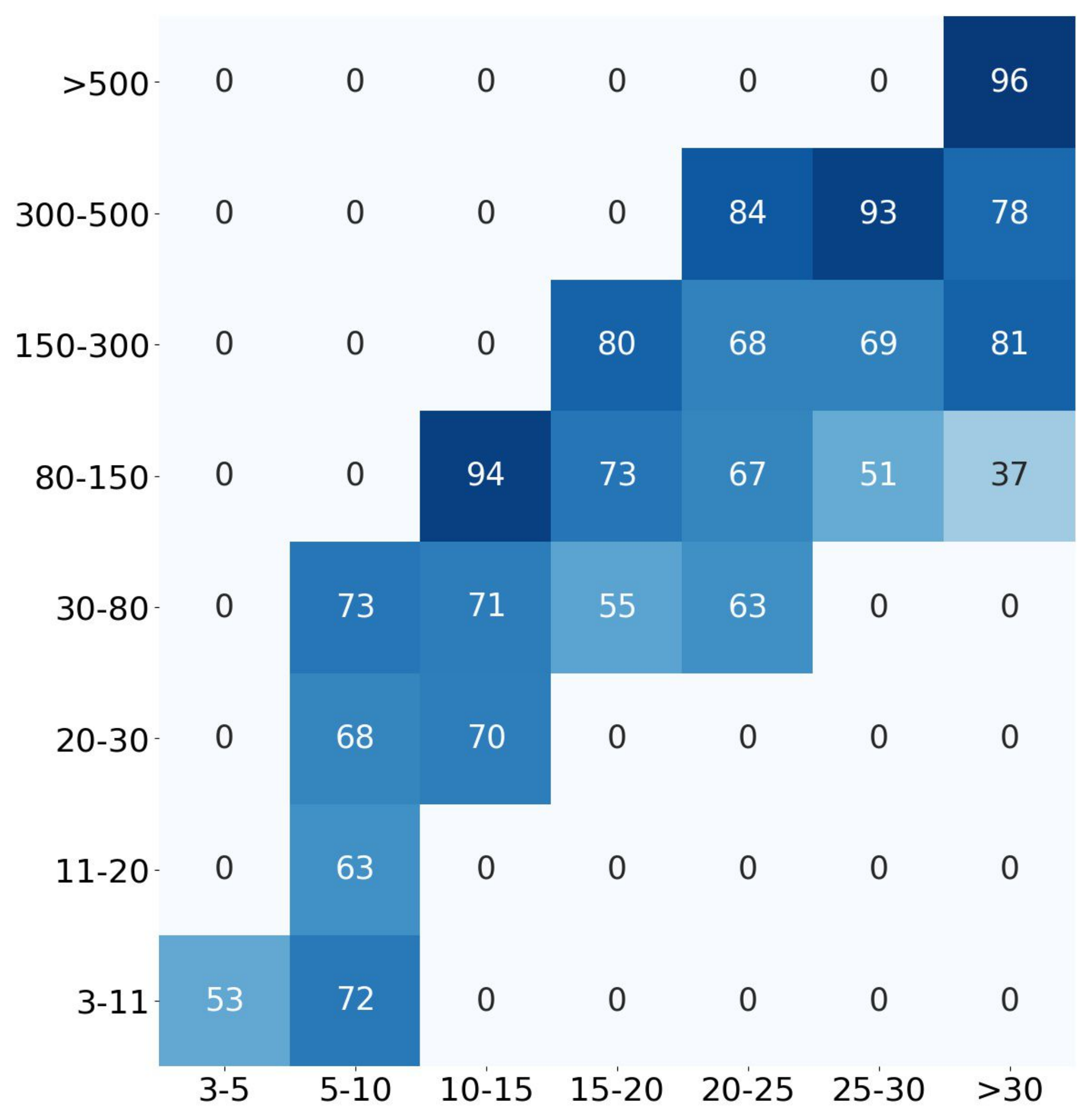}
  }
  \subfigure[$M_{43}$]{
    \includegraphics[width=0.17\textwidth]{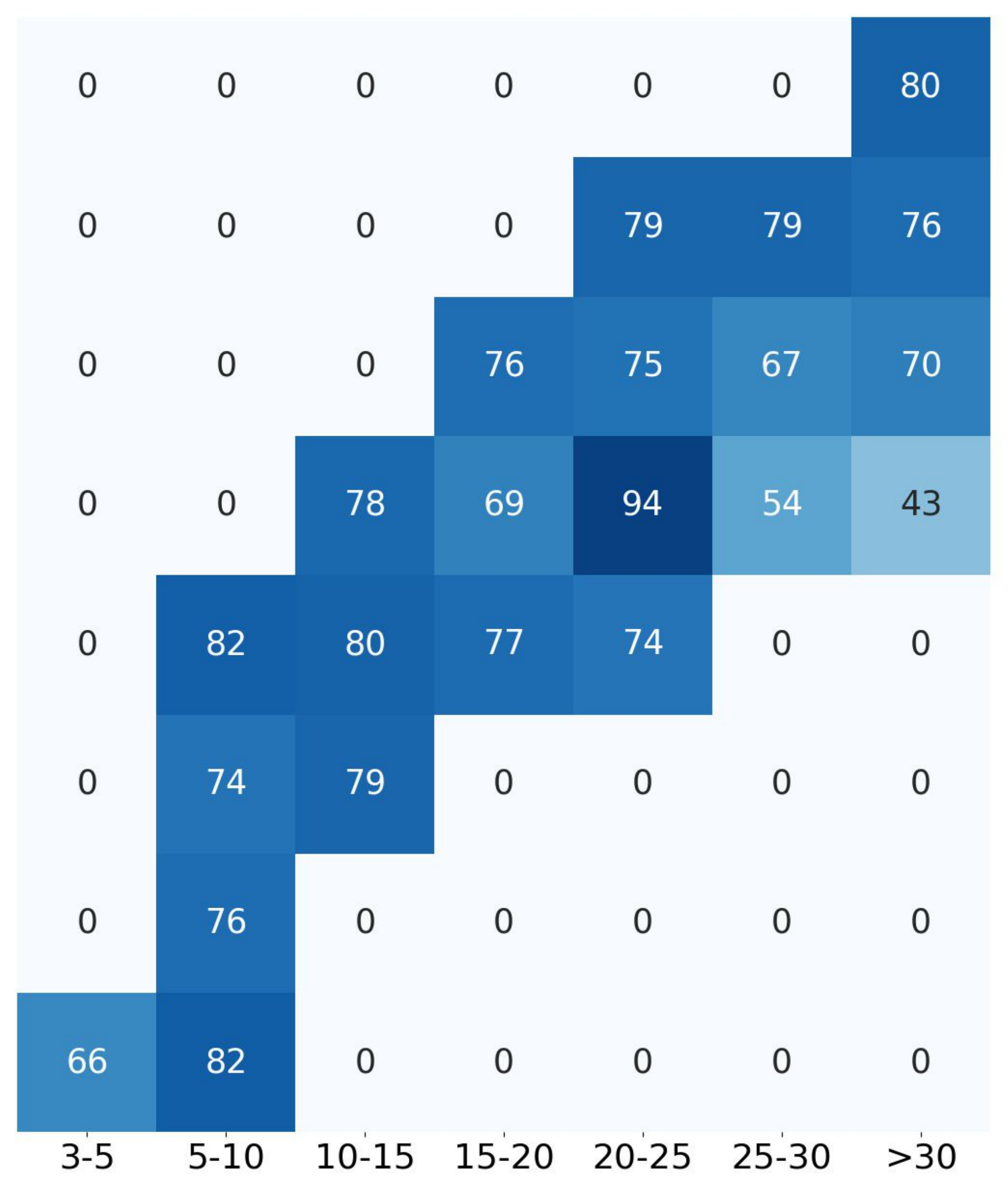}
  }
  \subfigure[$M_{44}$]{
    \includegraphics[width=0.17\textwidth]{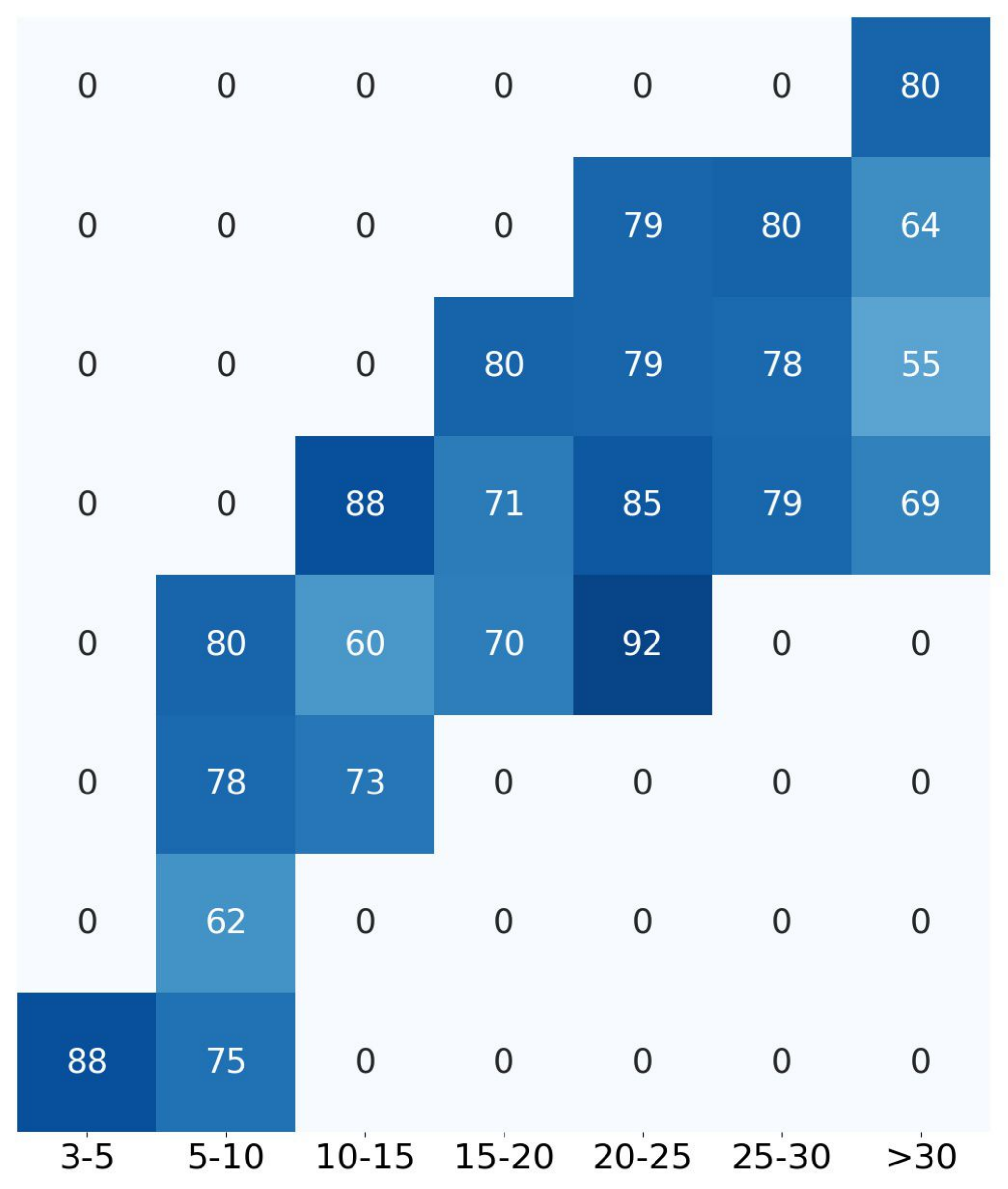}
  }
  \subfigure[$M_{45}$]{
    \includegraphics[width=0.17\textwidth]{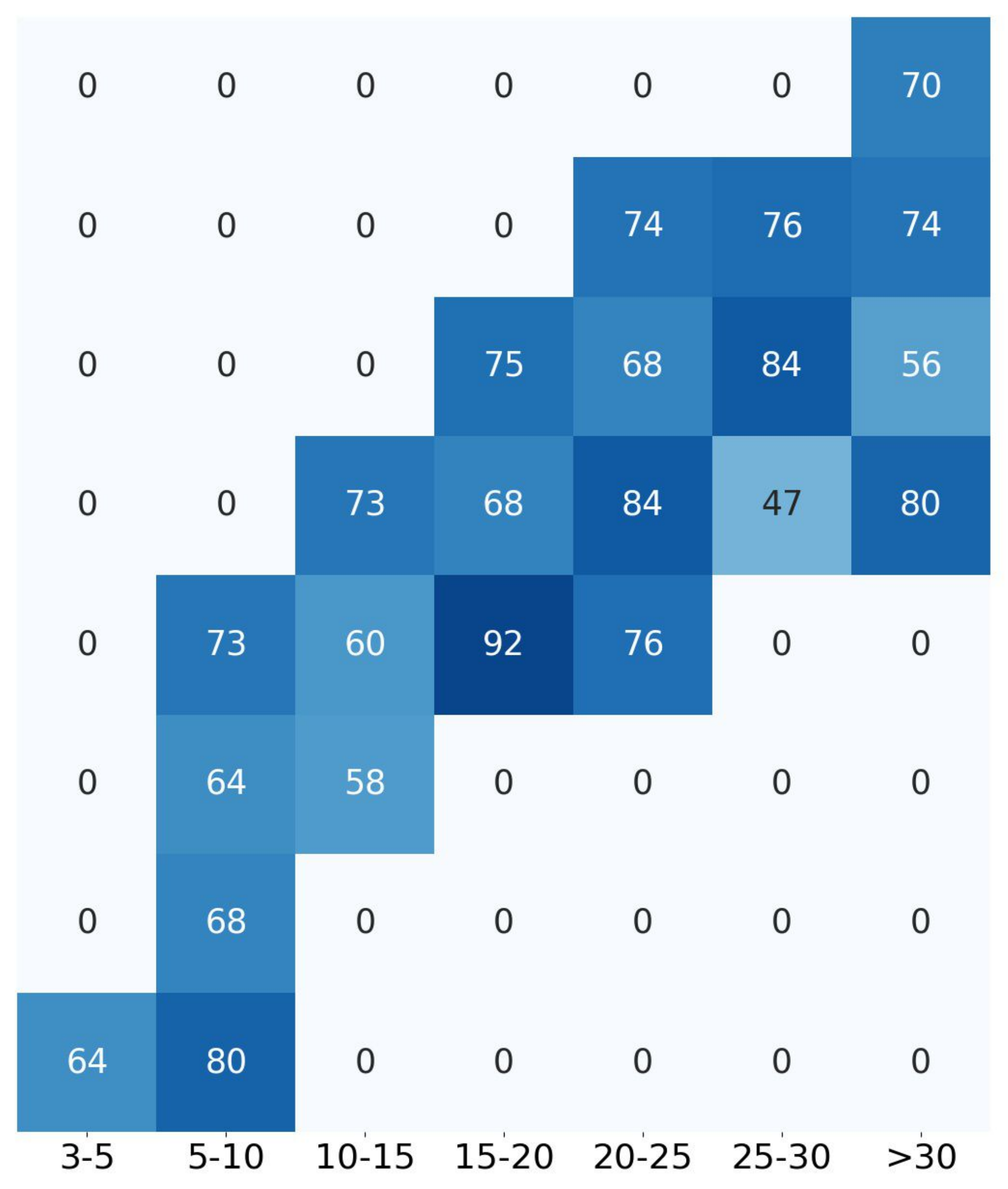}
  }
  \subfigure[$M_{46}$]{
    \includegraphics[width=0.17\textwidth]{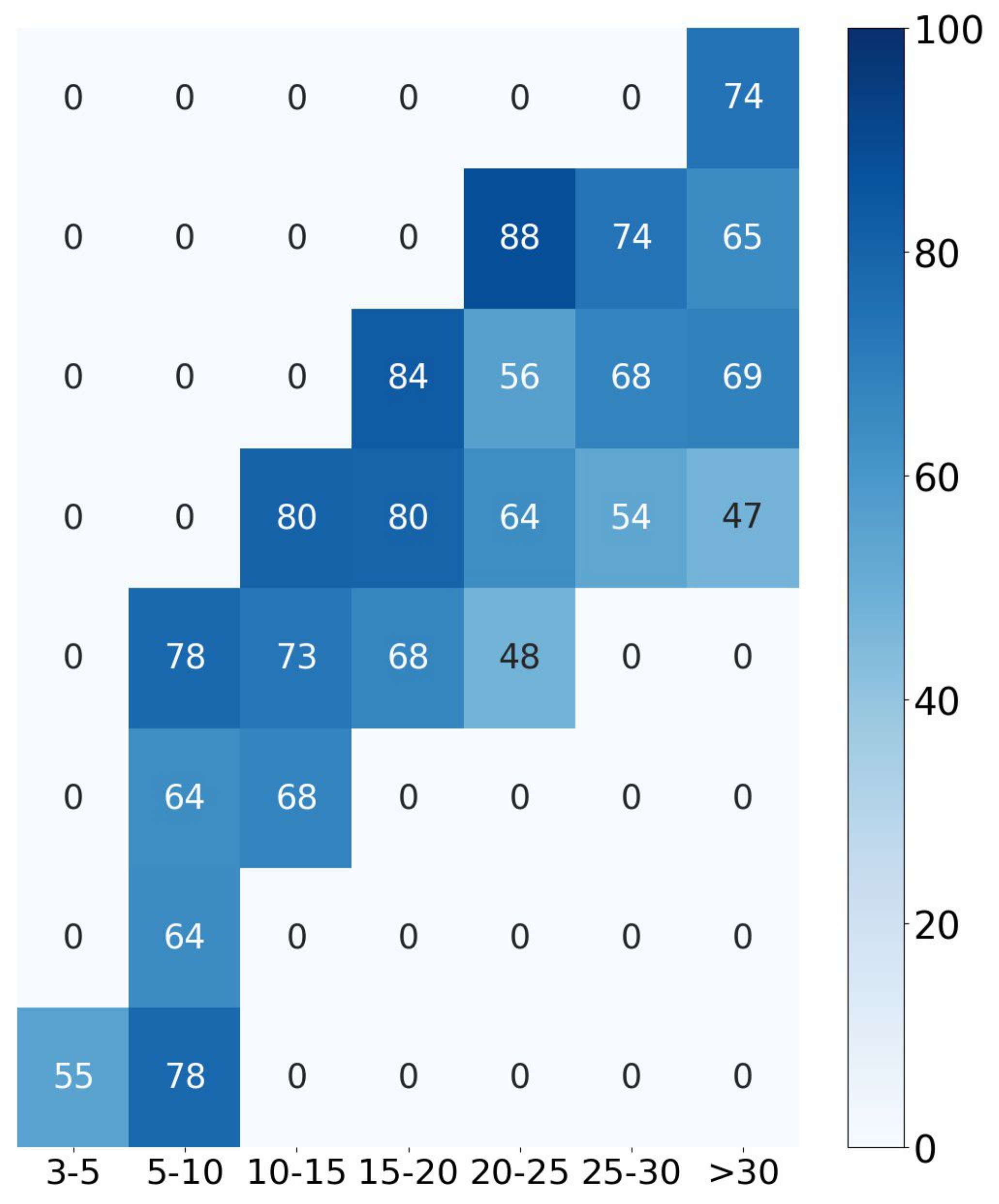}
  }
  \caption{Average CII in APS clusters and MAG clusters. Figure (a) to Figure (e) show the CII results in APS clusters and Figure (f) to Figure (j) show that in MAG clusters. The depth of the color indicates the number of clusters having specific vertices and edges. $X$ axis represent the number of vertices. $Y$ axis represents the number of edges. Each cell represents the ratio that how many scholars are with high CII.}
  \label{fig:ciiresult}
\end{figure*}
We achieve 10 groups of clustering results, wherein, 5 groups are for MAG while another 5 groups are for APS. To examine the clustering results in collaboration networks, we calculate average CII within each cluster. Considering the ``Rule of 150", we calculate CII between each couple of co-authored scholars and filter out scholars with top 20\% CII. We then cluster scholars with top 20\% CII. To examine whether the proposed algorithm achieves the best clustering results, we calculate the portion that how many scholars in high collaboration intensity are in the clustering results we achieved. The results show that our clustering results include all of scholars with top 20\% CII. Moreover, most ratios are larger than 60\%, which indicates that the clustering results achieved by CHIEF are both effective and accurate.

\begin{figure}[!b]
  \centering
  \subfigure[Academic Conference Networks]{
    \includegraphics[width=0.40\textwidth]{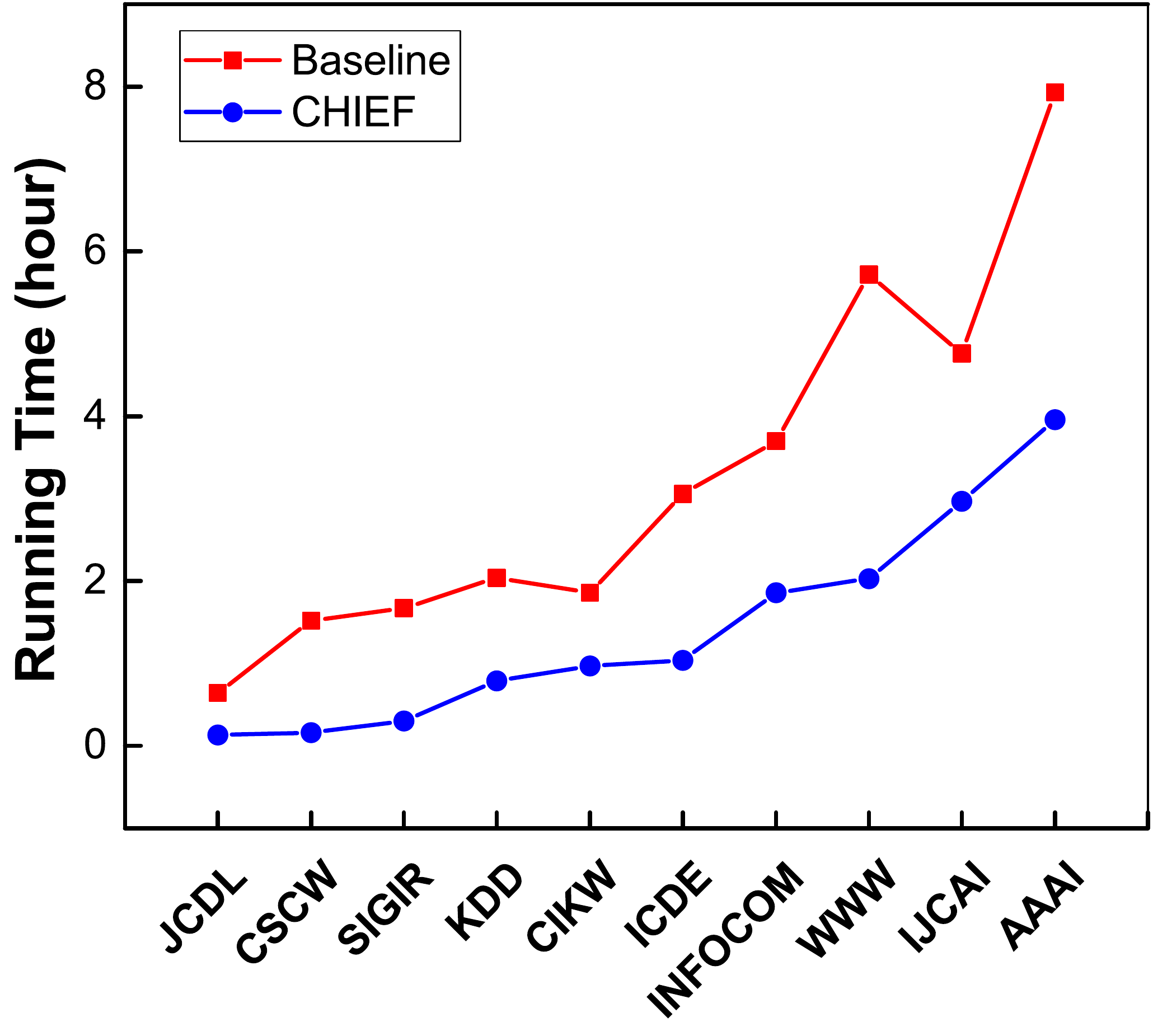}
  }
  \subfigure[Academic Journal Networks]{
    \includegraphics[width=0.40\textwidth]{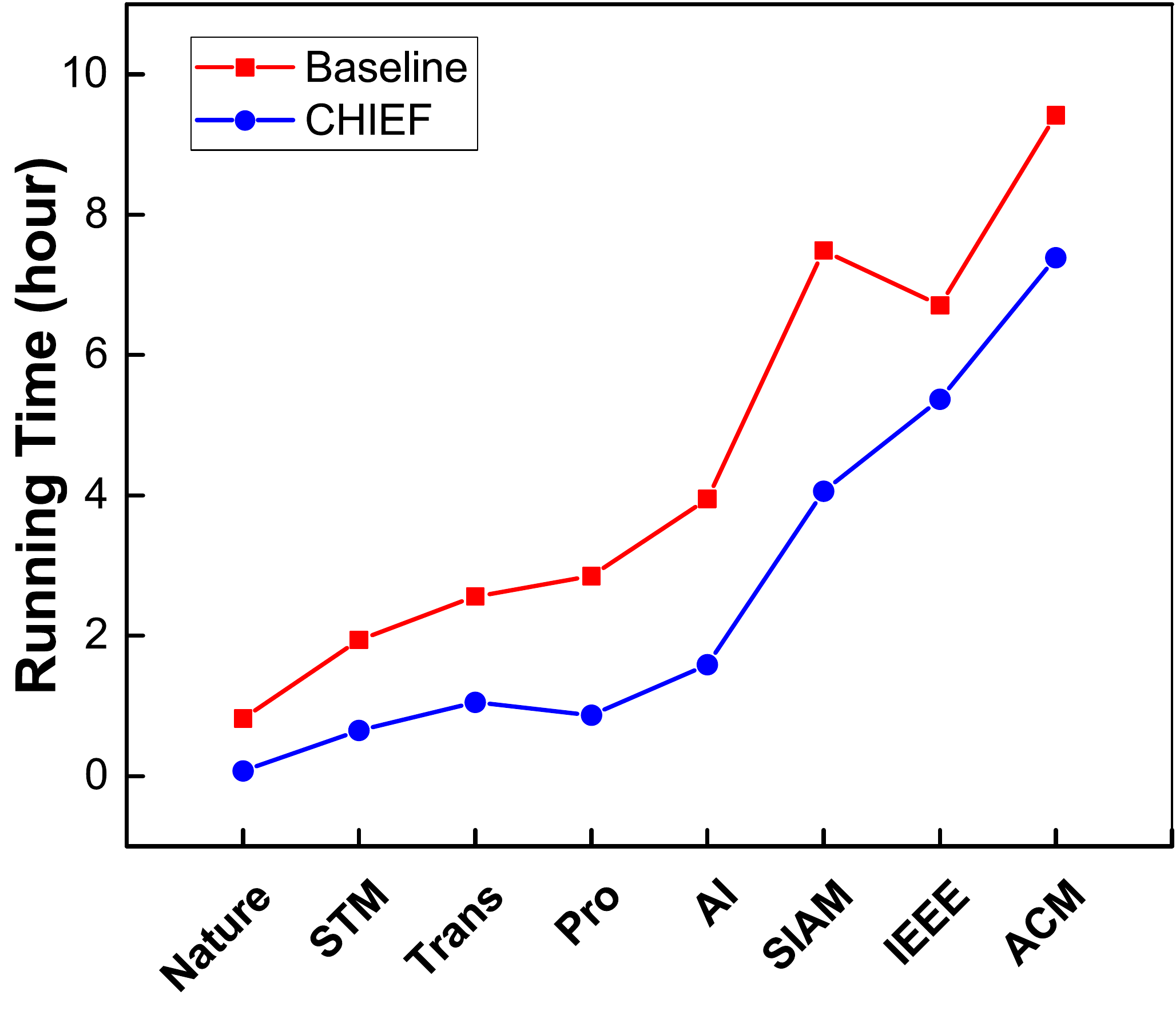}
  }
  \caption{Running time comparison of 18 academic networks in different scales.}
  \label{figmj}
\end{figure}

It turns out that all of the clusters are in high collaboration intensity. Experimental results are shown in Figure~\ref{fig:ciiresult}, wherein, the first 5 figures show CII in APS clusters and the latter 5 figures show that in MAG. To distinguish the results of two data sets, we use different colors in the two sets of graphs. In Figure~\ref{fig:ciiresult}, red series bars represent ratio of APS and blue series bars represent ratio of MAG.

We calculate the computational speed of CHIEF and a baseline method (motif clustering without $k$-connected subgraph finding process). We unprejudicedly select 10 top academic conferences and 8 top journals in MAG to generate 18 academic networks with different scales. Basic information of conferences is shown in Table~\ref{tab:m} and that of journals is shown in Table~\ref{tab:j}. Then we run both CHIEF and the baseline method on these 18 networks. Figure~\ref{figmj}(a) shows the running time comparison over the 10 academic conference networks. Figure~\ref{figmj}(b) shows that of the 8 top journal networks. It is noted that the operation time of IJCAI and IEEE are lower than that of other networks. The underlying reason is that these networks have different structures. Such difference might cause significant influence on operation time. Apparently, CHIEF is more efficient than the baseline algorithm. This indicates that $k$-connected subgraph finding process accelerates motif clustering. As mentioned previously, clustering results achieved from CHIEF-AP are in high collaboration intensity. Thus CHIEF-AP is both effective and efficient.

\begin{table}[!hb]
  \caption{Basic Information about Generated Synthetic Networks}
  \centering
  \begin{tabular}{ccccc}
    \toprule
    Network Label & NV & NE & RRP & Opt $k$\\
    \midrule
    N1 & $10^2$ & $3 \times 10^2$ & 0.2 & 5\\
    N2 & $10^3$ & $3 \times 10^3$ & 0.3 & 5\\
    N3 & $10^4$ & $3 \times 10^4$ & 0.4 & 4\\
    N4 & $10^5$ & $3 \times 10^5$ & 0.5 & 4\\
    N5 & $10^6$ & $3 \times 10^6$ & 0.6 & 3\\
    \bottomrule
  \end{tabular}
  \label{tab:an}
\end{table}

\subsection{Implementing CHIEF on Synthetic Networks}

To examine both the universality and the efficiency of the proposed algorithm, we generate synthesized networks at different scales. Table~\ref{tab:an} shows the details of generated synthetic networks. After generation, we calculate the optimal value of $k$ for each network. The last column of Table~\ref{tab:an} lists the values. In contrast to the real-world networks, synthetic networks are far denser than collaboration networks. The connectivity of a dense network is usually higher than a sparse one. Therefore, the optimal value of $k$ is 11 in MAG and APS collaboration networks, and $k<6$ in synthetic networks. In Table~\ref{tab:an}, NV represents the number of vertices, and NE represents the number of edges, RRP is randomization reconnection probability, and Opt $k$ represents the optimal value of $k$.

\begin{figure*}[tp]
  \centering
  \subfigure[N1]{
    \includegraphics[width=0.18\textwidth]{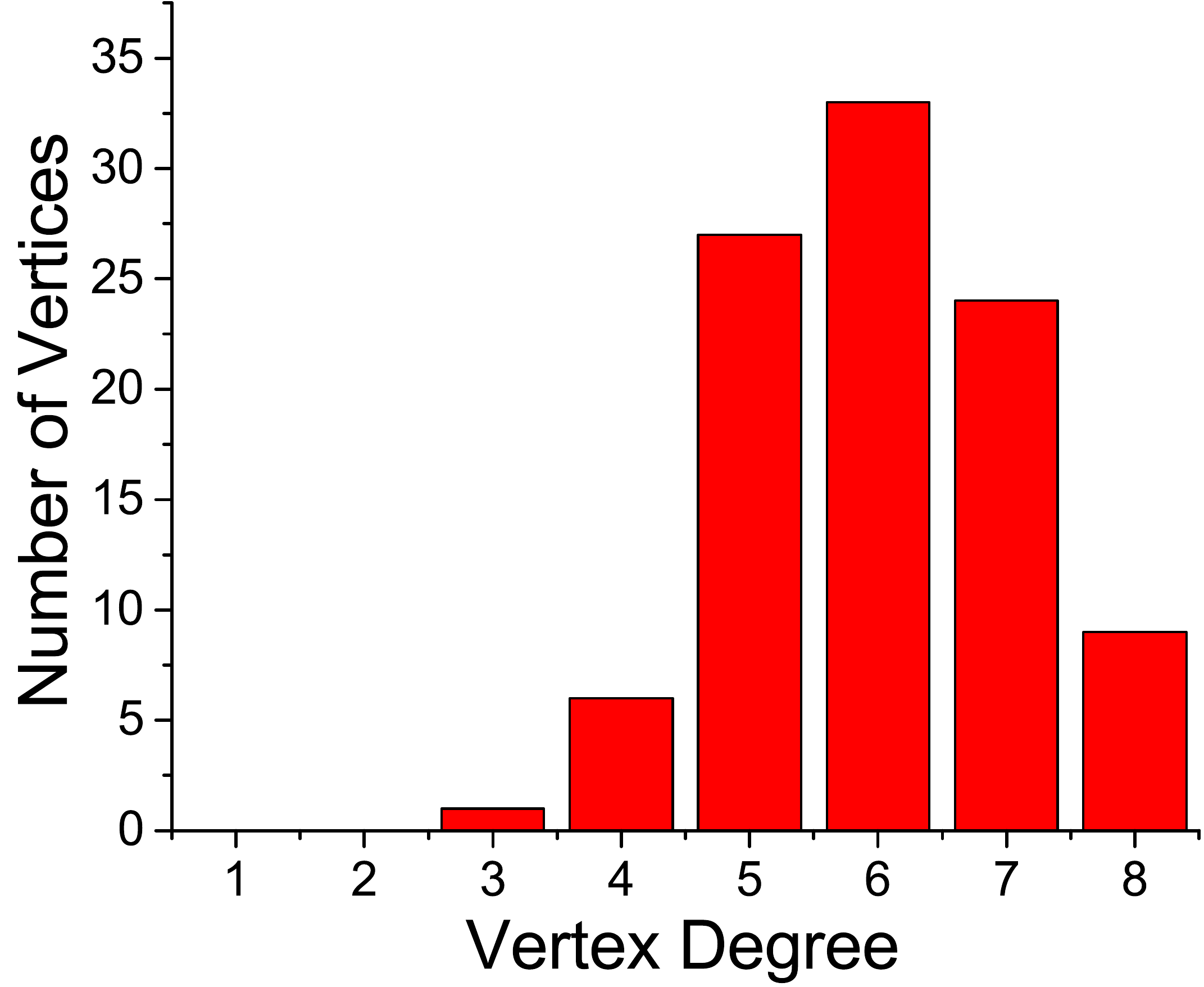}}
  \subfigure[N2]{
    \includegraphics[width=0.18\textwidth]{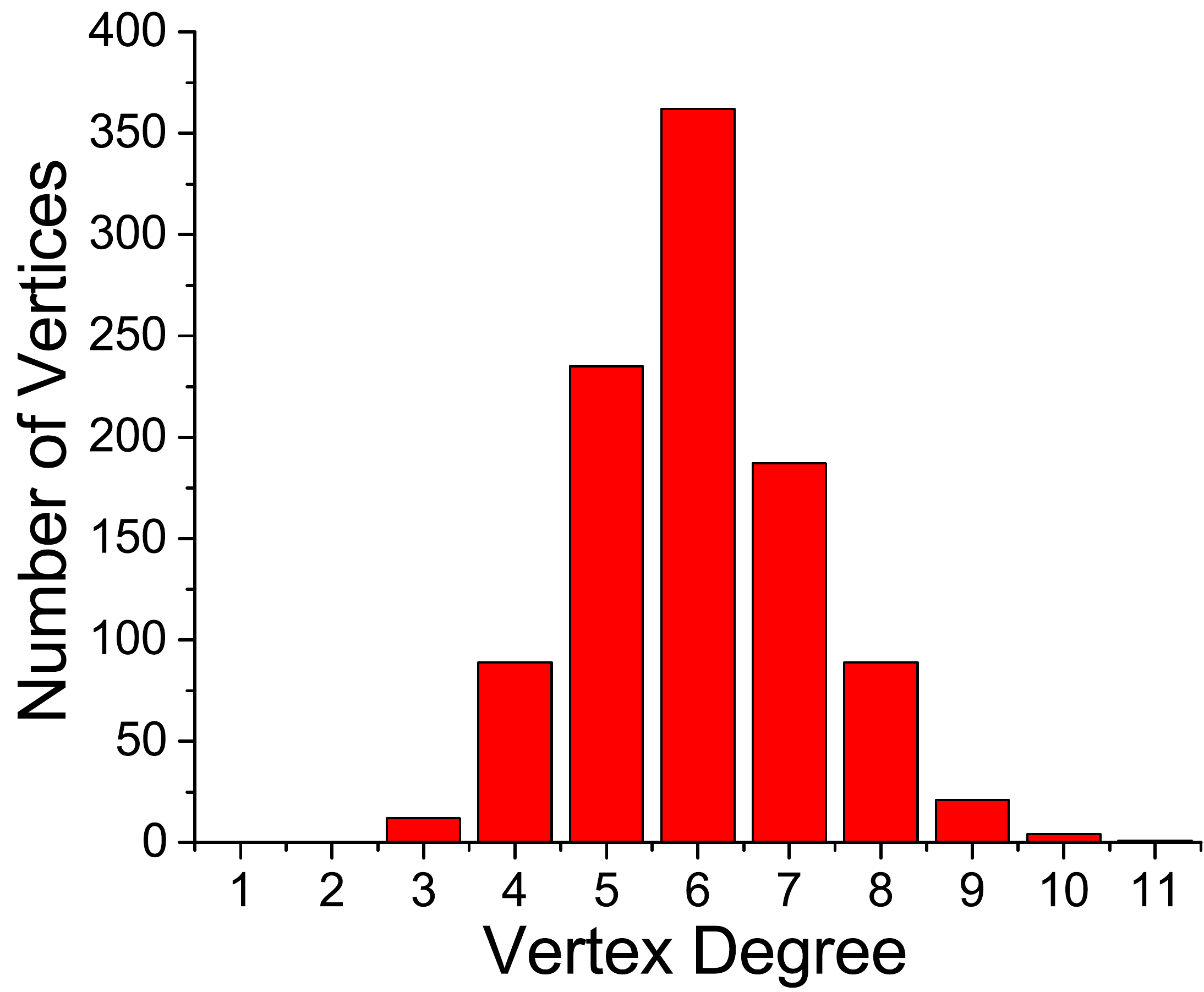}}
  \subfigure[N3]{
    \includegraphics[width=0.18\textwidth]{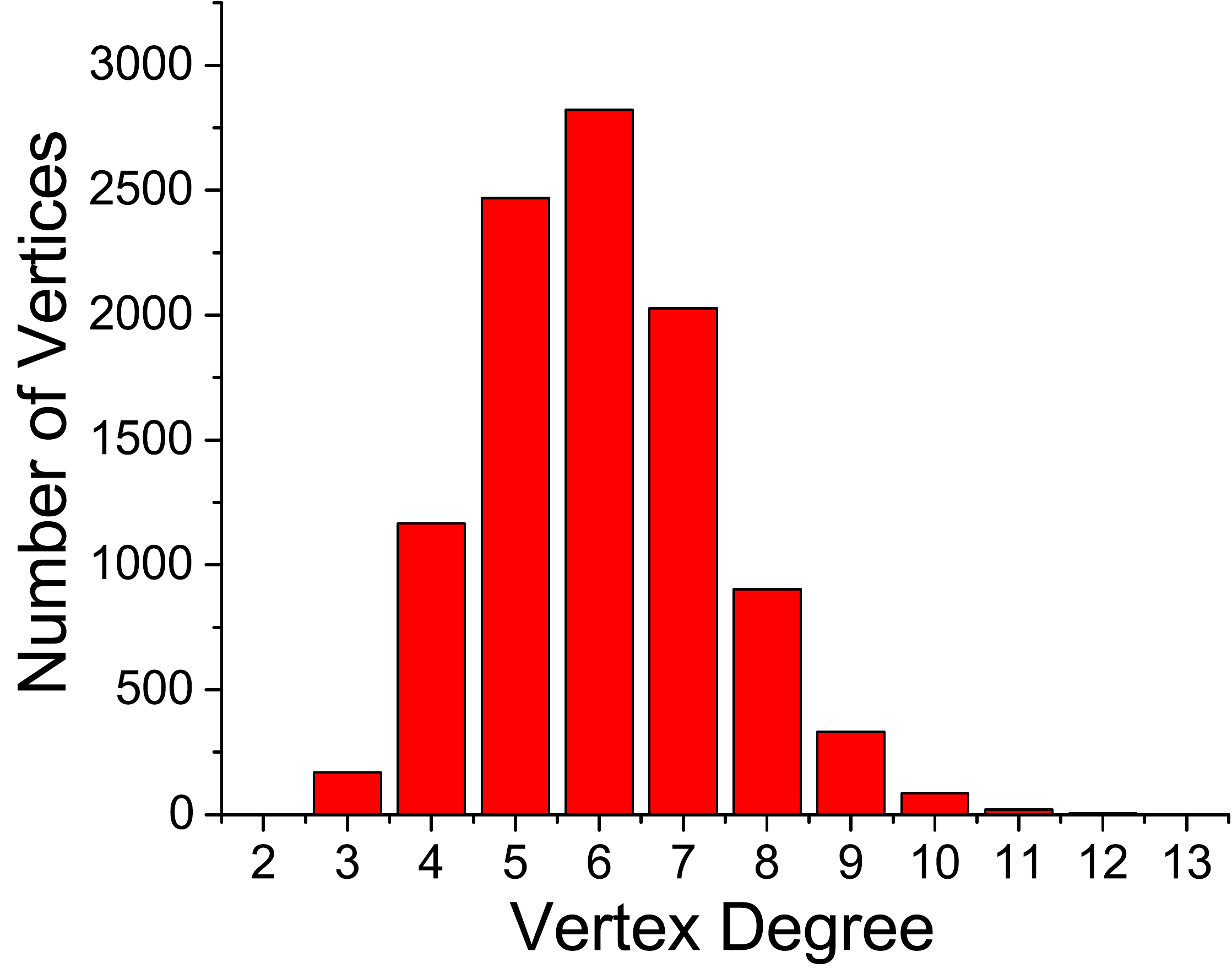}}
  \subfigure[N4]{
    \includegraphics[width=0.18\textwidth]{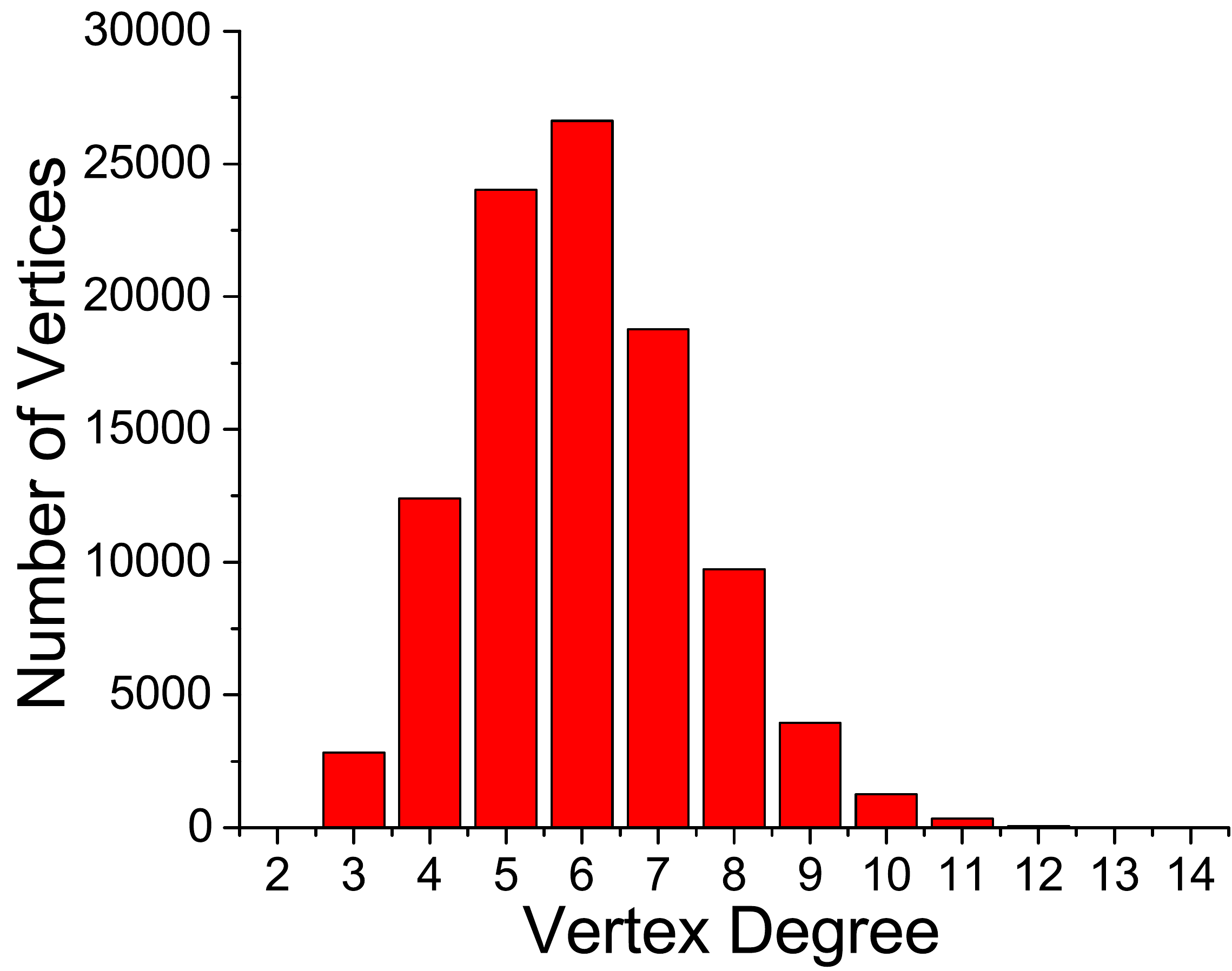}}
  \subfigure[N5]{
    \includegraphics[width=0.18\textwidth]{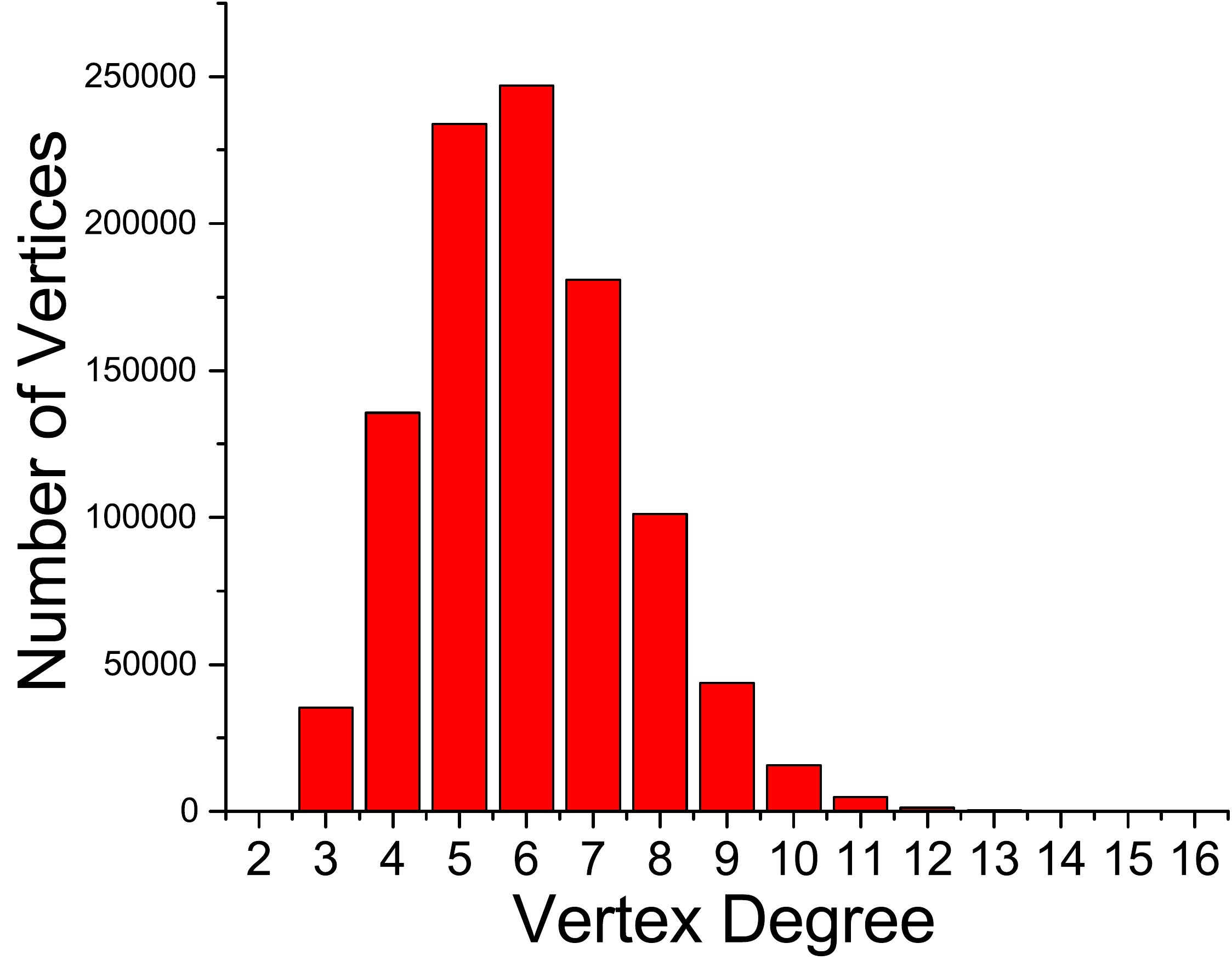}}
  \caption{The degree distribution of synthetic networks.}
  \label{fig:degree}
\end{figure*}

\begin{figure*}[tp]
  \centering
  \subfigure[N1]{
    \label{fig:an-6:1}
    \includegraphics[width=0.18\textwidth]{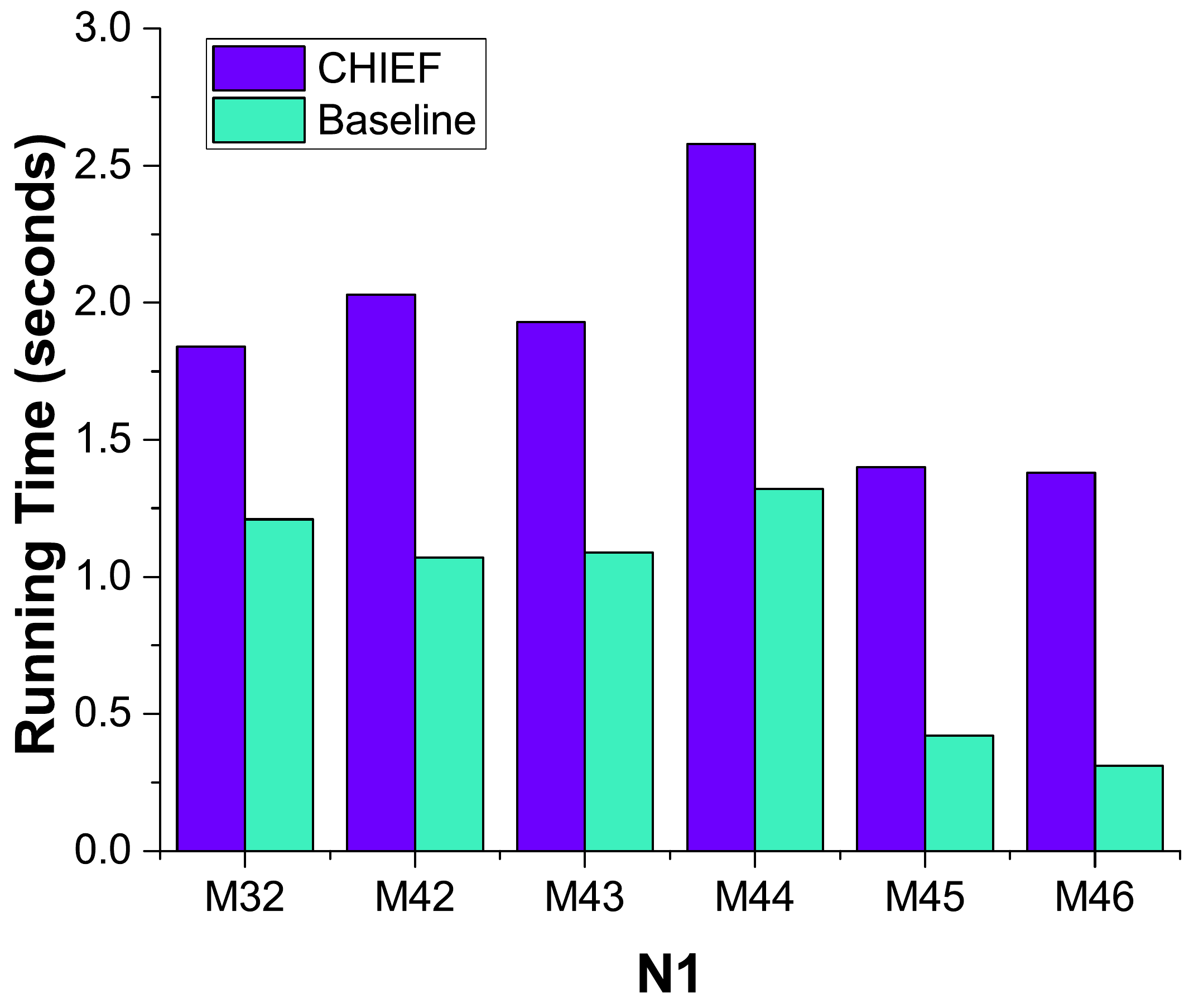}}
  \subfigure[N2]{
    \label{fig:an-6:2}
    \includegraphics[width=0.18\textwidth]{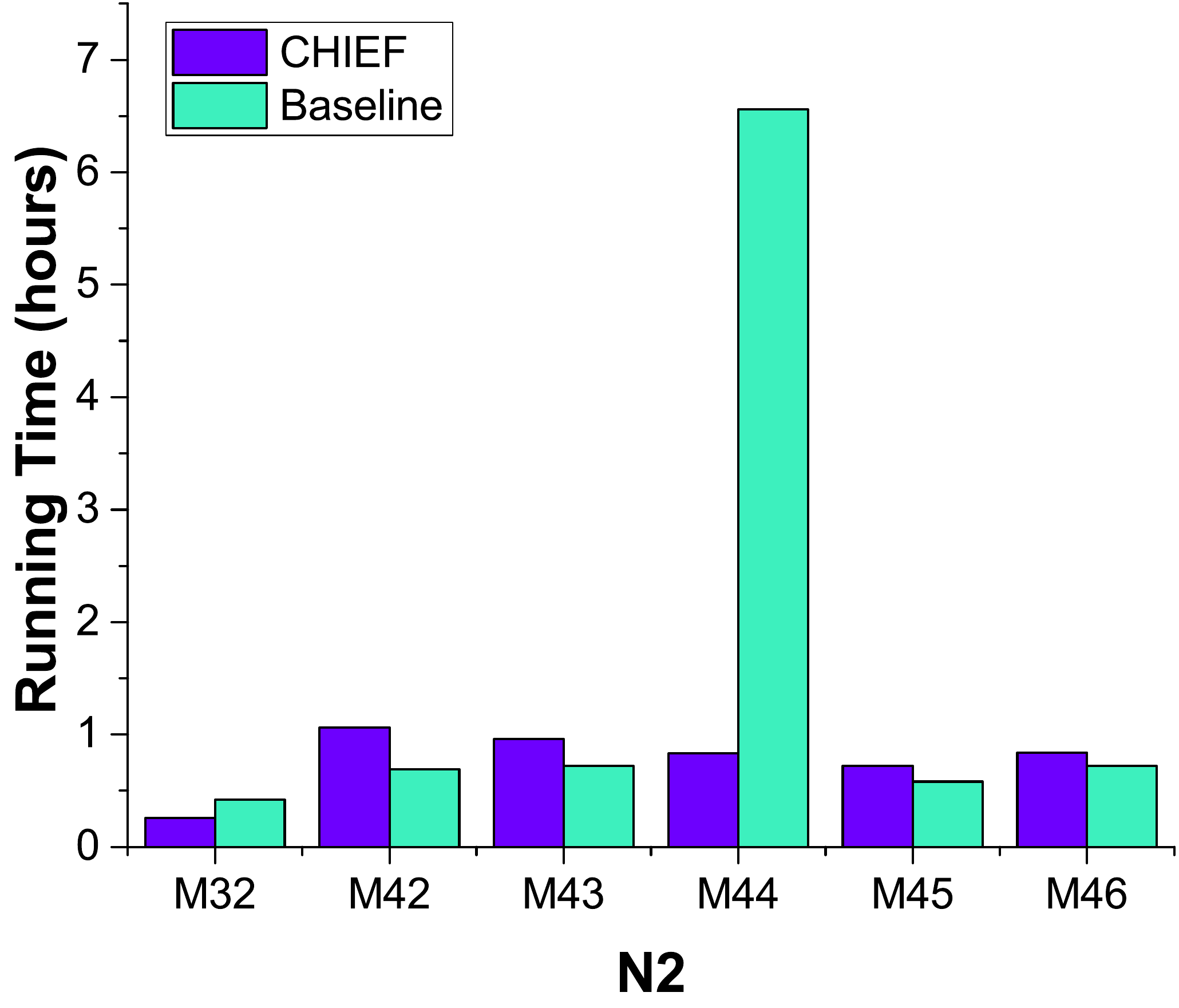}}
  \subfigure[N3]{
    \label{fig:an-6:3}
    \includegraphics[width=0.18\textwidth]{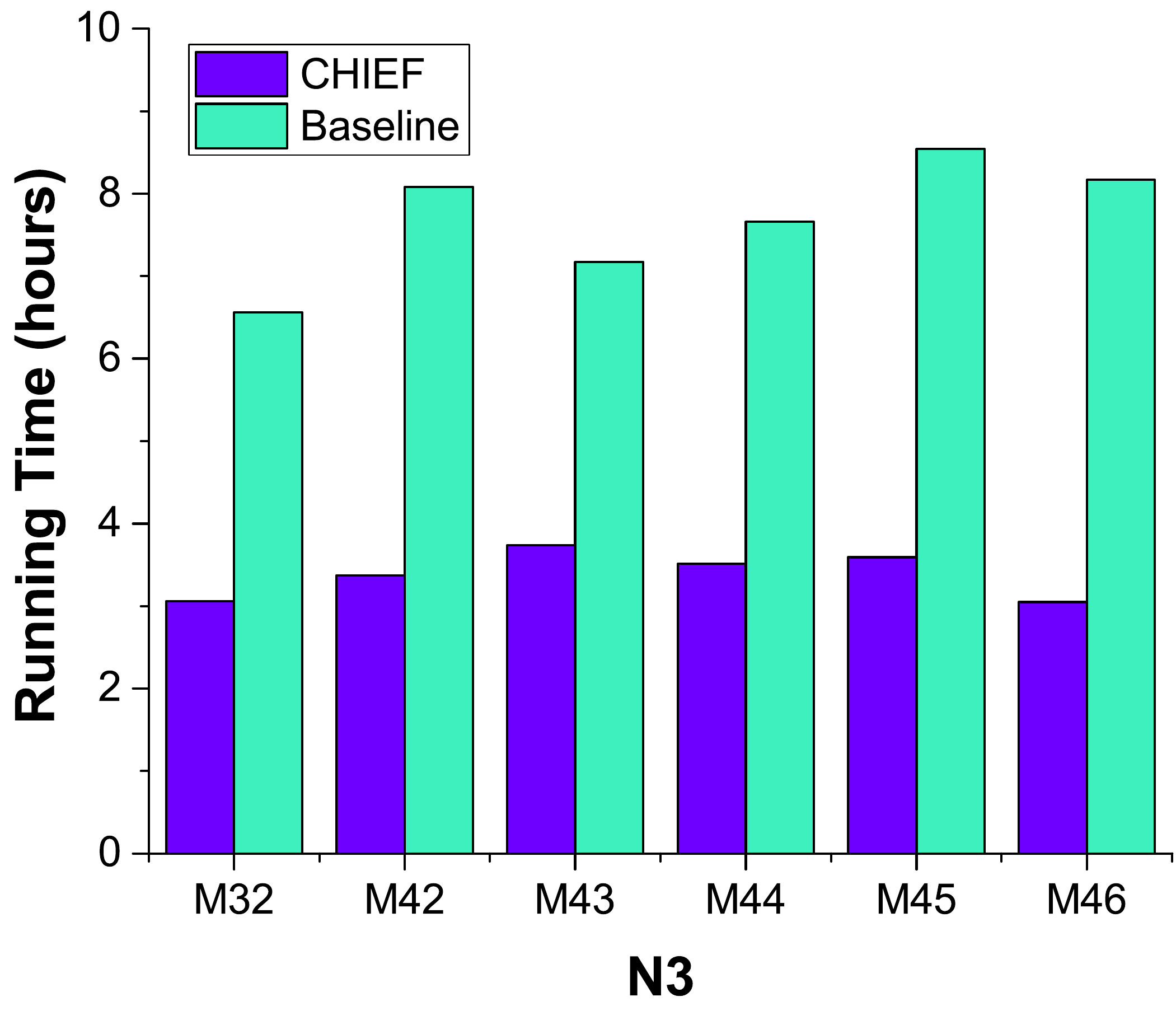}}
  \subfigure[N4]{
    \label{fig:an-6:4}
    \includegraphics[width=0.18\textwidth]{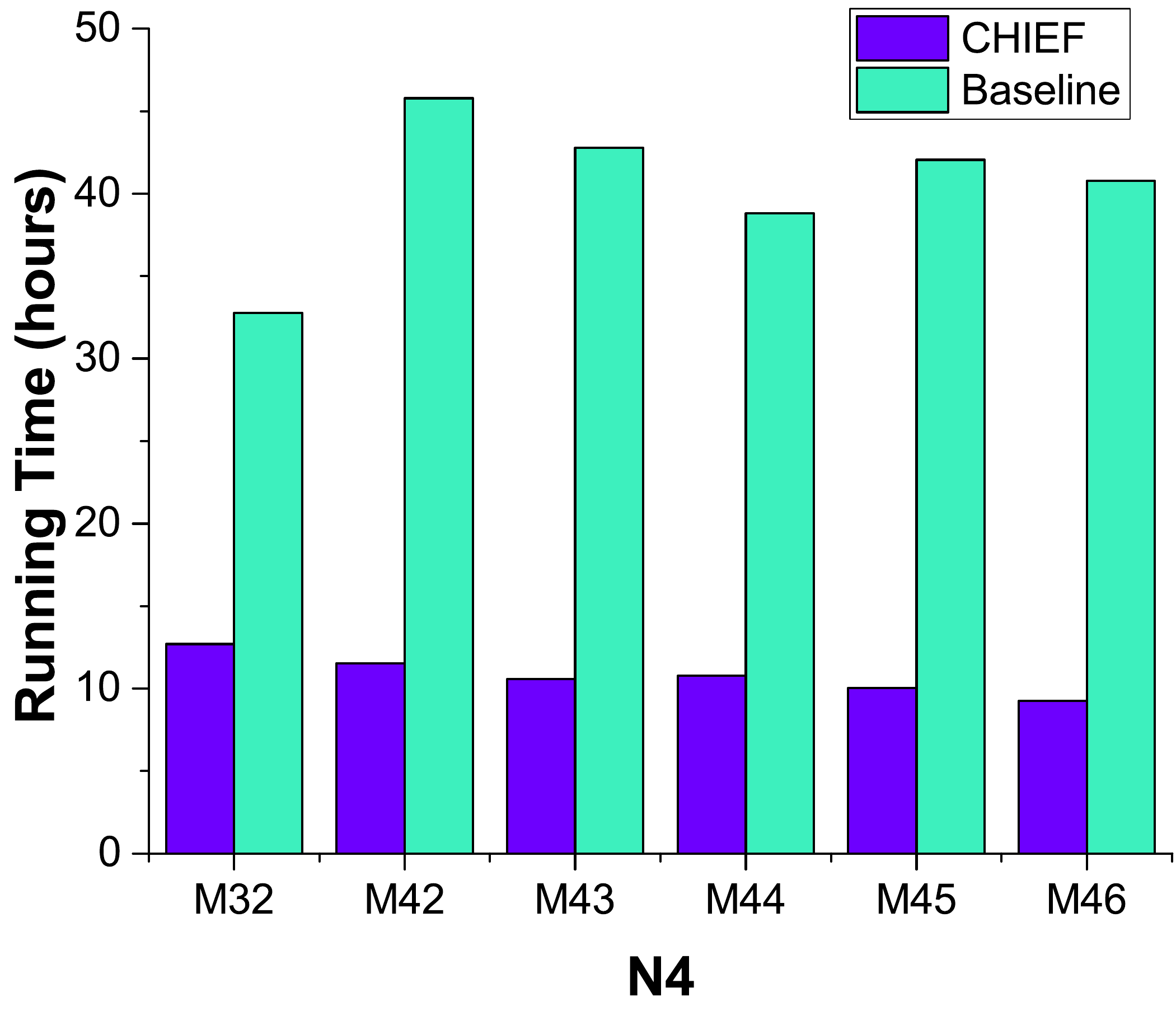}}
  \subfigure[N5]{
    \label{fig:an-6:5}
    \includegraphics[width=0.18\textwidth]{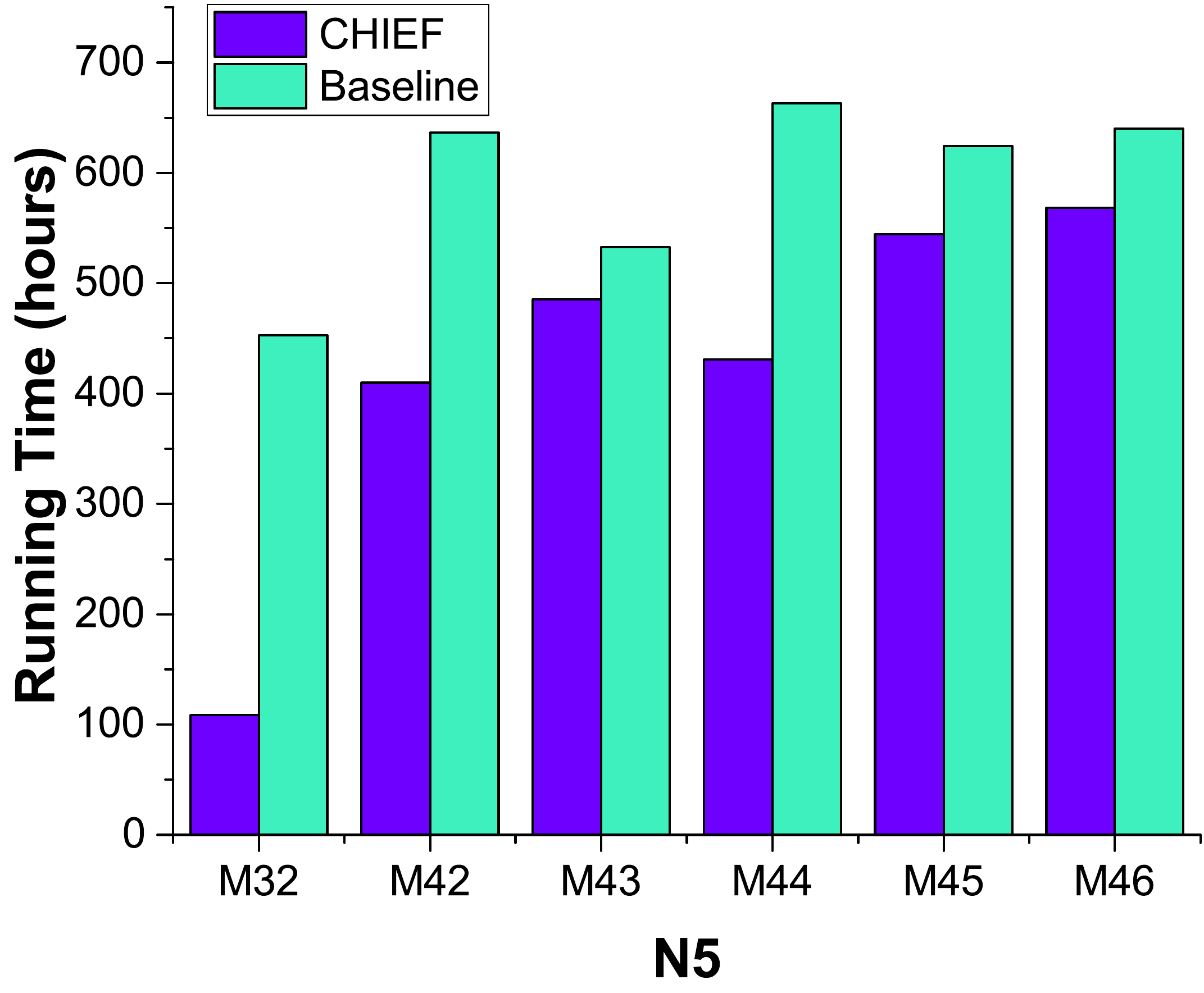}}
  \caption{Comparison of CHIEF and the baseline method in terms of the execution time, with subgraphs illustrating results under different motifs.}
  \label{fig:an-6}
\end{figure*}
\begin{table*}[tb]
  \centering
  \caption{Average CCP and CSP of Synthetic Networks}
  \begin{tabular}{c|cc|cc|cc}
    \toprule
    \multicolumn{1}{c|}{\multirow{2}[4]{*}{Network Label}} & \multicolumn{2}{c|}{$M_{32}$} & \multicolumn{2}{c|}{$M_{42}$} & \multicolumn{2}{c}{$M_{43}$} \\
    \cmidrule{2-7}          & Avg CCP & Avg CSP & Avg CCP & Avg CSP & Avg CCP & Avg CSP \\
    \midrule
    N1    & 1.3843  & 4.6932  & 1.3503  & 4.2036  & 1.6727  & 4.3645  \\
    N2    & 2.3797  & 5.7053  & 1.8268  & 5.2302  & 1.8552  & 5.4803  \\
    N3    & 2.0140  & 6.5789  & 1.9038  & 5.4463  & 1.8412  & 8.6171  \\
    N4    & 2.0527  & 7.5913  & 1.6070  & 8.4216  & 1.6553  & 6.6448  \\
    N5    & 1.8099  & 6.6005  & 1.7696  & 6.0328  & 1.6974  & 5.6340  \\
    \midrule
    \multicolumn{1}{c|}{\multirow{2}[4]{*}{Network Label}} & \multicolumn{2}{c|}{$M_{44}$} & \multicolumn{2}{c|}{$M_{45}$} & \multicolumn{2}{c}{$M_{46}$} \\
    \cmidrule{2-7}          & Avg CCP & Avg CSP & Avg CCP & Avg CSP & Avg CCP & Avg CSP \\
    \midrule
    N1    & 1.3055  & 4.9120  & 1.3724  & 4.1697  & 1.2326  & 4.5087  \\
    N2    & 1.6704  & 4.9110  & 2.4855  & 4.2562  & 1.5271  & 6.6375  \\
    N3    & 1.7426  & 4.8522  & 2.1669  & 5.1319  & 1.9385  & 7.4977  \\
    N4    & 1.8754  & 6.0901  & 1.7902  & 7.8963  & 1.7181  & 8.7147  \\
    N5    & 2.0034  & 5.2313  & 1.7754  & 7.6404  & 1.8618  & 8.3115  \\
    \bottomrule
  \end{tabular}
  \label{tab:cpsp}
\end{table*}

To imitate real-world social networks as much as possible, we use the same vertices degree distribution as small world. The vertices degree distribution of the five networks are shown in Figure~\ref{fig:degree}. It can be seen from Figure~\ref{fig:degree} that the degree distributions of artificial networks are relatively concentrated. We implement experiments without data preprocessing, which is different from experiments on collaboration networks. Actually, data preprocessing can remove infrequent vertices and edges, which makes it impossible to divide large component into several small connected components.

It is worth mentioning that though we generate synthetic networks to simulate real-world social networks, synthetic networks are still much denser than real world ones. Therefore, real social networks own stronger community characteristics than synthetic networks. However, the distinction has no effect on the experimental results in this subsection.

The experimental results are shown in Figure~\ref{fig:an-6}. Subgraphs of Figure~\ref{fig:an-6} show the comparisons of the running times consumed by CHIEF-AP and the baseline method, in which each subgraph shows the results for motif $M_{32}$, $M_{42}$, $M_{43}$, $M_{44}$, $M_{45}$, $M_{46}$, respectively. Generally, the efficiency of CHIEF precedes the baseline method significantly. This is because that when the scale of network increases, global motif cluster algorithm consumes much more time. Still, we can see something unusual from Figure~\ref{fig:an-6}. Though CHIEF-AP outperforms for the most time, the baseline method performs better under some circumstances. When implementing the two methods in network N1, the baseline method consumes less time than CHIEF-AP does. This is because that when the network scale is extremely small, finding maximal $k$-connected subgraphs is time-consuming, which makes the whole computing time higher than the baseline method. In the other four subgraphs, CHIEF-AP performs better than the baseline method no matter which motif is used in clustering. It is worth mentioning that in network N2, the baseline method consume an abnormal running time when using $M_{42}$ in clustering. This may be caused by two factors, i.e., motif structure and network structure. Motif structure of $M_{42}$ is complicate, which makes the clustering time extremely high. The abnormal running time may also be caused by the network structure, since baseline method did not partition network firstly. Generally, the efficiency of CHIEF-AP is much higher than baseline method, especially for large-scale networks. In the era of big data and large-scale networks, highly efficient algorithms like CHIEF is meaningful since high efficiency algorithms are in great demand.

The CCP and CSP values are shown in Table~\ref{tab:cpsp}. It can be seen that for all of motifs in this experiment, the minimum value of CCP equals 1.0000. According to statics, there are quite a number of clusters with CCP=1.0000. As we mentioned above, a lower CCP refers to a tighter cluster. This indicates that most of our clustering results are tight. Besides, maximum CCP of $M_{32}$ are relatively low. Even so, most CCP values are close to the minimum one, i.e., 1.0000. Take N3 and N4 as examples, the maximum CCP values of $M_{32}$ are 6.5853 and 6.1212. There are 2 clusters with CCP$>5$, while there are 145 clusters in N3 totally. For N4, there are only 5 clusters with CCP$>5$, while there are 202 clusters in total. This fact shows that most clusters formed by CHIEF are very tight. As for CSP, the average value of CSP is much higher than CCP, which indicates that the distances between clusters are much longer. In total, there are 7 clusters in N1, 68 in N2, 145 in N3, 205 in N4, and 437 in N5.

In general, triangle motif clustering consumes less time comparing to other motifs. This is because of the efficiency of processing triangle motif over other structure during the process of searching for homogeneous motif structures. On the other hand, higher-order motif clustering is practicable in certain networks, especially in large-scale networks, is considered beneficial, whereas the speed of its motif clustering procedure can be accelerated using the proposed algorithm.

\section{Conclusion}

The increasing scale of big (social) networks has brought in challenges for analyzing data due to its complicated structure. We solve this problem from two perspectives. First, we first partition the network to better break up the complicate structure of big networks. Second, we focus on higher-order network motifs in big (social) networks instead of triangle motifs. In this work, we have proposed CHIEF that offers efficient and effectiveness motif clustering for big (social) networks. CHIEF includes two acceleration components, CHIEF-ST for accurate acceleration and CHIEF-AP for approximate acceleration. The accuracy of both variants has been proved theoretically. To examine the efficiency of CHIEF-AP experimentally, we implement CHIEF-AP on 5 different synthetic networks and 18 conferences and journals networks. Experimental results show that CHIEF-AP outperforms baseline methods in big networks. To date, motif clustering still cannot be totally solved by parallel algorithm. Therefore, the high efficiency and effectiveness of CHIEF provide a new perspective of motif clustering in big (social) networks. In our future work, we will also plough deep into parallel solutions for motif clustering in big networks.

\bibliographystyle{ACM-Reference-Format}
\bibliography{ref}


\end{document}